\documentclass[11pt,a4paper]{article}
\usepackage{jheppub}
\usepackage[english]{babel}
\usepackage[utf8]{inputenc}
\usepackage{graphicx}
\usepackage{amsmath,amssymb,amsfonts,amsthm}
\usepackage{bm,bbm}
\usepackage{tikz,stackrel}
\usepackage{pstricks}
\usepackage{pifont} 
\usepackage{empheq}
\usepackage[makeroom]{cancel}
\newcommand{\be}{\begin{eqnarray}}
\newcommand{\ee}{\end{eqnarray}}

\newcommand{\bea}{\begin{eqnarray}}
\newcommand{\eea}{\end{eqnarray}}

\newcommand{\bc}{\begin{center}}
\newcommand{\ec}{\end{center}}
\def\bs{\begin{subequations}}
\def\es{\end{subequations}}
\newcommand{\ben}{\begin{equation*}}
\newcommand{\een}{\end{equation*}}
\newcommand{\ban}{\begin{eqnarray*}}
\newcommand{\ean}{\end{eqnarray*}}
\newcommand\nn{\nonumber\\}

\def\k{\kappa}

\def\om{\omega}

\def\cL{\mathcal{L}}

\def\cob{\color{blue}}

\newcommand{\au}[2]{#1.~#2}
\newcommand{\book}[5]{\emph{#1}, #2, #3, #4 (#5)}

\newcommand{\oarX}[1]{\href{http://arxiv.org/abs/#1}{{\ttfamily\cob arXiv:#1}}}
\newcommand{\arX}[1]{\href{http://arxiv.org/abs/#1}{{\ttfamily\cob arXiv:#1}}}
\newcommand{\doin}[6]{\href{http://dx.doi.org/#1}{{\cob {\it #2} {\bf #3 #4} (#6) #5}}}
\newcommand{\doinn}[5]{\href{http://dx.doi.org/#1}{{\cob {\it #2} {\bf #3} (#5) #4}}}
\newcommand{\doij}[5]{\href{http://dx.doi.org/#1}{{\cob {\it #2} {\bf #3} (#5) #4}}}

\newcommand{\ndoinn}[5]{\href{#1}{{\cob {\it #2} {\bf #3} (#5) #4}}}

\newcommand{\tia}[1]{\textit{#1},}

\def\Re{{\rm Re}}

\newcounter{listcounter}

\allowdisplaybreaks

\begin{document}

\title{Quantum field theory with ghost pairs}

\author[a]{Jiangfan Liu,}
\affiliation[a]{Department of Physics, Southern University of Science
and Technology, Shenzhen 518055, China}

\author[a]{Leonardo Modesto,}

\author[b]{Gianluca Calcagni}
\emailAdd{g.calcagni@csic.es (corresponding author)}
\affiliation[b]{Instituto de Estructura de la Materia, CSIC, Serrano 121, 28006 Madrid, Spain}

\abstract{We explicitly show that general local higher-derivative theories with \emph{only} complex conjugate ghosts and normal real particles are unitary at any perturbative order in the loop expansion. The proof presented here relies on integrating the loop energies on complex paths resulting from the deformation of the purely imaginary paths, when the external energies are continued from imaginary to real values. Contrary to the case of nonlocal theories, where the same integration path was first proposed, for the classes of theories studied here the same procedure is not analytic, but the resulting theory is unitary and unique when the complex ghosts are present in pairs. As an explicit application, a special class of higher-derivative super-renormalizable or finite gravitational and gauge theories turns out to be unitary at any perturbative order if we exclude the complex ghosts from the spectrum of the theory, as it is normally accepted for Becchi--Rouet--Stora--Tyutin (BRST) ghosts. Finally, we propose an analogy between confined gluons in quantum Yang--Mills theory and classical complex pairs in local higher-derivative theories. According to such interpretation, complex ghosts will not appear on shell as asymptotic states because confined in what is natural to name ``ghostballs.''}

\preprint{\doij{10.1007/JHEP02(2023)140}{JHEP}{2302}{140}{2023} [\arX{2208.13536}]}

\maketitle

\tableofcontents


\section{Introduction}

In $1977$, K.\ Stelle proposed and extensively studied the most general four-derivative theory for pure gravity in four dimensions, namely \cite{Stelle}
\be\label{stelle}
\cL\propto R + \alpha\, R^2 + \beta\, {\rm Ric}^2 , 
\ee
where Ric is the Ricci tensor. Such theory shows good quantum properties such as renormalizability and asymptotic freedom \cite{frts82,avbar86,Avr86,Shapirobook}, but the presence of a massive spin-two ghost instability at the classical level makes the theory non-unitary in its original quantization based on the Feynman prescription \cite{Stelle}. However, recently a new quantum prescription, based on the Cutkosky, Landshoff, Olive, and Polkinghorne \cite{CLOP} approach to Lee--Wick theories \cite{LW1, LW2}, was proposed by Anselmi and Piva \cite{AnselmiPiva1, AnselmiPiva2}. This new prescription allows one to get rid of the ghost instability and the unitarity problem is solved at any perturbative order in the loop expansion \cite{AnselmiPiva3}. At the classical level, the ghost (which Anselmi and Piva named ``fakeon'' since it can only appear as a virtual or fake particle) is removed from the spectrum by solving its equations of motion by means of advanced plus retarded Green's functions. Consistently, also the homogeneous solution \cite{ClassicalPrescription1, ClassicalPrescription2}
is fixed to zero. This is the classical equivalent of removal of ghosts in the quantum theory from the spectrum of allowed asymptotic states. In this respect, we will make shortly a comparison with the BRST ghosts. 

The prescription described above is very general and can be applied not only to real as well as complex ghosts, but also to normal real particles. In particular, it can be applied to achieve perturbative unitarity in the theories with only complex conjugate poles proposed in \cite{LM-Sh, ModestoLW} and  based on the general higher-derivative (with more than four derivatives) action advanced in \cite{shapiro3}.
Thus, we end up with a large class of super-renormalizable or finite and unitary higher-derivatives theories of quantum gravity. In order to guarantee tree-level unitarity, the theory originally proposed in \cite{LM-Sh, ModestoLW} was designed to have only complex conjugate massive poles in the propagator, besides the massless graviton. However, any higher-derivative theory turns out to be unitary according to the Anselmi--Piva prescription (AP). Similarly to Stelle's theory, at the classical level the real ghosts (fakeons) are removed from the spectrum by solving their equations of motion by means of advanced plus retarded Green's functions and by fixing to zero the homogeneous solution \cite{ClassicalPrescription1, ClassicalPrescription2}.

It is worth mentioning that Stelle's theory \cite{Stelle} with the AP prescription \cite{AnselmiPiva1, AnselmiPiva2} is the only strictly renormalizable (i.e., not super-renormalizable or finite) and unitary theory of gravity in four dimensions. On the other hand, the theories proposed in \cite{shapiro3,ModestoLW} represent an infinite countable class of super-renormalizable or finite models for quantum gravity.
However, a new approach to the early Universe recently proposed requires the fundamental theory to be finite and conformal invariant at the classical as well as at the quantum level \cite{ConCos1,ConCos2}. Therefore, we believe that the theories beyond Stelle's are favored in such respect. 

Let us here expand on the uniqueness issue in local as well as nonlocal quantum gravity. What really matters in quantum gravity is not the specific theory, but the universality class similarly to condense matter physics. Indeed, in the presence of integer-order derivatives we recognize three equivalence classes: (i) the four-derivative theory with AP prescription, which can be asymptotically free when coupled to higher-spin matter; (ii) local higher-derivative or nonlocal theories that could be super-renormalizable or (iii) finite. In the former case, the theories can be asymptotically free, while in the latter case they are certainly conformal invariant. Therefore, we have two possible scenarios: (i) all the fundamental interactions vanish in the ultraviolet regime or (ii) we end in a Weyl conformal phase at high energy. We can also distinguish local from nonlocal theories on the basis of the analyticity of scattering amplitudes. Indeed, contrary to the nonlocal theories the amplitudes in local theories show up cusps of nonanalyticity at the creation threshold of virtual particles.

From the physical point of view, we would like to present local quantum gravity with the AP prescription in analogy with the discovery of antimatter. If we take as our starting point the wave equation, in the beginning only matter was known to theoretical physicists because the Galilean relativity principle was perfectly compatible with Schr\"odinger's wave equation, which is linear in the time derivative.
However, Klein and Gordon first and Dirac afterwards proposed a two-derivative theory for particles compatible with the special relativity principle. The outcome was the discovery of antimatter. In our case, we are going beyond two-derivative theories with wave equations that have more than two derivatives. The outcome is the discovery of a new form of matter that exists only off-shell: the above-mentioned \emph{fakeons} (see also the end of section \ref{concl} about the absence of fakeons in the spectrum of asymptotic states). Hence, we end up with a clear picture of all the theories with any number of derivatives that were discarded in the past because of the Ostrogradsky instability. 

A last comment is about the analogy with gauge theories and BRST invariance. In any gauge-invariant theory, the gauge-fixing procedure leads to BRST invariance and, as an unavoidable consequence, to the presence of BRST ghosts. However, nobody questions the choice of fix the ghost number to be zero. In other words, we always assume that BRST ghosts do not go on-shell, i.e., that they never show up as asymptotic states. However, why this is actually correct? Because in the cut diagrams the BRST ghosts cancel with the longitudinal and temporal nonphysical degrees of freedom of the gauge fields, and, hence, they cannot be regenerated at any loop order in the loop expansion. Therefore, it is consistent to remove such ghost states from the spectrum because they will never reappear at any loop order. The same argument applies to fakeons, which can be consistently removed from the spectrum of the asymptotic states because the AP prescription guarantees that they will never appear again at any perturbative loop order. 

In this paper, we focus on theories with only complex conjugate poles and we prove unitarity at any perturbative order, implementing the same procedure successfully applied to nonlocal field theories \cite{Briscese:2018oyx,Briscese:2021mob}. Such procedure consists of integrating the internal loop energies on paths, in the complex hyperplane, that emerge analytically when external energies move from purely imaginary to real values. 
Contrary to the case of nonlocal theories, {which are a completely independent proposal we will not consider here,} the same deformation of the path in local theories turns out to be a nonanalytic procedure because of the presence of complex conjugate poles, that do not allow to deform analytically the integration path in the loop amplitudes. Indeed, the extra complex poles (in the energy hyperplane) coming from different propagators pinch the integration path when the energies are real. 
Nevertheless, such singularities in the loop amplitudes are integrable, so that unitarity is fulfilled exactly because the ghosts appear in complex pairs.

\section{Special classes of higher-derivative theories}
In this section, we review three examples of local higher-derivative theories: a simple scalar, a gauge, and a gravitational higher-derivative field. However, for the sake of simplicity we will derive the Cutkosky rules only for the scalar field theory, while the generalization to the gauge and gravitational cases is straightforward making use of the Ward identities. The simplest higher-derivative scalar field theory with one real degree of freedom and a pair of complex ghosts is described by the following Lagrangian, 
\be
\mathcal{L}_{\phi} = -\frac{1}{2}  \phi \,  \left[ \left( \frac{\Box}{M^2} \right)^2 + 1 \right] 
\left(\Box +  m^2 \right)\phi  - \lambda \sum_{n=4}^{N} \frac{c_n}{n !} \phi^n\, , \quad N \in \mathbb{N} \, , 
\label{phin1}
\ee
where $M$ is a fundamental mass scale. A generalization of the theory is obtained by means of the following replacement in (\ref{phin1}), 
\be
\left[ \left( \frac{\Box}{M^2} \right)^2 + 1 \right] \quad \longrightarrow \quad {\rm P}(\Box) \, , 
\label{Pbox}
\ee
where ${\rm P}$ is a polynomial of integer degree $\gamma$ with only complex-conjugate zeros. The quantum theory becomes more convergent when increasing the degree $\gamma$ of the polynomial ${\rm P}$. Moreover, the tree-level unitarity is secured as proved in \cite{LM-Sh,ModestoLW}.  

A local super-renormalizable gauge theory with complex ghosts was proposed in \cite{ModestoLW, LM-Sh} as a special polynomial limit of nonlocal gauge and/or gravitational theories \cite{Krasnikov, kuzmin, modesto, review, modestoLeslaw, Calcagni:2014vxa}, 
and afterwards a generalization of such theory was shown to be finite in \cite{Universally, FiniteGaugeTheory}. The action for the finite nonlocal Yang--Mills theory in flat spacetime reads
\be \!\! 
{S}_{\rm YM} = -\frac{1}{2 g_{\rm YM}^2} \int d^D x \,  {\rm tr}\Big[ {\bf F} \, {\rm P}( {\cal D}^2_M)
{\bf F} + \frac{s_g}{M^4} {\bf F}^2 \left( {\cal D}^2_{M} \right)^{\gamma-2} {\bf F}^2 \Big]  \, ,  \quad \gamma \in 
\mathbb{N} \, , 
\label{gauge}
\ee
where we used the gauge-covariant box operator defined by ${\cal D}^2_M={\cal D}_\mu{\cal D}^\mu/M^2$, in which ${\cal D}_\mu$ is a gauge-covariant derivative (in the adjoint representation) acting on gauge-covariant field strength $ {\bf F}: = {F}_{\rho\sigma} = F_{\rho\sigma}^a T^a$ of the gauge potential $A_{\mu}$ (where $T^a$ are the generators of the gauge group in the adjoint representation). 
Moreover, $M$ is an extra, likely large, mass scale,  $s_g$ is a dimensionless parameter, and $\gamma$ is an integer that can be selected in order to have divergences only at one loop for a suitable choice of the degree $\gamma$ of ${\rm P}$. 

The action for gravity \cite{kuzmin,modesto,modestoLeslaw} consists of the Einstein--Hilbert term, a local operator that is quadratic in the Ricci tensor and Ricci scalar, and a potential at least cubic in the Riemann tensor, namely 
\be
S_{\rm g}  =   - \frac{1}{2\k^{2}}  \int  d^D x \sqrt{|g|} \left[ R + G_{\mu\nu} \gamma(\Box) R^{\mu\nu} + V(\mathcal{R}) \right] \, .
\label{gravity}
\ee
where $\k^2=8\pi G$ and $R$, $ R_{\mu\nu}$, and $G_{\mu\nu}$  are the Ricci scalar, Ricci curvature, and the  Einstein tensor, respectively. Moreover, $V(\mathcal{R})$ is a generalized potential and $\mathcal{R}$ stands for scalar, Ricci, or Riemann curvatures, and derivatives thereof. We work in signature $(+,-,\dots,-)$. The local form factor $\gamma(\Box)$ depends on the scale $M$ and is defined in terms of a polynomial ${\rm P}(\Box)$ by
\be
\gamma(\Box) = \frac{ {\rm P }(\Box) -1}{\Box} \, .
\ee
The minimal six-derivative super-renormalizable theory has been proposed for the first time in \cite{ModestoLW} and extensively studied at quantum level in \cite{Rachwal:2021bgb}. The action reads
\be
S_{\rm g}  =   - \frac{1}{2\k^{2}} \int  d^D x \sqrt{|g|} \left[ R + \frac{ G_{\mu\nu} \Box R^{\mu\nu}}{M^4}  + V(\mathcal{R}) \right] \, .
\label{gravityMin}
\ee

The propagators of the gauge and graviton fields are modified by the local form factors, but introducing only extra complex conjugate poles besides the real gauge fields and the graviton. {In the case of gauge theory, one adds a ghost action and a gauge-fixing term to the non-Abelian action (\ref{gauge}) in such a way as to guarantee BRST invariance \cite{Shapirobook}. Integrating out one of the fields, the total action reads
\be
{S}_{\rm YM} &=& -\frac{1}{2 g_{\rm YM}^2} \int d^D x \,  {\rm tr}\Big[F_{\mu\nu} \, {\rm P}({\cal D}^2_M)
F^{\mu\nu} + \frac{s_g}{M^4} F_{\mu\nu}F^{\mu\nu} ( {\cal D}^2_{M})^{\gamma-2} F_{\sigma\tau}F^{\sigma\tau}  \Big]\nonumber\\
&&+\frac{1}{g_{\rm YM}^2}\int d^D x \, \Big[\bar C_a {\rm P}({\cal D}^2_M)\partial^\mu {\cal D}^{ab}_\mu C_b-\frac{1}{2\xi_{\textsc{ym}}}(\partial^\mu A_\mu^a) {\rm P}({\cal D}^2_M)(\partial^\nu A_{\nu\, a})\Big]  \, , \label{acym}
\ee
where $C^a$ is a complex ghost and $\xi_{\textsc{ym}}$ is a gauge-fixing parameter. Therefore, in momentum space ($k_\mu$) the propagator for the gauge vector is 
\be
G(k)^{\rm YM}_{\mu\nu \, ab } = - i  \delta_{ab} \frac{1}{{\rm P}(k^2) \left( k^2 +  i \epsilon \right) } \left( \eta_{\mu\nu} -     \frac{k_\mu k_\nu}{k^2} \right) + {(-   \delta_{ab})} i \xi_{\textsc{ym}} \frac{k_\mu k_\nu}{\omega_{\textsc{ym}}(k^2) (k^2 { + i \epsilon})}
 \, , \label{NLPym}
\ee
where $\omega_{\textsc{ym}}(k^2)$ is a gauge-fixing weight function, while the ghost propagator is
\be
G(k)^C_{ab}=\delta_{ab}\frac{i}{{\rm P}(k^2) \left( k^2 +  i \epsilon \right) }\,. 
\label{GYM}
\ee
The case of the gravitational theory (\ref{gravity}) is very similar but with a more complicated index structure and the graviton propagator reads \cite{HigherDG}}
\be
G(k)^{\rm g} = i \frac{1}{{\rm P}(k^2) \left( k^2 +  i \epsilon \right) }  \left( P^{(2)} - \frac{1}{D-2} P^{(0)} \right)
+ 
i  \frac{\xi_{\rm g} (2P^{(1)} + \bar{P}^{(0)} ) }{2 (k^2  {+ i \epsilon}) \, \omega_{\rm g}( k^2)}  \, ,
\label{NLPg}
\ee
where $\xi_{\rm g}$ and $\omega_{\rm g}(k^2)$ are a gauge-fixing parameter and weight function, respectively \cite{modesto}. For the sake of simplicity, we omitted the four tensorial indices of the projectors $P^{(2)}$, $P^{(0)}$,  $P^{(1)}$ \cite{modesto}. The vertices in the gauge theory (\ref{gauge}) and in the gravitational theory (\ref{gravity}) are very involved in order to preserve gauge and general coordinate invariance. However, they are harmless to unitarity because both theories are gauge-invariant or BRST-invariant. Therefore, it will be straightforward to prove their unitarity once it is proved for the higher-derivative scalar field theory (\ref{phin1}).


\section{Cutkosky rules for a local scalar field theory with complex ghost pairs} \label{Cutkosky rules}
In this section, we compute the discontinuity of the amplitude for the scalar theory (\ref{phin1}) with $n=3$, namely, we focus on cubic interactions, at one and two loops, to finally state the general Cutkosky rules. The Cutkosky rules are an intermediate step towards the proof of unitarity and to formulate them we will need to calculate the imaginary part of the amplitudes.

The propagator in momentum space for the theory (\ref{phin1}) is easily obtained inverting the kinetic operator, and it reads
\be
G(k) = 
i \Delta_{\rm F}(k) = \frac{i M^4}{(k^2-m^2+i\epsilon) \left[ (k^2)^2+M^4 \right] } , 
\label{ScalarG}
\ee
where we have introduced the usual Feynman prescription for the real pole. We do not introduce any prescription for the complex conjugate poles, but we simply modify the integration contour on the energies (one by one) from a purely imaginary one when the external energies are purely imaginary to a deformed complex path when the external energies are changed to real ones.

\subsection{One-loop bubble diagram}
For the one-loop bubble diagram, the amplitude reads:
\be
\mathcal{M} &=& - \frac{i \lambda^2}{2}
\left[\frac{M^4}{M^4 + m^4}\right]^2
\int \frac{d^4 k}{(2\pi)^4}\frac{M^4}{(k^2-m^2+i\epsilon) \left[ (k^2)^2+M^4 \right]}\nonumber\\
&&\hspace{2.5cm} \qquad\times\frac{M^4}{[(p-k)^2-m^2+i\epsilon] \{[(p-k)^2]^2+M^4 \}} \, ,
\label{bubble}
\ee
where the two overall factors $M^4/(M^4 + m^4)$ come from the two external legs when the LSZ formula is employed. We will omit such constant from now on. 

\begin{figure}[h] 
\centering 
\includegraphics[width=1\textwidth]{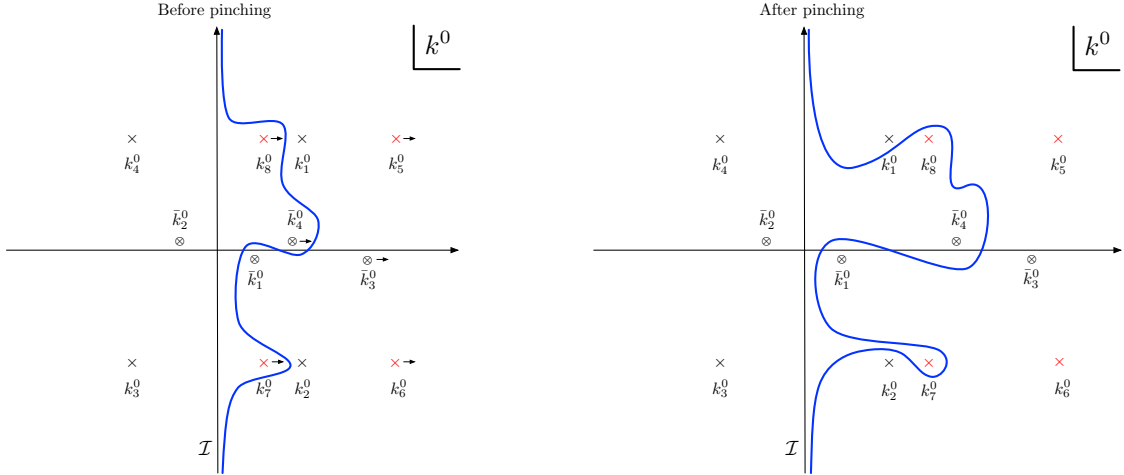}
\caption{Integration path for the one-loop bubble diagram in the complex energy plane $k^0$ before and after the pinching. The poles $k^0_8$ and $k^0_7$ move from the left to the right (left panel) till they pinch the integration path and they swap with the poles $k^0_1$ and $k^0_2$, respectively (right panel). As proven in the main text, the pinching singularity is integrable and does not give any contribution to the imaginary part of the amplitude, but it contributes to the real part, as evident in Figs.~\ref{Real_amplitude} and \ref{m=M=1}. Notice that also $\bar{k}^0_3$ and $\bar{k}^0_4$ move towards the right.}
\label{integration_path}
\end{figure}

The one-loop integral (\ref{bubble}) shows a total number of twelve poles in the $k^0$-plane. 
Six poles come from the first propagator in (\ref{bubble}), and they read 
\be
\bar{k}_{1,2}^0 = \pm\sqrt{\bm{k}^2+m^2-i\epsilon} \, , \qquad 
k_{1,2}^0  = \sqrt{\bm{k}^2\pm iM^2} \, , \qquad 
k_{3,4}^0 = - \sqrt{\bm{k}^2\pm iM^2} \, ,
\label{fixedPoles}
\ee
where normal real poles are denoted by $\bar{k}_i^0$ and complex poles are denoted simply by $k_i^0$. 
All the above poles (\ref{fixedPoles}) are fixed in the $k^0$-plane when we change the external energy $p^0$ from purely imaginary to real. On the other hand, the poles corresponding to the second propagator move in the $k^0$-plane when $p^0$ becomes real. Such poles are
\be
&&\bar{k}_{3,4}^0 = p^0\pm\sqrt{(\bm{p}-\bm{k})^2+m^2-i\epsilon} \, , \nonumber\\ 
&&k_{5,6}^0 = p^0+\sqrt{(\bm{p}-\bm{k})^2\pm iM^2} \,  , \\
&&k_{7,8}^0 = p^0-\sqrt{(\bm{p}-\bm{k})^2\pm iM^2} \, .\nonumber 
\ee
For the sake of simplicity, we can assume the external momentum to be $\bm{p} = 0$.
In the continuation of the external energy from purely imaginary to real values, at some point the pole $k_7^0$ coincides with the pole $k_2^0$ and, at the same time, the pole $k_8^0$ coincides with the pole $k_1^0$. At such moment, the integration path $\mathcal{C}$ is pinched and the amplitude becomes nonanalytic. In the left panel of figure \ref{integration_path}, we show the positions of the poles before the pinching, while in the right panel we display the poles after the pinching. For $\bm{p} \neq 0$, the position of the poles 
changes, but the pinching is unavoidable. 


%
According to figure\ref{integration_path} and Cauchy's theorem, the integral along the path in the figure consists of three contributions: the integration along the imaginary axis, one residue contribution coming from one of the moving normal real poles, namely $\bar{k}_{4}^0$, and the residue contribution coming from the left moving complex pair $k_{7,8}^0$.  Indeed, 
\be
&& \int_{+ i \infty}^{- i \infty} \left( \, \dots \right)  +  \int_{\mathcal C} \left( \, \dots \right) = (2 \pi i ) \, {\rm Res} (\bar{k}_{4}^0) 
+ (2 \pi i ) \, {\rm Res} ({k}_{7}^0) + (2 \pi i ) \, {\rm Res} ({k}_{8}^0) \, , \nonumber \\
&& 
 \int_{\mathcal C} \left( \, \dots \right) = \int_{- i \infty}^{+ i \infty} \left(\, \dots \right) + 
 (2 \pi i ) \, {\rm Res} (\bar{k}_{4}^0) 
+ (2 \pi i ) \, {\rm Res} ({k}_{7}^0) + (2 \pi i ) \, {\rm Res} ({k}_{8}^0) \, .
\ee
Schematically, we can write the amplitude as
\be
\mathcal{M}=\mathcal{M}_{\rm I} +\mathcal{M}_{\rm ResR}+\mathcal{M}_{\rm ResC} \, . 
\label{Cauchy1L}
\ee

\subsubsection{Imaginary part}

Let us now compute the imaginary part of the above integral (\ref{Cauchy1L}). 

\underline{\bf Amplitude $\mathcal{M}_{\rm I}$} --- The integral $\mathcal{M}_{\rm I}$ is purely real because all the poles are far from the imaginary axis. For the sake of simplicity, we take $\bm{p} = 0$ and we use indiscriminately $p^0 = p_0$ thanks to our choice of metric signature ($\eta_{00}=+1$). Moreover, we replace $k^0 = i k_4$. Hence, the integral $\mathcal{M}_{\rm I}$ reads
%
%
%
\be
\mathcal{M}_{\rm I}   &=&  - \frac{i \lambda^2 M^8}{2(2\pi)^4}\int i d k_4 d^3{k} \frac{1}{(-k_4^2-\bm{k}^2-m^2+i\epsilon)\left[ (-k_4^2-\bm{k}^2)^2+M^4 \right] } \nonumber \\
&&\times \frac{1}{p_0^2-k_4^2-\bm{k}^2-m^2+i(\epsilon-2 p_0 k_4)}\nonumber\\
&&\times \frac{1}{(p_0^2 - k_4^2-\bm{k}^2)^2-(2  p_0 k_4)^2+M^4-i \left[ 2(p_0^2-k_4^2-\bm{k}^2)(2 p_0 k_4) \right]}\,.
\label{Im01}
\ee
Taking the limit for $\epsilon \rightarrow 0$ (we will come back to this point after showing the result),
\be
\mathcal{M}_{\rm I}  
& \underset{\epsilon\rightarrow 0}{=} & \frac{\lambda^2 M^8}{2(2\pi)^4}\int d k_4 d^3{k} \frac{1}{(- k_4^2-\bm{k}^2-m^2)\left[(-k_4^2-\bm{k}^2)^2 + M^4 \right]} \nonumber\\
&& \hspace{0.1cm} \times 
 \frac{p_0^2 - k_4^2-\bm{k}^2-m^2+i(2 p_0 k_4)}{(p_0^2-k_4^2-\bm{k}^2-m^2)^2+(2 p_0 k_4)^2} 
\nonumber 
\\
&&
\hspace{0.1cm} 
%
\times
\frac{(p_0^2-k_4^2-\bm{k}^2)^2 - (2 p_0 k_4)^2+M^4+i \left[ 2( p_0^2-k_4^2 - \bm{k}^2)(2 p_0 k_4) \right]}{\left[ (p_0^2-k_4^2-\bm{k}^2)^2-(2 p_0 k_4)^2+M^4 \right]^2+ \left[2(p_0^2-k_4^2-\bm{k}^2)(2 p_0 k_4) \right]^2} \, . 
\label{Again}
\ee
We notice that the imaginary part of the integral above is an odd function of $k_4$. Therefore, for $\epsilon = 0$ 
\be
{\rm Im} \,\mathcal{M}_{\rm I}= 0
\ee
 because we are integrating an odd function of $k_4$ on an even domain. For $\epsilon \neq 0$ the imaginary part is non zero, but we can safely take the limit $\epsilon\rightarrow 0$ in (\ref{Im01}). Indeed, the first denominator is never zero because $k_4^2 +\bm{k}^2 + m^2$ is always positive, while the third denominator for $k_4 \neq 0$ is never zero because it has a nonvanishing imaginary part. On the other hand, the $k_4 = 0$ case is a zero-measure contribution to the integral. Let us expand on the latter case and focus on the third denominator in (\ref{Im01}), which we can rewrite in the following way:
 \be
 && \int_{- \infty}^{-\tilde{\epsilon} } \frac{dk_4  \,  (  \, \dots )}{p_0^2-k_4^2-\bm{k}^2-m^2+i(\epsilon-2 p_0 k_4)} 
 + \int_{- \tilde{\epsilon} }^{+ \tilde{\epsilon}}  \frac{dk_4 \,  (  \, \dots )}{p_0^2-k_4^2-\bm{k}^2-m^2+i(\epsilon-2 p_0 k_4)} 
 \nonumber\\
&&
 + \int_{+ \tilde{\epsilon} }^{+ 
\infty }  \frac{dk_4 \,  (  \, \dots )}{p_0^2-k_4^2-\bm{k}^2-m^2+i(\epsilon-2 p_0 k_4)}  \, . 
\label{IntegralJB}
 \ee
In the first and third integral, we can take $\epsilon = 0$ because $k_4$ is nonvanishing and, then, the denominator is never zero, too, since it has a nonvanishing imaginary contribution $- 2  i \epsilon  p_0 k_4$. Hence, the integral above (\ref{IntegralJB}) 
simplifies to
 \be
&&  
\int_{- \infty}^{-\tilde{\epsilon} }  \frac{dk_4 \, ( \, \dots ) }{p_0^2-k_4^2-\bm{k}^2-m^2 - i 2 p_0 k_4} 
  + \int_{- \tilde{\epsilon} }^{+ \tilde{\epsilon}}  \frac{dk_4 \, ( \, \dots ) }{p_0^2-k_4^2-\bm{k}^2-m^2+i(\epsilon-2 p_0 k_4)}  
 \nonumber \\
&&
 + \int_{+ \tilde{\epsilon} }^{+ 
\infty }  \frac{dk_4 \, ( \, \dots )  }{p_0^2-k_4^2-\bm{k}^2-m^2 - i 2 p_0 k_4}  . 
\label{ZeroM}
 \ee
On the other hand, for $\epsilon \rightarrow 0$ the second integral in (\ref{ZeroM}) reads
\be
&& \lim_{\tilde{\epsilon} \rightarrow 0 } 
\lim_{{\epsilon} \rightarrow 0}  \int_{- \tilde{\epsilon} }^{+ \tilde{\epsilon}}  \frac{dk_4 \,  ( \, \dots )}{p_0^2-k_4^2-\bm{k}^2-m^2+i(\epsilon-2 p_0 k_4)} \nonumber \\
&&\qquad
=  \lim_{\tilde{\epsilon} \rightarrow 0 } 
\lim_{{\epsilon} \rightarrow 0}  \int_{- \tilde{\epsilon} }^{+ \tilde{\epsilon}} d k_4 \frac{ p_0^2-k_4^2-\bm{k}^2-m^2 - i(\epsilon-2 p_0 k_4) }{(p_0^2-k_4^2-\bm{k}^2-m^2)^2 + (\epsilon-2 p_0 k_4)^2}  ( \, \dots ) \nonumber \\
&& \qquad=  
\lim_{\tilde{\epsilon} \rightarrow 0 } 
 \, 2 \tilde{\epsilon} 
  \left[ {\rm PV}  \frac{1}{p_0^2 -\bm{k}^2-m^2} - i \pi \delta(p_0^2 - \bm{k}^2-m^2)
  \right] 
  \, ( \dots )\Big|_{k_4 =0} = 0 \, , 
\ee
where in the next-to-last step we evaluated the integrand in $k_4=0$ for $\tilde{\epsilon}  \rightarrow 0 $, and at the last step we made use of the following identity for distributions: $\lim_{\tilde{\epsilon}  \rightarrow 0} \tilde{\epsilon} \, \delta(z) = 0$. 
Indeed, in the space of distributions for finite $f(0)$
\be
\lim_{\tilde{\epsilon}  \rightarrow 0}  \int dz \, \tilde{\epsilon} \, \delta(z) \, f(z) = \lim_{\tilde{\epsilon}  \rightarrow 0} \tilde{\epsilon} \, f(0) = 0 \, .
\ee
Therefore, the second integral does not contribute to (\ref{ZeroM}), and we can replace (\ref{ZeroM}) with:
\be
\int_{- \infty }^{+ \infty}  \frac{dk_4 }{p_0^2-k_4^2-\bm{k}^2-m^2 - i 2 p_0 k_4} ( \, \dots ) \, .
\ee
In other words, we can take $\epsilon =0$ in the third denominator of (\ref{Im01}). Coming back to the first denominator in
(\ref{Im01}), but for $m=0$, in a similar manner we can show that the contribution to the integral in $k_4$ is of zero measure. 

\underline{\bf Amplitude $\mathcal{M}_{\rm ResR}$} --- 
We now evaluate the contribution of 
$\mathcal{M}_{\rm ResR}$ to the imaginary part of the integral. 
For that purpose, we have to pick up the residue at the pole $\bar{k}_{4}^0$, 
namely 
\be
\mathcal{M}_{\rm ResR} &=& 
(2\pi i) \, {\rm Res} (\bar{k}_{4}^0) \, \sigma(\Re\,\bar{k}_{4}^0) \nonumber \\
&=& 
\frac{\lambda^2 M^8}{2} \!\! \int \!\! \frac{d^3 k}{(2\pi)^3} 
\,
\frac{\sigma(\Re\,\bar{k}_{4}^0)}{p_0^2 - 2 p_0 \sqrt{\bm{k}^2 + m^2-i\epsilon}} 
\,
\frac{1}{(p_0^2-2 p_0 \sqrt{\bm{k}^2+m^2}+m^2)^2+M^4}\nonumber \\
  &&\times
 \frac{1}{-2\sqrt{\bm{k}^2+m^2}}
 \,\,
  \frac{1}{m^4+M^4} \nonumber \\
  &=&
 \frac{\lambda^2 M^8}{2} \!\! \int \! \! \frac{4 \pi {\rm k}^2 d {\rm k}}{(2\pi)^3} 
\,
\frac{\sigma(\Re\,\bar{k}_{4}^0)}{p_0^2 - 2 p_0 \sqrt{\bm{k}^2+m^2-i\epsilon}} 
\,\,
\frac{1}{(p_0^2 - 2 p_0 \sqrt{\bm{k}^2+m^2}+m^2)^2+M^4}\nonumber \\
  &&\times
 \frac{1}{- 2\sqrt{\bm{k}^2+m^2}}
 \,\,
  \frac{1}{m^4 + M^4}  ,
  \label{MResR0}
  \ee
  where $\sigma$ is the Heaviside step function and in the last line we used spherical coordinates and defined the radial coordinate ${\rm k} \equiv |\bm{k}|$ in momentum space. 
Making the following change of variables in the integral (\ref{MResR0}), 
\be
E=\sqrt{\bm{k}^2+m^2} \, , \quad dE= \frac{{\rm k}}{E}\,d {\rm k} \, ,  \quad {\rm k} \equiv |\bm{k} | \, ,
\label{CoV}
\ee
the integral (\ref{MResR0}) turns into
\be
\mathcal{M}_{\rm ResR} 
 &=& - \frac{\lambda^2 M^8}{2(2\pi)^2} \frac{1}{m^4+M^4}   \!\!    \int_m^\infty \!\! dE \frac{\sqrt{E^2-m^2}}{ p_0 ( p_0 - 2 E + i \epsilon)} \,\, \frac{\sigma(p_0 - E)}{(p_0^2 - 2 p_0 E+ m^2)^2+M^4} \nonumber \\
&=& - \frac{\lambda^2 M^8}{2(2\pi)^2} \frac{1}{m^4+M^4}  \!\!   \int_m^{p_0} \!\!\!\!   dE \left\{ {\rm PV} \! 
\left[ \frac{\sqrt{E^2 -  m^2}}{p_0 (p_0 - 2E)} \,\,  \frac{1}{(p_0^2- 2 p_0 E+m^2)^2+M^4} \right]\right. \nonumber \\
  &&\left.- i \pi \frac{\sqrt{E^2 - m^2}}{p_0} \frac{\delta(p_0 - 2E)}{(p_0^2 -2 p_0 E + m^2)^2+M^4}\right\}
\nonumber \\
&=& - \frac{\lambda^2 M^8}{2(2\pi)^2} \frac{1}{m^4 + M^4}  \!\!   \int_m^{p_0} \!\!\!\!   dE \left\{ {\rm PV} \! 
\left[ \frac{\sqrt{E^2 -  m^2}}{p_0 (p_0 - 2E)} \,\,  \frac{1}{(p_0^2 - 2 p_0 E + m^2)^2 + M^4} \right]\right. \nonumber \\
  &&\left.  
- i \pi \frac{\sqrt{E^2 - m^2}}{p_0} \frac{\delta[- 2 (E - p_0/2)]}{(p_0^2 -2 p_0 E + m^2)^2 + M^4}\right\}  .
\nonumber 
\ee
Evaluating the second integrand in $E$,
\be
\mathcal{M}_{\rm ResR} 
& = & - \frac{\lambda^2 M^8}{2(2\pi)^2} \frac{1}{m^4+M^4}  \!\!  \int_m^{p_0} \!\!\!\!   dE \,\, {\rm PV} \! 
\left[ \frac{\sqrt{E^2 -  m^2}}{p_0(p_0 - 2E)} \,\,  \frac{1}{(p_0^2 - 2 p_0 E+m^2)^2+M^4} \right]  
\nonumber \\
&&
\hspace{0cm} 
+ 
\left(- \frac{\lambda^2 M^8}{2(2\pi)^2} \frac{1}{m^4+M^4} \right) 
(- i \pi ) \frac{\sqrt{\left(\frac{p_0}{2} \right)^2 - m^2}}{2 p_0} 
\frac{1}{ \left( \cancel{p_0^2} - \cancel{2 p_0 \, p_0/2 }+ m^2 \right)^2+M^4} \, .\nonumber\\
\ee
Finally, 
\be
\mathcal{M}_{\rm ResR} 
& = & - \frac{\lambda^2 M^8}{2(2 \pi )^2} \frac{1}{m^4+M^4} \!\! 
\int_m^{p_0} \!\!\!\!  dE  \,  {\rm PV}\! \left[ \frac{\sqrt{E^2-m^2}}{p_0(p_0-2E)} \frac{1}{(p_0^2-2 p_0 E+m^2)^2+M^4} \right]  
\nonumber \\
&&
+ i  \frac{\lambda^2}{32\pi}\frac{M^8}{(m^4+M^4)^2}\sqrt{1-\frac{4 m^2}{p_0^2}}\sigma(p-2m). 
\label{DiscRImplicit}
\ee
Therefore, the imaginary part of $\mathcal{M}_{\rm ResR}$ can be written as:
\be
{\rm Disc} \,\mathcal{M}_{\rm ResR} &\equiv& 2 i\, {\rm Im} \,\mathcal{M}_{\rm ResR}\nonumber\\
&=& - \frac{i \lambda^2}{2}\frac{M^8}{(m^4+M^4)^2} \int \frac{d^4 k}{(2\pi)^4} (-2\pi i) \delta^{(4)}(k^2-m^2) (-2\pi i) \delta^{4}[(p-k)^2-m^2] \,,\nonumber\\
\label{DiscR}
\ee
where ``Disc'' stands for discontinuity. For the sake of completeness, we now show that the second contribution to (\ref{DiscRImplicit}) and (\ref{DiscR}) are equivalent, i.e.,
\be
{\rm Disc} \,\mathcal{M}_{\rm ResR} & = & - \frac{i \lambda^2}{2}\frac{M^8}{(m^4+M^4)^2} \int \frac{d^4 k}{(2\pi)^4} (-2\pi i) \delta(k^2-m^2) (-2\pi i) \delta[(p-k)^2-m^2]
\nonumber \\
& = & - \frac{i \lambda^2}{2}\frac{M^8}{(m^4+M^4)^2}  \int \frac{d^4 k}{(2\pi)^4} (-2\pi i)^2 \delta(k_0^2 - E^2) \delta[(p_0-k_0)^2 - E^2]
\nonumber \\
&= & \frac{i \lambda^2}{2}\frac{4 \pi^2 M^8}{(m^4+M^4)^2}  \int \frac{d^4 k}{(2\pi)^4} \frac{1}{4 E^2}\nonumber\\
&&\qquad\times \left[ \delta(k_0 + E) + \delta(k_0 - E) \right] \left[ \delta(p_0 - k_0 + E) + \delta(p_0 - k_0 - E) \right]
\nonumber \\
&= & \frac{i \lambda^2}{2}\frac{4 \pi^2 M^8}{(m^4+M^4)^2}  \int \frac{d^4 k}{(2\pi)^4} \frac{1}{4 E^2}\nonumber\\
&&\qquad\times  \Big[ \delta(k_0 + E) \delta(p_0 - k_0 + E) + \delta(k_0 + E) \delta(p_0 - k_0 - E) 
\nonumber \\
&&\qquad\quad
 + \delta(k_0 - E) \delta(p_0 - k_0 + E) + \delta(k_0 - E) \delta(p_0 - k_0 - E)  \Big] \, . 
\ee
Since $p_0 > 0$, the above integral simplifies to:
\be
{\rm Disc} \,\mathcal{M}_{\rm ResR} & = & \frac{i \lambda^2}{2}\frac{4 \pi^2 M^8}{(m^4+M^4)^2}  \int \frac{d^3 k}{(2\pi)^4} \frac{1}{4 E^2}\nonumber\\
&&\qquad\times \left[ \cancel{\delta(p_0 + 2 E)} + \cancel{\delta(p_0)} + \cancel{\delta(p_0)} + \delta(p_0 - 2 E)  \right] 
\nonumber \\
& \underset{ (\ref{CoV}) }{=} & \frac{i \lambda^2}{2}\frac{4 \pi^2 M^8}{(m^4+M^4)^2}  \int \frac{d E}{(2\pi)^4} \frac{4 \pi \sqrt{E^2 - m^2}}{4 E} \frac{\delta(p_0 /2 - E)}{2}
\nonumber \\
&= &  i  \frac{\lambda^2}{16\pi}\frac{M^8}{(m^4+M^4)^2}\sqrt{1-\frac{4 m^2}{p_0^2}}\sigma(p_0 - 2m)
\nonumber \\
&= & 2 i\, {\rm Im} \,\mathcal{M}_{\rm ResR} \, . 
\label{DiscResR}
\ee

It deserves to be noticed that the real part of the amplitude $\mathcal{M}_{\rm ResR}$ [namely the first integral in (\ref{DiscRImplicit})] is singular in $E = p_0/2$. However, such singularity can be integrated splitting the integration domain into three regions: (here $\varepsilon$ is not the same of the Feynman prescription at the real pole):
\be
 \int_m^\infty dE \left( \dots \right) 
 = \int_m^{p_0/2 - \varepsilon} dE \left( \dots \right) 
 + \int_{p_0/2 -\varepsilon}^{p_0/2 +\varepsilon} dE  \left( \dots \right) 
 +\int_{p_0/2+ \varepsilon}^\infty dE  \left( \dots \right) \, .
 \label{IntSingR}
 \ee
It turns out that the second integral above is zero when properly computed by the mean of the principal value. Indeed, we can explicitly perform the first integral in (\ref{DiscRImplicit}):
\be
\int_{p_0/2 -\varepsilon}^{p_0/2 +\varepsilon} dE  \left( \dots \right) = \int_{p_0/2 - \varepsilon}^{p_0/2 + \varepsilon} dE \frac{\sqrt{E^2 -  m^2}}{p_0(p_0 - 2E)} \,\,  \frac{1}{(p_0^2 - 2 p_0 E + m^2)^2 + M^4} \, , 
\ee
which for $E \simeq p_0/2$ can be approximated with (we make the replacement $E \simeq p_0/2$ everywhere except when the integral is singular)
\be
\int_{p_0/2 -\varepsilon}^{p_0/2 +\varepsilon} dE  \left( \dots \right) 
& \underset{E \simeq p_0/2}{ = }& \int_{p_0/2 - \varepsilon}^{p_0/2 + \varepsilon} dE \frac{\sqrt{(p_0/2)^2 -  m^2}}{p_0(p_0 - 2E)} \,\,  \frac{1}{(\cancel{p_0^2} - \cancel{2p_0(p_0/2)} + m^2)^2 + M^4}
\nonumber \\
& \!\! = \!\! & \frac{\sqrt{(p_0/2)^2 -  m^2}}{-2 p_0 (m^4 +M^4)} \int_{p_0/2 - \varepsilon}^{p_0/2 + \varepsilon} dE \frac{1}{E - p_0/2}
\nonumber \\
& \!\! = \!\! & \frac{\sqrt{(p_0/2)^2 -  m^2}}{-2 p_0 (m^4 +M^4)}  \ln \frac{| \varepsilon |}{| - \varepsilon |} = 0 \, .
\ee

\underline{\bf Amplitude $\mathcal{M}_{\rm ResC}$} --- 
In order to get $\mathcal{M}_{\rm ResC}$, we have to pick up the residues at the poles in $k_{7,8}^0$:
\be
\mathcal{M}_{\rm ResC} &=  - \frac{i \lambda^2}{2} \,  (2 \pi i )
 \left[ {\rm Res}(k_{7}^0) \, \sigma(\Re\,k_{7}^0)+{\rm Res} (k_{8}^0) \, \sigma(\Re\,k_{8}^0) \right] \, . 
\ee
Let us now compute the two contributions above one by one. The first residue reads
\be
&& \hspace{-1cm}
 {\rm Res} (k_{7}^0) =M^8 \int \frac{d^3 k}{(2\pi)^4} \frac{1}{p_0^2-2 p_0 \sqrt{\bm{k}^2+iM^2}+iM^2-m^2}\nonumber \\
&&
\qquad\times \frac{1}{\left( p_0^2-2 p_0\sqrt{\bm{k}^2+ i M^2} + i M^2 \right)^2+M^4} 
\,\, \frac{1}{iM^2-m^2} 
\,\, \frac{1}{-4iM^2\sqrt{\bm{k}^2+iM^2}} \, .
\label{ReSette}
\ee
The second residue reads
\be
&& \hspace{-1cm}
 {\rm Res} (k_{8}^0) = M^8 \int \frac{d^3{k}}{(2\pi)^4} \frac{1}{p_0^2- 2 p_0 \sqrt{\bm{k}^2 - i M^2}-i M^2 - m^2}
\nonumber \\
&&\qquad \times \frac{1}{\left( p_0^2 - 2 p_0 \sqrt{\bm{k}^2 - i M^2} - i M^2 \right)^2 + M^4} 
\,\, \frac{1}{- i M^2-m^2}
\,\,  \frac{1}{4 i M^2\sqrt{\bm{k}^2-iM^2}} \, . 
\label{ReOtto}
\ee
Since ${\rm Res}(k_{7}^0)$ and ${\rm Res}(k_{8}^0)$ are complex conjugate functions, 
it turns out that the sum of the two residues is purely real, namely 
\be
\mathcal{M}_{\rm ResC} = \frac{ \lambda^2}{2} \,(2 \pi) \, 
2 \, {\rm Re}  \,  {\rm Res} ( k_{7}^0 ) \quad {\rm or} \quad 
\mathcal{M}_{\rm ResC} = \frac{ \lambda^2}{2} \,(2 \pi) \, 
2 \, {\rm Re} \,  {\rm Res} ( k_{8}^0 ). 
\ee
Finally, 
\be
\hspace{-1cm} 
 \mathcal{M}_{\rm ResC} &=& \lambda^2M^8 \, {\rm Re} \Bigg\{\int \frac{d^3 k}{(2\pi)^3} 
 \frac{\sigma(\Re\,k_{7}^0)}{p_0^2 - 2 p_0 \sqrt{\bm{k}^2 + i M^2} + i M^2-m^2} 
 \nonumber \\
\hspace{-1cm}&& \times \frac{1}{\left( p_0^2-2 p_0 \sqrt{\bm{k}^2+iM^2} + i M^2 \right)^2+M^4} 
\,\, \frac{1}{iM^2-m^2}
\,\,
 \frac{1}{-4 i M^2\sqrt{\bm{k}^2 + i  M^2}} \Bigg\}.
 \label{MC}
\ee

 So far we have showed that only the real ``normal'' particle contributes to the imaginary part of the amplitude:
\be
\hspace{-0.8cm}{\rm Disc} \,\mathcal{M} &=& 2 i {\rm Im} \,\mathcal{M}  =
 2 i {\rm Im}(  \mathcal{M}_{\rm I} +\mathcal{M}_{\rm ResR}+\mathcal{M}_{\rm ResC}  ) =
 2 i {\rm Im} \,\mathcal{M}_{\rm ResR}
 \nonumber \\
& = &  - i \frac{\lambda^2}{2}\frac{M^8}{(m^4+M^4)^2} \int \frac{d^4 k}{(2\pi)^4} (-2\pi i)\, \delta(k^2-m^2) (-2\pi i) \delta[(p-k)^2-m^2] \, .
\ee
However, the amplitudes (\ref{ReSette}) and (\ref{ReOtto}) are not analytic because of the pinching singularity that occurs for $k_7^0 = k_2^0$ or $k_8^0 = k_1^0$. 

Since ${\rm k}$ and $p_0$ are real, the first and third denominators in (\ref{MC}) are regular everywhere, while the second denominator in (\ref{MC}) is singular in ${\rm k} = |\bm{k}| =\sqrt{p_0^4-4M^4}/2p_0$ (pinching singularity), as we are going to show. The second denominator reads:
\be
\hspace{-0.4cm}0&=& \left( p_0^2-2 p_0 \sqrt{\bm{k}^2+ i M^2} + i M^2 \right)^2 + M^4 \nonumber \\
&=& \left[ \left( p_0^2-2 p_0 \sqrt{\bm{k}^2 + i M^2} + i M^2 \right)  + i M^2  \right]
\left[ \left( p_0^2-2 p_0 \sqrt{\bm{k}^2+ i M^2} + \cancel{i M^2} \right)  - \cancel{i M^2}  \right].\nonumber\\
\ee
Therefore, the only real solution for ${\rm k}$ is
\be
{\rm k} = \frac{\sqrt{p_0^4-4 M^4}}{2 p_0}. 
\ee
However, we can prove that such singularity is integrable by dividing the whole integration region in (\ref{MC}) into three sectors (here $\varepsilon$ is not the same of the Feynman prescription):
 \be
 \int_0^\infty d {\rm k} \left( \dots \right) 
 = \int_0^{\alpha-\varepsilon } d {\rm k} \left( \dots \right) 
 + \int_{\alpha-\varepsilon}^{\alpha+\varepsilon} d {\rm k}  \left( \dots \right) 
 +\int_{\alpha+\varepsilon}^\infty d {\rm k}  \left( \dots \right) \, , \qquad \alpha = \frac{\sqrt{p_0^4-4 M^4}}{2 p_0} \, .\nonumber\\
 \label{IntSingC}
 \ee
The integrand is well defined in the first and third integration domain, but it meets a singularity in the second region. 
However, when we send $\varepsilon$ to zero, such integrand is antisymmetric with respect to the point ${\rm k} = \alpha$, and, thus, it vanishes. On the other hand, the first and third integrals can be computed numerically taking $\varepsilon$ to be smaller and smaller and showing that the result is independent of such parameter (we got the same result varying in the range $\varepsilon= 10^{-10}\!-\!10^{-20}$. 

Let us now show that the integral $\int_{\alpha-\varepsilon}^{\alpha+\varepsilon} d {\rm k}  \left( \dots \right)$ in (\ref{IntSingC}) 
is zero. 
Around the pinching at ${\rm k} = \alpha$, the radial variable ${\rm k}$ in the integrand can be replaced by $\alpha$ everywhere except in the second denominator of (\ref{MC}), which turns out to be singular. Therefore, the integral simplifies to
\be
 \int_{\alpha-\varepsilon}^{\alpha+\varepsilon} d {\rm k} ( \dots ) &\propto& 
 \int_{\alpha-\varepsilon}^{\alpha+\varepsilon} d {\rm k} \frac{1}{\left( p_0^2-2 p_0\sqrt{\bm{k}^2+iM^2} + i M^2 \right)^2 + M^4} 
 \nonumber \\
&=&\int_{\alpha-\varepsilon}^{\alpha+\varepsilon} d {\rm k} \,
\frac{1}{\left( p_0^2 - 2 p_0 \sqrt{\bm{k}^2 + i M^2} + i M^2 \right)  + i M^2}\nonumber\\
&&\qquad\qquad\times\frac{1}{\left( p_0^2-2 p_0 \sqrt{\bm{k}^2+ i M^2} + \cancel{i M^2} \right)  - \cancel{i M^2}} \, .
\label{InterC}
%
\ee
Rationalizing the first term at the denominator of (\ref{InterC}), 
\be
&& \hspace{-1.2cm} 
 \int_{\alpha-\varepsilon}^{\alpha+\varepsilon} d {\rm k} ( \dots ) \propto 
 \int_{\alpha-\varepsilon}^{\alpha+\varepsilon} d {\rm k} \, 
\frac{1}{p_0^2 - 2 p_0 \sqrt{{\rm k}^2+iM^2}+2iM^2} 
\,\, 
\frac{1}{p_0^2 - 2 p_0 \sqrt{{\rm k}^2+iM^2}}
\nonumber \\
&& \hspace{1cm}
= \int_{\alpha-\varepsilon}^{\alpha+\varepsilon} d {\rm k} \, 
\frac{p_0^2 + 2 i M^2 + 2 p_0 \sqrt{{\rm k}^2 + i M^2}}{(p_0^2 + 2 i M^2)^2 - 4 p_0^2 ( {\rm k}^2 + i M^2 )} 
\, \,
\frac{1}{p_0^2 - 2 p_0 \sqrt{{\rm k}^2+iM^2}}    
\nonumber \\
&&\hspace{1cm}
 = \int_{\alpha-\varepsilon}^{\alpha+\varepsilon} d {\rm k} \, \frac{p_0^2 + 2 i M^2 + 2 p_0 \sqrt{{\rm k}^2+iM^2}}{p_0^4 - 4 M^4 - 4 p_0^2 {\rm k}^2} 
\,\, 
\frac{1}{p_0^2 - 2 p_0\sqrt{{\rm k}^2+iM^2}} \, .
\ee
Performing the change of variables ${\rm k} \rightarrow {\rm k}' = {\rm k}^2$, $d {\rm k}' = 2 \, {\rm k} \,d {\rm k}$, and introducing the new small parameter $\varepsilon'=2\alpha\varepsilon > 0$, 
\be
\int_{\alpha-\varepsilon}^{\alpha+\varepsilon} d {\rm k} \, ( \dots ) & \propto &
 \int_{\alpha^2 - \varepsilon'}^{\alpha^2+\varepsilon'} 
 \frac{d {\rm k}'}{2\sqrt{{\rm k}'}} \frac{p_0^2 + 2 i M^2 + 2 p_0 \sqrt{{\rm k}' + i M^2}}{p_0^4 - 4 M^4 - 4 p_0^2 {\rm k}'} \frac{1}{p_0^2 - 2 p_0 \sqrt{{\rm k}' + i M^2}}
\nonumber \\
&\underset{ {\rm k}' = \alpha^2}{\simeq}& \int_{\alpha^2 - \varepsilon'}^{\alpha^2 + \varepsilon'} \frac{d {\rm k}'}{2 \alpha} \frac{p_0^2 + 2 i M^2 + 2 p_0 \sqrt{\alpha^2 + i M^2}}{- 4 p_0^2 ({\rm k}' - \alpha^2)} \frac{1}{p_0^2 - 2 p_0 \sqrt{\alpha^2 + i M^2}}
\nonumber 
\\
&=& \frac{p_0}{\sqrt{p_0^4 -4 m^4}} \frac{p_0^2 +2 i M^2 + p_0^2 +2 i M^2}{-4 p_0^2} \frac{1}{p_0^2 - (p_0^2 +2 i M^2)} \int_{\alpha^2-\varepsilon'}^{\alpha^2+\varepsilon'} \frac{d {\rm k}'}{{\rm k}'-\alpha^2}
\nonumber 
\\
&=& -i \frac{p_0^2+2iM^2}{4 p_0 M^2 \sqrt{p_0^4 - 4 M^4}} \int_{\alpha^2-\varepsilon'}^{\alpha^2+\varepsilon'} \frac{d {\rm k}'}{{\rm k}'-\alpha^2}\nonumber \\
&=& -i \frac{p_0^2+2iM^2}{4p_0 M^2 \sqrt{p_0^4-4M^4}} \ln \left( \frac{|\varepsilon'|}{|-\varepsilon'|} \right) =0 \, . 
\ee

\subsubsection{Real part}

\begin{figure}[h] 
	\centering 
	\hspace{-1cm}
	\includegraphics[width=0.53\textwidth]{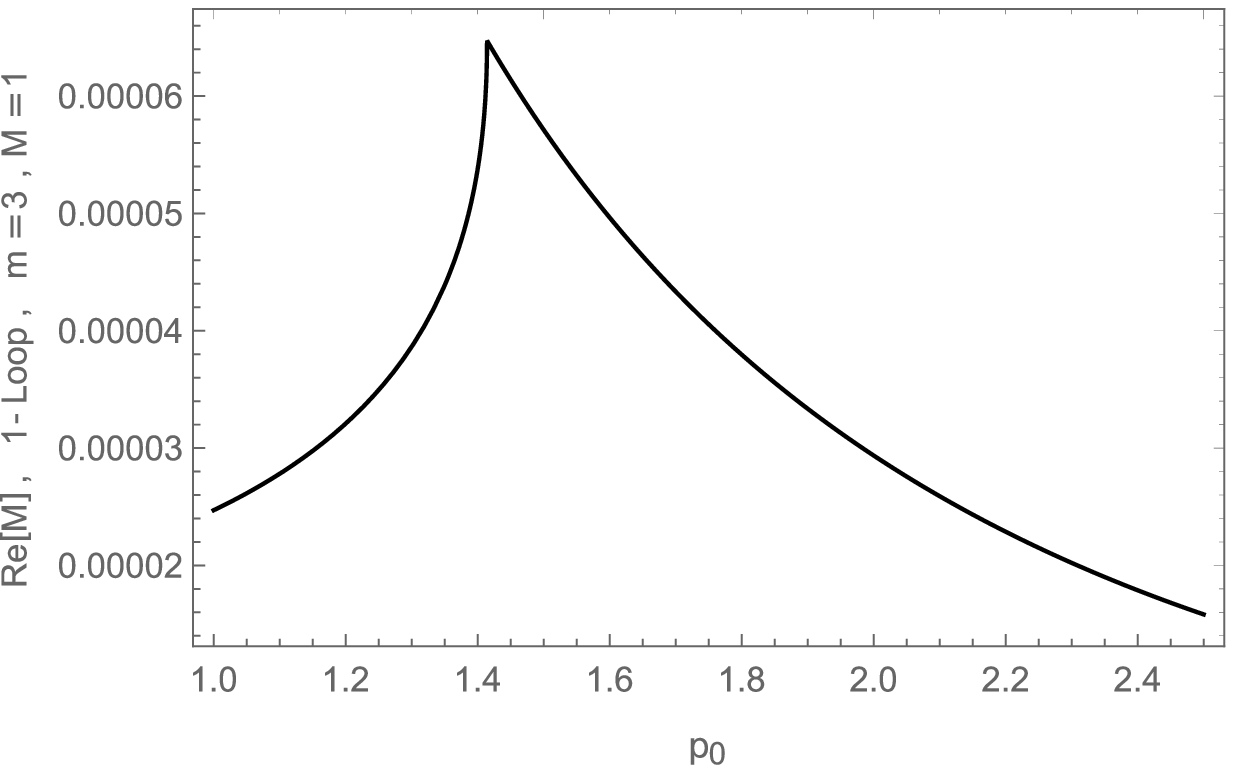}
	\hspace{1.3cm} 
	\includegraphics[width=0.35\textwidth]{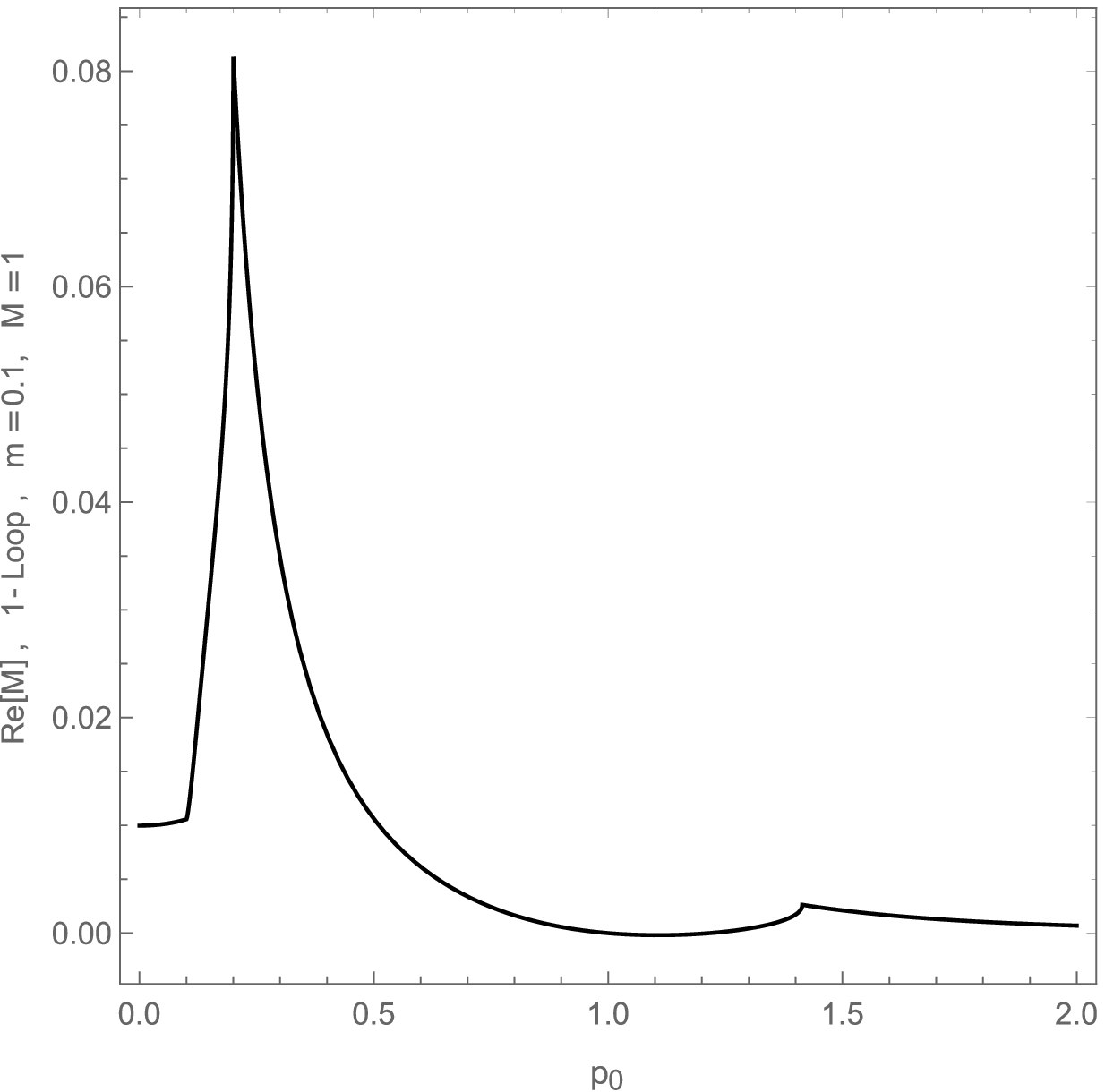}
	\caption{Left panel: real part of the amplitude around the pinching at $k_7^0 = k_2^0$ (or $k_8^0 = k_1^0$) for $m=3$ and $M=1$. Right panel: real part of the amplitude for $m=0.1$, $M=1$. The plot explicitly shows that the amplitude is nonanalytic also at the threshold $p_0 = \sqrt{2} M = \sqrt{2}$.}
	\label{Real_amplitude}
\end{figure}

\begin{figure}[h] 
	\centering 
	\includegraphics[width=0.404\textwidth]{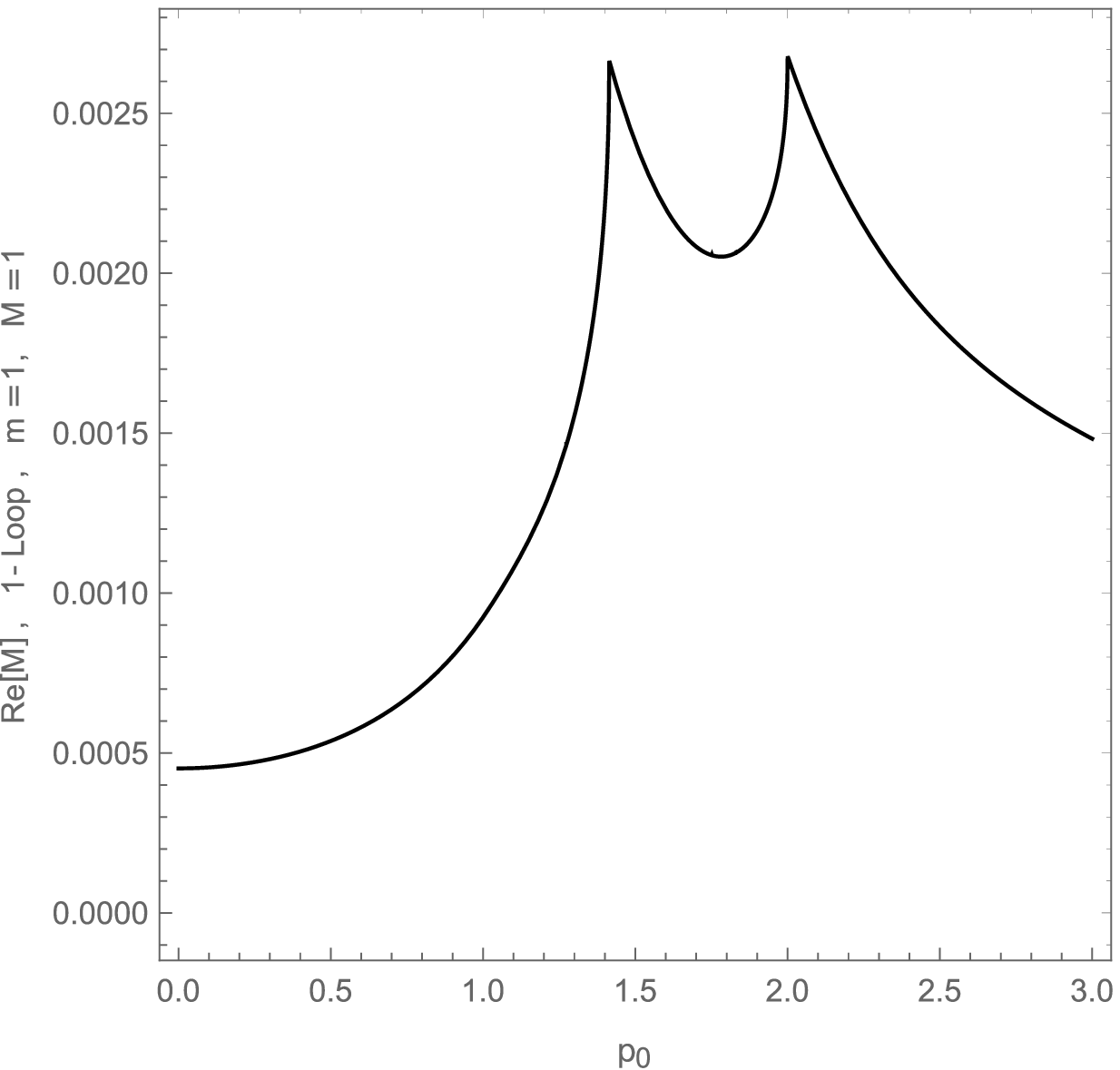}
	\hspace{1.5cm} 
	\includegraphics[width=0.4\textwidth]{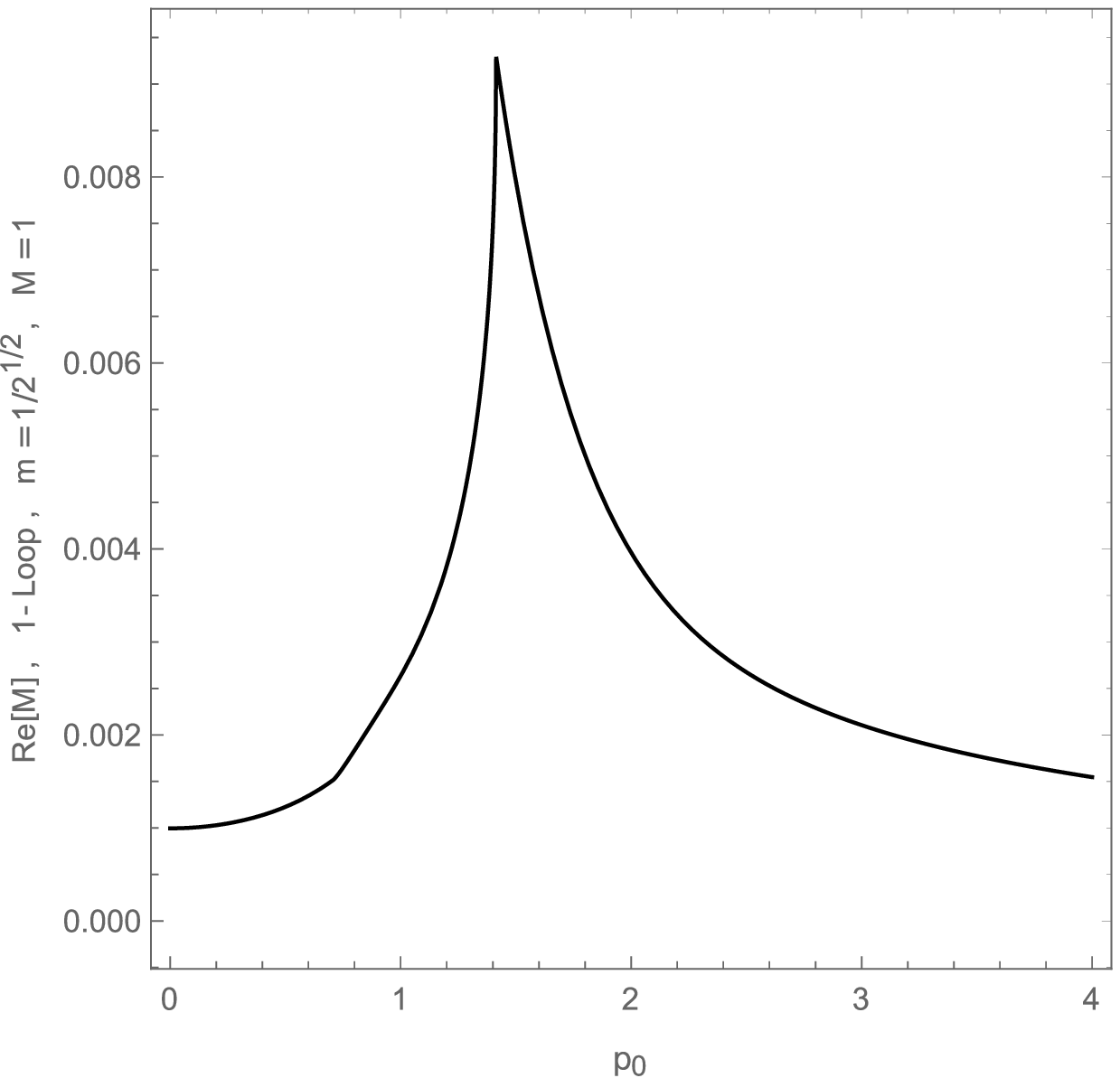}
	\caption{Left panel: real part of the amplitude with $m=M=1$. The real physical threshold is at $p_0 =  2$ and the virtual threshold at $p_0 = \sqrt{2}$. 
	Right panel: real part of the amplitude with $m=1/\sqrt{2}$ and $M=1$. For such values of the masses, the two cusps overlap.}
	\label{m=M=1}
\end{figure}

The real part of the amplitudes is evaluated implementing numerically the prescriptions (\ref{IntSingR}) and (\ref{IntSingC}) in the expression 
\be
{\rm Re} \, \mathcal{M} = {\rm Re} \, \mathcal{M}_{\rm I} + {\rm Re} \, \mathcal{M}_{\rm ResR} + {\rm Re} \, \mathcal{M}_{\rm ResC} \, . 
\ee
The real part of the total amplitude around the pinching is plotted in figure \ref{Real_amplitude}. The cusp located at $p_0=2m$ corresponds to the normal threshold, while the little cusp at $p_0=\sqrt{2}M$ corresponds to the threshold at the pinching singularity. 
In the left panel of figure \ref{m=M=1}, we consider the case $m=M=1$, in which the two mass scales coincide. The two cusps are now of the same of magnitude, but are not equal. Finally, in the right panel of figure \ref{m=M=1} the two thresholds coincide for $m=1/\sqrt{2}$ and $M=1$. 
%
%
%

For the sake of completeness, in Appendix \ref{TriBox} we also explicitly compute the discontinuity of the one-loop triangular and square amplitudes.

\subsection{Two-loop amplitude}
In this section, we study the two-loop bubble amplitude for the $\phi^4$ theory. 
The two-loop amplitude consists of three internal propagators and two vertices:
\be
\mathcal{M} &=& -\frac{\lambda^2 M^{12}}{2} 
\int_{(\mathcal{I}\times \mathbb{R}^3)^2} \frac{i d^4 k_2}{(2\pi)^4} \frac{i d^4 k_1}{(2\pi)^4} \frac{1}{(k_1^2-m^2+i\epsilon)(k_1^4+M^4)} \frac{1}{(k_2^2-m^2+i\epsilon)(k_2^4+M^4)}
\nonumber \\
&&\qquad\qquad\times \frac{1}{[(p-k_1-k_2)^2-m^2+i\epsilon][(p-k_1-k_2)^4+M^4]}\,, 
\ee
where, according to the definition of the quantum theory, we assumed purely imaginary external energy and integrated both the internal energies $k_1^0$ and $k_2^0$ along the imaginary axis. The integral is real because all poles are far away from the integration path but, when we continue the external energy to real values, the pole structure is significantly affected because the position of the poles depends on $p^0$. 
In particular, by changing the value of $p^0$, the poles may move across the original integration path along the imaginary axis $\mathcal{I}$. Hence, in order to preserve analyticity, the integration path must be deformed to a new path $\mathcal{C}$ that avoids such poles. However, the amplitude turns out to be nonanalytic anyway because the complex conjugate poles can overlap pinching the integration path. This is analogous to what we have already seen for the one-loop amplitude. Hence, in order to prove unitarity we should compute the imaginary part of the two-loop amplitude after the continuation from $\mathcal{I}$ to $\mathcal{C}$. This is tantamount to derive the Cutkosky rules. 

The continued amplitude described above is defined integrating $k^0_1$ and $k^0_2$ along the paths $\mathcal{C}_1$ and $\mathcal{C}_2$, respectively,
\be
\label{newamplitude}
\mathcal{M} &=& -\frac{\lambda^2 M^{12}}{2}\int_{\mathcal{C}_2\times \mathbb{R}^3} \frac{i d^4 k_2}{(2\pi)^4} \int_{\mathcal{C}_1\times \mathbb{R}^3} \frac{i d^4 k_1}{(2\pi)^4} \frac{1}{(k_1^2-m^2+i\epsilon)(k_1^4+M^4)}\nonumber \\
&&\times \frac{1}{(k_2^2-m^2+i\epsilon)(k_2^4+M^4)}
\frac{1}{[(p-k_1-k_2)^2-m^2+i\epsilon][(p-k_1-k_2)^4+M^4]} \, .\nonumber\\ 
\label{two-loops}
\ee
%
Let us first consider the twelve poles in the $k_1^0$-plane, i.e.,
\begin{equation}
\begin{aligned}
\bar{k}_{1; 1,2}^0 & =\pm\sqrt{\bm{k}_1^2+m^2-i\epsilon} \, , 
&\quad\bar{k}_{1; 3,4}^0 & =p^0 - k_2^0 \pm\sqrt{\bm{k}_1^2+m^2-i\epsilon} \, , \\
k_{1; 1,2}^0 & =\sqrt{\bm{k}_1^2\pm iM^2} \, , 
&\quad k_{1; 3,4}^0 & =-\sqrt{\bm{k}_1^2\pm iM^2} \, ,  \\
k_{1; 5,6}^0 & =p^0 -k_2^0 +\sqrt{(\bm{k}_1+\bm{k}_2)^2 \pm iM^2} \, ,  
&\quad k_{1; 7,8}^0 & =p^0 -k_2^0 -\sqrt{(\bm{k}_1+\bm{k}_2)^2 \pm iM^2} \, ,  
\end{aligned}
\end{equation}
where normal poles are denoted by $\bar{k}_{1;i}^0$ and complex poles are denoted by $k_{1;i}^0$. Moreover, we again take $\bm{p}=0$. As in the case of the one-loop amplitude, only the poles $\bar{k}_{1;4}^0$, $k_{1;7}^0$ and $k_{1;8}^0$ move across the imaginary axis when we move the external energy from purely imaginary to real values. 
Therefore, $\bar{k}_{1;4}^0$, $k_{1;7}^0$, and $k_{1;8}^0$ will contribute to the residues, and, accordingly, the amplitude (\ref{two-loops}) can be divided into four terms:
\be
\mathcal{M}  \equiv  \mathcal{M}_1+\mathcal{M}_2+\mathcal{M}_3+\mathcal{M}_4 \, . 
\label{Emme}
\ee
When we implement the Cauchy theorem in the above formula, the $\mathcal{M}_1$ contribution is the integral along the imaginary axis:
\be
\hspace{-.8cm}\mathcal{M}_1 &=& -\frac{\lambda^2 M^{12}}{2} \int_{\mathcal{C}_2 \times \mathbb{R}^3}  \frac{i d^4 k_2}{(2 \pi)^4} 
\frac{1}{(k_2^2 - m^2 + i \epsilon)(k_2^4+M^4)}\int_{\mathcal{I}\times \mathbb{R}^3} \frac{i d^4 k_1}{(2\pi)^4} \nn
\hspace{-.8cm}&&\times \frac{1}{(k_1^2-m^2+i\epsilon)(k_1^4 + M^4)}\frac{1}{[(p - k_1 - k_2)^2-m^2+i\epsilon][(p - k_1 - k_2)^4+M^4]} \, .
\label{Emme1}
\ee
The $\mathcal{M}_2$ term is the contribution of the residue evaluated in $\bar{k}_{1;4}^0$: 
\be
\mathcal{M}_2 &=& -\frac{\lambda^2 M^{12}}{2} 
\int_{\mathcal{C}_2 \times  \mathbb{R}^3}  \frac{i d^4 k_2}{(2\pi)^4} \frac{1}{(k_2^2 - m^2 + i \epsilon)(k_2^4 + M^4)} \int_{\mathbb{R}^3} \frac{id^3 k_1}{(2\pi)^4} \, (2\pi i) \nn
&&\times\frac{1}{(\bar{k}_{1;4}^0)^2 - \bm{k}_1^2 - m^2 +i\epsilon}  \frac{1}{[(\bar{k}_{1;4}^0)^2 - \bm{k}_1^2]^2 +M^4} \,\, \frac{1}{ - 2\sqrt{(\bm{k}_1+\bm{k}_2)^2 +m^2 - i \epsilon}}
\nn
&&\times \frac{1}{m^4 +M^4} \sigma[\Re(\bar{k}_{1;4}^0)] \, .
\label{Emme2}
\ee
The $\mathcal{M}_3$ term comes from the residue evaluated in $k_{1;7}^0$:
\be
\mathcal{M}_3 &=& 
- \frac{\lambda^2 M^{12}}{2}
\int_{\mathcal{C}_2 \times \mathbb{R}^3}  \frac{i d^4 k_2}{(2\pi)^4} \frac{1}{(k_2^2-m^2+i\epsilon)(k_2^4+M^4)} 
\int_{\mathbb{R}^3} \frac{id^3 k_1}{(2\pi)^4} \, (2\pi i) \nn
&&\times \frac{1}{(k_{1;7}^0)^2 - \bm{k}_1^2 - m^2 +i\epsilon} \frac{1}{[(k_{1;7}^0)^2 - \bm{k}_1^2]^2 +M^4} 
\,\,
\frac{1}{iM^2 -m^2}
\nn
&&\times 
\frac{1}{-4iM^2 \sqrt{(\bm{k}_1+\bm{k}_2)^2 +iM^2}} \sigma[\Re(k_{1;7}^0)] \, .
\label{Emme3}
\ee
The $\mathcal{M}_4$ term comes from the residue evaluated in $k_{1;8}^0$:
\be
\mathcal{M}_4 &=& -\frac{\lambda^2 M^{12}}{2}
\int_{\mathcal{C}_2 \times \mathbb{R}^3}  \frac{i d^4 k_2}{(2\pi)^4} \frac{1}{(k_2^2 - m^2 + i \epsilon)(k_2^4 + M^4)} \int_{\mathbb{R}^3} \frac{id^3 k_1}{(2\pi)^4} \, (2\pi i) \nn
&&\times \frac{1}{(k_{1;8}^0)^2 - \bm{k}_1^2 - m^2 +i\epsilon}  \frac{1}{[(k_{1;8}^0)^2 - \bm{k}_1^2]^2 +M^4} \,\,
\frac{1}{-iM^2 -m^2} 
\nn
&&\times \frac{1}{4iM^2 \sqrt{(\bm{k}_1+\bm{k}_2)^2 -iM^2}} \sigma[\Re(k_{1;8}^0)].
\label{Emme4}
\ee
Notice that the pole structure in the $k_2^0$-plane for each of the above four terms is different. Hence, we have to discuss each term $\mathcal{M}_i$ ($i=1,2,3,4$) separately. 


\underline{\bf Amplitude $\mathcal{M}_1$} --- The amplitude $\mathcal{M}_1$ has twelve poles in the $k_2^0$ plane, namely 
\begin{equation}
\begin{aligned}
\bar{k}_{2; 1,2}^0 & =\pm\sqrt{\bm{k}_2^2+m^2-i\epsilon} \, , 
&\quad \bar{k}_{2; 3,4}^0 & =p^0 - k_1^0 \pm\sqrt{\bm{k}_1^2+m^2-i\epsilon};\\
k_{2; 1,2}^0 & =\sqrt{\bm{k}_2^2\pm iM^2} \, , 
&\quad k_{2; 3,4}^0 & =-\sqrt{\bm{k}_2^2\pm iM^2} \, ,  \\
k_{2; 5,6}^0 & =p^0 -k_1^0 +\sqrt{(\bm{k}_1+\bm{k}_2)^2 \pm iM^2} \, , 
&\quad k_{2; 7,8}^0 & =p^0 -k_1^0 -\sqrt{(\bm{k}_1+\bm{k}_2)^2 \pm iM^2} \, . 
\end{aligned}
\end{equation}
As we expected, only the poles $\bar{k}_{2;4}^0$, $k_{2;7}^0$, and $k_{2;8}^0$ move across the imaginary axis, contributing to the residues. Therefore, $\mathcal{M}_1$ can be subdivided into four terms, too,
\be
\mathcal{M}_1=\mathcal{M}_{11}+\mathcal{M}_{12}+\mathcal{M}_{13}+\mathcal{M}_{14} \, .
\label{Emme1B}
\ee
$\mathcal{M}_{11}$ is the integral of $k_2^0$ along the imaginary axes:
\be
\hspace{-.9cm}\mathcal{M}_{11}
&=& -\frac{\lambda^2 M^{12}}{2} \int_{\mathcal{I} \times \mathbb{R}^3}  \frac{i d^4 k_1}{(2\pi)^4} \frac{1}{(k_1^2-m^2+i\epsilon)(k_1^4+M^4)} \int_{\mathcal{I}\times \mathbb{R}^3} \frac{i d^4 k_2}{(2\pi)^4}\nn
\hspace{-.9cm}&&\times \frac{1}{(k_2^2-m^2+i\epsilon)(k_2^4+M^4)} \frac{1}{[(p-k_1-k_2)^2-m^2+i\epsilon][(p-k_1-k_2)^4+M^4]} \, . 
\label{M11}
\ee
The $\mathcal{M}_{12}$ term is the contribution at the pole $\bar{k}_{2;4}^0$:
\be
\mathcal{M}_{12} &=&  -\frac{\lambda^2 M^{12}}{2}\int_{\mathcal{I}\times \mathbb{R}^3}  \frac{i d^4 k_1}{(2\pi)^4} \frac{1}{(k_1^2-m^2+i\epsilon)(k_1^4+M^4)} \int_{\mathbb{R}^3} \frac{id^3 k_2}{(2\pi)^4} (2\pi i)\nn
&&\times \frac{1}{(\bar{k}_{2;4}^0)^2 - \bm{k}_2^2 - m^2 +i\epsilon} \frac{1}{[(\bar{k}_{2;4}^0)^2 - \bm{k}_2^2]^2 +M^4} \nn
&&\times\frac{1}{-2\sqrt{(\bm{k}_1+\bm{k}_2)^2 +m^2 -i\epsilon}} \,\, \frac{1}{m^4 +M^4} \sigma[\Re(\bar{k}_{2;4}^0)] \, . 
\label{M12}
\ee
The $\mathcal{M}_{13}$ term is the contribution at the pole $k_{2;7}^0$: 
\be
\mathcal{M}_{13} &=& 
- \frac{\lambda^2 M^{12}}{2} \int_{\mathcal{I}\times \mathbb{R}^3}  \frac{i d^4 k_1}{(2\pi)^4} \frac{1}{(k_1^2-m^2+i\epsilon)(k_1^4+M^4)} \int_{\mathbb{R}^3} \frac{id^3 k_2}{(2\pi)^4} (2\pi i)\nn
&&\times \frac{1}{(k_{2;7}^0)^2 - \bm{k}_2^2 - m^2 +i\epsilon}\frac{1}{[(k_{2;7}^0)^2 - \bm{k}_2^2]^2 +M^4} 
\,\, \frac{1}{iM^2 -m^2} \nn
&&\times\frac{1}{-4iM^2 \sqrt{(\bm{k}_1+\bm{k}_2)^2 +iM^2}} \sigma[\Re(k_{2;7}^0)] \, .
\label{M13}
\ee
The $\mathcal{M}_{14}$ term is the contribution at the pole $k_{2;8}^0$:
\be
\mathcal{M}_{14} &=& 
-\frac{\lambda^2 M^{12}}{2} \int_{\mathcal{I}\times \mathbb{R}^3}  \frac{i d^4 k_1}{(2\pi)^4} \frac{1}{(k_1^2-m^2+i\epsilon)(k_1^4+M^4)} \int_{\mathbb{R}^3} \frac{id^3 k_2}{(2\pi)^4} (2\pi i)\nn
&&\times \frac{1}{(k_{2;8}^0)^2 - \bm{k}_2^2 - m^2 +i\epsilon}  \frac{1}{[(k_{2;8}^0)^2 - \bm{k}_2^2]^2 +M^4} 
\,\, \frac{1}{-iM^2 -m^2} \nn
&&\times
\frac{1}{4iM^2 \sqrt{(\bm{k}_1+\bm{k}_2)^2 -iM^2}} \sigma[\Re(k_{2;8}^0)] \, .
\label{M14}
\ee

\underline{\bf Amplitude $\mathcal{M}_2$} --- The amplitude $\mathcal{M}_2$ has nine poles in the $k_2^0$ plane, namely 
\be
&& \bar{k}_{2; 1,2}^0  =\pm\sqrt{\bm{k}_2^2+m^2-i\epsilon} \, , \\
&& \bar{k}_{2; 3}^0  =p^0 -\sqrt{\bm{k}_1^2 +m^2 -i\epsilon} -\sqrt{(\bm{k}_1+\bm{k}_2)^2+m^2-i\epsilon} \, , \\
&& k_{2; 1,2}^0  =\sqrt{\bm{k}_2^2\pm iM^2} \, ,  \\
&& k_{2; 3,4}^0  =-\sqrt{\bm{k}_2^2\pm iM^2} \, ,  \\
&& k_{2; 5,6}^0  =p^0 -\sqrt{\bm{k}_1^2 \pm iM^2} -\sqrt{(\bm{k}_1+\bm{k}_2)^2 +m^2 -i\epsilon} \, ,  
\ee
of which the poles $\bar{k}_{2;3}^0$, $k_{2;5}^0$, and $k_{2;6}^0$ move across the imaginary axis $k_2^0$ and contribute to the residues when applying Cauchy's theorem. Therefore, $\mathcal{M}_2$ can be divided into four terms,
\be
\mathcal{M}_2=\mathcal{M}_{21}+\mathcal{M}_{22}+\mathcal{M}_{23}+\mathcal{M}_{24} \, . 
\label{Emme2B}
\ee
$\mathcal{M}_{21}$ is the integral of $k_2^0$ along the imaginary axis:
\be
\mathcal{M}_{21}  
&=& -\frac{\lambda^2 M^{12}}{2} \int_{\mathbb{R}^3} \frac{id^3 k_1}{(2\pi)^4} (2\pi i) \int_{\mathcal{I} \times \mathbb{R}^3}  \,\, \frac{i d^4 k_2}{(2\pi)^4} \frac{1}{k_2^2-m^2+i\epsilon} 
\,\,
\frac{1}{k_2^4+M^4} \nn
&&\times
\frac{1}{(\bar{k}_{1;4}^0)^2 - \bm{k}_1^2 - m^2 + i \epsilon}  \frac{1}{[(\bar{k}_{1;4}^0)^2 - \bm{k}_1^2]^2 +M^4} 
\nn
&&\times
\frac{1}{-2\sqrt{(\bm{k}_1+\bm{k}_2)^2 +m^2 -i\epsilon}} \,\,
\frac{1}{m^4 +M^4} \sigma[\Re(\bar{k}_{1;4}^0)] \, .
\label{Emme21} 
\ee
The $\mathcal{M}_{22}$ term is the contribution at the pole $\bar{k}_{2;3}^0$: 
\be
\mathcal{M}_{22} &=& 
-\frac{\lambda^2 M^{12}}{2} \int_{\mathbb{R}^3} \frac{id^3 k_1}{(2\pi)^4} (2\pi i) \int_{\mathbb{R}^3}  \frac{i d^3 k_2}{(2\pi)^4} (2\pi i) \frac{1}{(\bar{k}_{2;3}^0)^2-\bm{k}_2^2-m^2+i\epsilon} 
\nn
&&\times
\frac{1}{[(\bar{k}_{2;3}^0)^2-\bm{k}_2^2]^2+M^4} \frac{1}{-2\sqrt{\bm{k}_1^2 +m^2 -i\epsilon}}\nn
&&\times \frac{1}{\left[\left(p^0 -\bar{k}_{2;3}^0 -\sqrt{(\bm{k}_1+\bm{k}_2)^2 +m^2 -i\epsilon}\right)^2 - \bm{k}_1^2\right]^2 +M^4} \nn
&&\times
\frac{1}{-2\sqrt{(\bm{k}_1+\bm{k}_2)^2 +m^2 -i\epsilon}}
\,\,
 \frac{\sigma[\Re(\bar{k}_{2;3}^0)]}{m^4 +M^4} 
\, . 
\label{Emme22}
\ee
The $\mathcal{M}_{23}$ term is the contribution at the pole $k_{2;5}^0$:
\be
\mathcal{M}_{23} &=& 
 -\frac{\lambda^2 M^{12}}{2} \int_{\mathbb{R}^3} \frac{id^3 k_1}{(2\pi)^4} (2\pi i) \int_{\mathbb{R}^3}  \frac{i d^3 k_2}{(2\pi)^4} (2\pi i) \frac{1}{(k_{2;5}^0)^2-\bm{k}_2^2-m^2+i\epsilon}
\nn
&&\times  \frac{1}{[(k_{2;5}^0)^2-\bm{k}_2^2]^2+M^4}
 \,\,
  \frac{1}{iM^2 -m^2}\,\,
\frac{1}{-4iM^2 \sqrt{\bm{k}_1^2 +iM^2}} 
\nn
&&\times
 \frac{1}{-2\sqrt{(\bm{k}_1+\bm{k}_2)^2 +m^2 -i\epsilon}} 
 \,\,
 \frac{1}{m^4 +M^4} \sigma[\Re(k_{2;5}^0)] \, .
\label{Emme23}
\ee
The $\mathcal{M}_{24}$ term is the contribution at the pole $k_{2;6}^0$:
\be
\mathcal{M}_{24} &=& 
-\frac{\lambda^2 M^{12}}{2} \int_{\mathbb{R}^3} \frac{id^3 k_1}{(2\pi)^4} (2\pi i) \int_{\mathbb{R}^3}  
\frac{i d^3 k_2}{(2\pi)^4} (2\pi i) \frac{1}{(k_{2;6}^0)^2-\bm{k}_2^2-m^2+i\epsilon} \nn
&&\times
\frac{1}{[(k_{2;6}^0)^2-\bm{k}_2^2]^2+M^4} 
\,\, \frac{1}{-iM^2 -m^2}\,\,\frac{1}{4iM^2 \sqrt{\bm{k}_1^2 -iM^2}} 
\nn
&&\times
 \frac{1}{-2\sqrt{(\bm{k}_1+\bm{k}_2)^2 +m^2 -i\epsilon}} 
 \,\,
 \frac{1}{m^4 +M^4} \sigma[\Re(k_{2;6}^0)] \, . 
\label{Emme24}
\ee

\underline{\bf Amplitude $\mathcal{M}_3$} --- The amplitude $\mathcal{M}_3$ has also nine poles in the $k_2^0$ plane, namely 
%
%
\be
&& \bar{k}_{2; 1,2}^0  =\pm\sqrt{\bm{k}_2^2+m^2-i\epsilon} \, , \nonumber \\
&& \bar{k}_{2; 3}^0  =p^0 -\sqrt{\bm{k}_1^2 +m^2 -i\epsilon} -\sqrt{(\bm{k}_1+\bm{k}_2)^2+iM^2} \, , \nonumber\\
&& k_{2; 1,2}^0  =\sqrt{\bm{k}_2^2\pm iM^2} \, ,\nonumber \\
&& k_{2; 3,4}^0  =-\sqrt{\bm{k}_2^2\pm iM^2} \, ,  \nonumber\\
&& k_{2; 5,6}^0  =p^0 -\sqrt{\bm{k}_1^2 \pm iM^2} -\sqrt{(\bm{k}_1+\bm{k}_2)^2 +iM^2} \, ,  
\ee
of which only the poles $\bar{k}_{2;3}^0$, $k_{2;5}^0$, and $k_{2;6}^0$ can move across the imaginary axis. 
Therefore, $\mathcal{M}_3$ can be divided into four contributions, 
\be
\mathcal{M}_3=\mathcal{M}_{31}+\mathcal{M}_{32}+\mathcal{M}_{33}+\mathcal{M}_{34} \, .
\label{Emme3B}
\ee
In $\mathcal{M}_{31}$, we integrate $k_2^0$ along the imaginary axis:
\be
\mathcal{M}_{31}
&=& -\frac{\lambda^2 M^{12}}{2} \int_{\mathbb{R}^3} \frac{id^3 k_1}{(2\pi)^4} (2\pi i) \int_{\mathcal{I}\times \mathbb{R}^3}  \frac{i d^4 k_2}{(2\pi)^4} \frac{1}{k_2^2-m^2+i\epsilon}
 \,\,
  \frac{1}{k_2^4+M^4} 
\nn
&&\times
  \frac{1}{(k_{1;7}^0)^2 - \bm{k}_1^2 - m^2 +i\epsilon} \,\, \frac{1}{[(k_{1;7}^0)^2 - \bm{k}_1^2]^2 +M^4} 
\,\,
 \frac{1}{iM^2 -m^2} 
\nn
&&\times
 \frac{1}{-4iM^2\sqrt{(\bm{k}_1+\bm{k}_2)^2 +iM^2}}  \sigma[\Re(k_{1;7}^0)] \,.
\label{Emme31}
\ee
The $\mathcal{M}_{32}$ term is the contribution at the pole $\bar{k}_{2;3}^0$:
\be
\mathcal{M}_{32} &=& 
-\frac{\lambda^2 M^{12}}{2} \int_{\mathbb{R}^3} \frac{id^3 k_1}{(2\pi)^4} (2\pi i) \int_{\mathbb{R}^3}  \frac{i d^3 k_2}{(2\pi)^4} (2\pi i) \frac{1}{(\bar{k}_{2;3}^0)^2-\bm{k}_2^2-m^2+i\epsilon} 
\nn
&&\times
 \frac{1}{[(\bar{k}_{2;3}^0)^2-\bm{k}_2^2]^2+M^4}\,\,\frac{1}{-2\sqrt{\bm{k}_1^2 +m^2 -i\epsilon}}\,\,
\frac{1}{m^4 +M^4} 
\,\,
\frac{1}{iM^2 -m^2} 
\nn
&&\times
\frac{1}{-4iM^2\sqrt{(\bm{k}_1+\bm{k}_2)^2 +iM^2}}  \sigma[\Re(\bar{k}_{2;3}^0)] \, .
\label{Emme32}
\ee
The $\mathcal{M}_{33}$ term is the contribution at the pole ${k}_{2;5}^0$:
\be
\mathcal{M}_{33} &=& 
-\frac{\lambda^2 M^{12}}{2} \int_{\mathbb{R}^3} \frac{id^3 k_1}{(2\pi)^4} (2\pi i) \int_{\mathbb{R}^3}  \frac{i d^3 k_2}{(2\pi)^4} (2\pi i) \frac{1}{(k_{2;5}^0)^2-\bm{k}_2^2-m^2+i\epsilon} 
\nn
&&\times
 \frac{1}{[(k_{2;5}^0)^2-\bm{k}_2^2]^2+M^4}\,\, \frac{1}{iM^2 -m^2}\,\,\frac{1}{-4iM^2 \sqrt{\bm{k}_1^2 +iM^2}} 
\,\, 
\frac{1}{iM^2 -m^2} 
\nn
&&\times
 \frac{1}{-4iM^2\sqrt{(\bm{k}_1+\bm{k}_2)^2 +iM^2}} \sigma[\Re(k_{2;5}^0)] \, .
\label{Emme33}
\ee
The $\mathcal{M}_{34}$ term is the contribution at the pole ${k}_{2;6}^0$:
\be
\mathcal{M}_{34} &=& 
-\frac{\lambda^2 M^{12}}{2} \int_{\mathbb{R}^3} \frac{id^3 k_1}{(2\pi)^4} (2\pi i) \int_{\mathbb{R}^3}  \frac{i d^3 k_2}{(2\pi)^4} (2\pi i) \frac{1}{(k_{2;6}^0)^2-\bm{k}_2^2-m^2+i\epsilon} 
\nn
&&\times
\frac{1}{[(k_{2;6}^0)^2-\bm{k}_2^2]^2+M^4}\,\, \frac{1}{-iM^2 -m^2} \,\, \frac{1}{4iM^2 \sqrt{\bm{k}_1^2 -iM^2}} 
\,\,
\frac{1}{iM^2 -m^2} 
\nn
&&\times
\frac{1}{-4iM^2\sqrt{(\bm{k}_1+\bm{k}_2)^2 +iM^2}} \sigma[\Re(k_{2;6}^0)] \, . 
\label{Emme34}
\ee

\underline{\bf Amplitude $\mathcal{M}_4$} --- The amplitude $\mathcal{M}_2$ has nine poles in the $k_2^0$ plane, namely 
%
%
\be
&& \bar{k}_{2; 1,2}^0  =\pm\sqrt{\bm{k}_2^2+m^2-i\epsilon} \, , \nonumber \\
&& \bar{k}_{2; 3}^0  =p^0 -\sqrt{\bm{k}_1^2 +m^2 -i\epsilon} -\sqrt{(\bm{k}_1+\bm{k}_2)^2-iM^2} \, , \nonumber\\
&& k_{2; 1,2}^0  =\sqrt{\bm{k}_2^2\pm iM^2} \, , \nonumber\\
&& k_{2; 3,4}^0  =-\sqrt{\bm{k}_2^2\pm iM^2} \, ,  \nonumber\\
&& k_{2; 5,6}^0  =p^0 -\sqrt{\bm{k}_1^2 \pm iM^2} -\sqrt{(\bm{k}_1+\bm{k}_2)^2 -iM^2} \, , 
\ee
of which the poles $\bar{k}_{2;3}^0$, $k_{2;5}^0$, and $k_{2;6}^0$ can move across the imaginary axis. Hence, $\mathcal{M}_4$ can be divided into four terms:
\be
\mathcal{M}_4 = \mathcal{M}_{41}+\mathcal{M}_{42}+\mathcal{M}_{43}+\mathcal{M}_{44}\, . 
\label{Emme4B}
\ee
In $\mathcal{M}_{41}$, we integrate $k_2^0$ along the imaginary axis:
\be
\mathcal{M}_{41} &=& 
 -\frac{\lambda^2 M^{12}}{2} \int_{\mathbb{R}^3} \frac{id^3 k_1}{(2\pi)^4} (2\pi i) \int_{\mathcal{I}\times \mathbb{R}^3}  \frac{i d^4 k_2}{(2\pi)^4} \frac{1}{k_2^2-m^2+i\epsilon} 
 \,\,
 \frac{1}{k_2^4+M^4} \nn
&&\times
 \frac{1}{(k_{1;8}^0)^2 - \bm{k}_1^2 - m^2 +i\epsilon} \,\,\frac{1}{[(k_{1;8}^0)^2 - \bm{k}_1^2]^2 +M^4}
\,\,
 \frac{1}{-iM^2 -m^2}
\nn
&&\times
  \frac{1}{4iM^2\sqrt{(\bm{k}_1+\bm{k}_2)^2 -iM^2}}  \sigma[\Re(k_{1;8}^0)] \, .
\label{Emme41}
\ee
The $\mathcal{M}_{42}$ term is the contribution at the pole $\bar{k}_{2;3}^0$:
\be
\mathcal{M}_{42} &=& 
-\frac{\lambda^2 M^{12}}{2} \int_{\mathbb{R}^3} \frac{id^3 k_1}{(2\pi)^4} (2\pi i) \int_{\mathbb{R}^3}  \frac{i d^3 k_2}{(2\pi)^4} (2\pi i) \frac{1}{(\bar{k}_{2;3}^0)^2-\bm{k}_2^2-m^2+i\epsilon} 
\nn
&&\times
 \frac{1}{[(\bar{k}_{2;3}^0)^2-\bm{k}_2^2]^2+M^4}\,\,\frac{1}{-2\sqrt{\bm{k}_1^2 +m^2 -i\epsilon}}\times \,\, \frac{1}{m^4 +M^4} \frac{1}{-iM^2 -m^2}
\nn
&&\times
\frac{1}{4iM^2\sqrt{(\bm{k}_1+\bm{k}_2)^2 -iM^2}}  \sigma[\Re(\bar{k}_{2;3}^0)].
\label{Emme42}
\ee
The $\mathcal{M}_{43}$ term is the contribution at the pole $k_{2;5}^0$:
\be
\mathcal{M}_{43} &=& 
-\frac{\lambda^2 M^{12}}{2} \int_{\mathbb{R}^3} \frac{id^3 k_1}{(2\pi)^4} (2\pi i) \int_{\mathbb{R}^3}  \frac{i d^3 k_2}{(2\pi)^4} (2\pi i) \frac{1}{(k_{2;5}^0)^2-\bm{k}_2^2-m^2+i\epsilon} 
\nn
&&\times
 \frac{1}{[(k_{2;5}^0)^2-\bm{k}_2^2]^2+M^4}\,\,\frac{1}{iM^2 -m^2} \,\, \frac{1}{-4iM^2 \sqrt{\bm{k}_1^2 +iM^2}}
\,\,
 \frac{1}{-iM^2 -m^2} 
\nn
&&\times
 \frac{1}{4iM^2\sqrt{(\bm{k}_1+\bm{k}_2)^2 -iM^2}} \sigma[\Re(k_{2;5}^0)]. 
\label{Emme43}
\ee
The $\mathcal{M}_{44}$ term is the contribution at the pole $k_{2;6}^0$:
\be
\mathcal{M}_{44} &=& 
 -\frac{\lambda^2 M^{12}}{2} \int_{\mathbb{R}^3} \frac{id^3 k_1}{(2\pi)^4} (2\pi i) \int_{\mathbb{R}^3}  \frac{i d^3 k_2}{(2\pi)^4} (2\pi i) \frac{1}{(k_{2;6}^0)^2-\bm{k}_2^2-m^2+i\epsilon} 
\nn
&&\times
 \frac{1}{[(k_{2;6}^0)^2-\bm{k}_2^2]^2+M^4} 
 \,\,
 \frac{1}{-iM^2 -m^2} \,\, 
 \frac{1}{4iM^2 \sqrt{\bm{k}_1^2 -iM^2}}
\,\,
 \frac{1}{-iM^2 -m^2} 
\nn
&&\times
 \frac{1}{4iM^2\sqrt{(\bm{k}_1+\bm{k}_2)^2 -iM^2}} \sigma[\Re(k_{2;6}^0)]. 
\label{Emme44}
\ee

In the limit $\epsilon \rightarrow 0$, the following amplitudes form six complex conjugate pairs:
\be
&&\mathcal{M}_{13}  = \mathcal{M}_{14}^* \, , \qquad 
\mathcal{M}_{23}  = \mathcal{M}_{24}^* \, , \qquad 
\mathcal{M}_{31}  = \mathcal{M}_{41}^* \, , \nn
&&\mathcal{M}_{32}  = \mathcal{M}_{42}^* \, , \qquad 
\mathcal{M}_{33} = \mathcal{M}_{44}^* \, , \qquad 
\mathcal{M}_{34} = \mathcal{M}_{43}^* \, .
\label{complexConj}
\ee
Therefore, their sum does not contribute to the imaginary part of the amplitude at two loops.
However, this statement requires to check that all the denominators containing $i\epsilon$ 
in the amplitudes (\ref{complexConj}) are never zero. 
\begin{itemize}
\item\underline{In $\mathcal{M}_{13}$} (\ref{M13}), the parameter $\epsilon$ appears in the first and second denominators. The first denominator $k_1^2-m^2+i\epsilon$ is nonzero for the integration variable $k^0_1\neq 0$ because $k^0_1 \in \mathbb{C}$, while for $k^0_1 =0$ the contribution to the integral is of zero measure.
The second denominator $(k_{2;7}^0)^2 - \bm{k}_2^2 - m^2 +i\epsilon$ is zero only when the integration variable $k_1^0 = -{\rm Im}\sqrt{(\bm{k}_1+\bm{k}_2)^2 +iM^2}$. However, it is again a zero-measure subset when we integrate in $k_1^0$.
\item\underline{In $\mathcal{M}_{23}$} (\ref{Emme23}), the parameter $\epsilon$ appears in the first and fifth denominators.
Integrating in the real variables $\bm{k}_1$ and $\bm{k}_2$, the first denominator $(k_{2;5}^0)^2 - \bm{k}_2^2 - m^2 +i\epsilon$ is never zero because $\bm{k}_1$ and $\bm{k}_2$ are real while the imaginary part of the denominator $i M^2$ is nonzero (see $k_{2;5}^0$). 
The fifth denominator is nonzero because of the mass term $m^2$. For $m=0$, the integral vanishes anyway because the subset of singularities at $\bm{k}_1+\bm{k}_2=0$ has zero measure. 
\item\underline{In $\mathcal{M}_{31}$} (\ref{Emme31}), the parameter $\epsilon$ appears in the first and third denominators.
The first denominator $k_2^2 - m^2 + i\epsilon$ is nonzero for the integration variable $k^0_2\neq 0$ because $k^0_2 \in \mathbb{C}$. For $k^0_2 = 0$, the denominator vanishes, but the contribution to the integral is of zero measure. The denominator $(k_{1;7}^0)^2 - \bm{k}_1^2 - m^2 + i \epsilon$ is zero only when the integration variable takes the value 
$k_2^0 = - {\rm Im} \sqrt{(\bm{k}_1+ \bm{k}_2)^2 +iM^2}$, which is a zero-measure subset for the integration domain of $k_2^0$. 
\item\underline{In $\mathcal{M}_{32}$} (\ref{Emme32}), the parameter $\epsilon$ appears in the first, second, and third denominators. Integrating in $\bm{k}_1$ and $\bm{k}_2$, the denominators $(\bar{k}_{2;3}^0)^2 - \bm{k}_2^2 - m^2 + i \epsilon$ and $[(\bar{k}_{2;3}^0)^2 -\bm{k}_2^2]^2 +M^4$ are never zero because $\bar{k}_{2;3}^0$ always takes complex values (see the $iM^2$ contribution in $\bar{k}_{2;3}^0$), while the integration variables $\bm{k}_1$ and $\bm{k}_2$ are real. The third denominator $-2 \sqrt{\bm{k}_1^2 +m^2 -i\epsilon}$ is never zero for $m^2>0$, while for $m=0$ and $\bm{k}_1 \neq 0$ such denominator is nonzero because $\bm{k}_1$ is real. For $m=0$, the $\bm{k}_1=0$ subset is of zero measure in the integration domain of $\bm{k}_1$.
\item\underline{In $\mathcal{M}_{33}$} (\ref{Emme33}), the parameter $\epsilon$ appears in the first and second denominators. 
Both the denominators $(k_{2;5}^0)^2 - \bm{k}_2^2 - m^2 + i \epsilon$ and 
$[(k_{2;5}^0)^2-\bm{k}_2^2]^2+M^4$ never vanish because 
$k_{2;5}^0$ always takes complex values (see the $i M^2$ contribution), while the integration variables $\bm{k}_1$ and $\bm{k}_2$ are real. 
\item\underline{In $\mathcal{M}_{34}$} (\ref{Emme33}), the parameter $\epsilon$ appears in the first and second denominators.
The first denominator $(k_{2;6}^0)^2 - \bm{k}_2^2 - m^2 +i\epsilon$ reaches zero only for $\bm{k}_1^2 = (\bm{k}_1 +\bm{k}_2)^2$. Im fact, the imaginary part of the first denominator is $- {\rm Im}\sqrt{\bm{k}_1^2 - i M^2} - {\rm Im}\sqrt{(\bm{k}_1 +\bm{k}_2)^2 + i M^2}$, which is zero when $\bm{k}_1^2 = (\bm{k}_1 +\bm{k}_2)^2$. Once we make sure the imaginary part vanishes, we can further manipulate the value of $p^0$ to make the real part of the first denominator vanish, too. However, for fixed $\bm{k}_1$, the $\bm{k}_2$ the condition above forms a two-dimensional sphere, which is a zero-measure subset in the three-dimensional integration domain over $\bm{k}_2$. 
The second denominator 
$[(\bar{k}_{2;6}^0 -\bm{k}_2^2)]^2 +M^4$ is never zero because $\bar{k}_{2;6}^0$ always takes complex values (see the $iM^2$ contribution in $\bar{k}_{2;6}^0$), while the integration variables $\bm{k}_1$ and $\bm{k}_2$ are real.
\end{itemize}
Therefore, the complex conjugate pairs (\ref{complexConj}) do not give any imaginary contribution to the two-loop amplitude. 

Next, we have to check whether $\mathcal{M}_{11}$, $\mathcal{M}_{12}$, $\mathcal{M}_{21}$, and $\mathcal{M}_{22}$ give any imaginary contribution. 
The imaginary part of the integrand of $\mathcal{M}_{11}$ is an odd function of $(k_1^0 +k_2^0)$. Let us replace $k_1^0$ and $k_2^0$ with $i k_1^4$ and $i k_2^4$, respectively. We know that the first and second denominators are real, so that the imaginary contribution comes from the third denominator. We can rewrite the latter as
{\small
\be
\hspace{-.2cm}&&
\frac{1}{[(p-k_1-k_2)^2-m^2+i\epsilon][(p-k_1-k_2)^4+M^4]} \nonumber \\
\hspace{-.2cm}&& \quad\,
= \frac{1}{[(p^0 - i k_1^4 - i k_2^4)^2 - (\bm{p} - \bm{k}_1 - \bm{k}_2)^2 - m^2 + i\epsilon][((p^0 - i k_1^4-i k_2^4)^2 - (\bm{p} - \bm{k}_1 - \bm{k}_2)^2)^2 + M^4]} 
\nonumber \\
\hspace{-.2cm}&& \quad\,
= 
\frac{1}{(p^0)^2 -  (k_1^4 + k_2^4)^2 - (\bm{p} - \bm{k}_1 - \bm{k}_2)^2 - m^2 + i [\epsilon - 2 p^0 (k_1^4 + k_2^4)]}\,\,\frac{1}{f(p,k)} \nonumber \\
\hspace{-.2cm}&& \quad\,
= 
\frac{(p^0)^2 -  (k_1^4 + k_2^4)^2 - (\bm{p} - \bm{k}_1 - \bm{k}_2)^2 - m^2 - i [\epsilon - 2 p^0 (k_1^4 + k_2^4)]}{[(p^0)^2 -  (k_1^4 + k_2^4)^2 - (\bm{p} - \bm{k}_1 - \bm{k}_2)^2 - m^2]^2 + [\epsilon - 2 p^0 (k_1^4 + k_2^4)]^2}\,\,\frac{f^*(p,k)}{|f(p,k)|^2}\nonumber
\ee}
where
\be
f(p,k)&=&[(p^0)^2 - (k_1^4 + k_2^4)^2 - (\bm{p} - \bm{k}_1 - \bm{k}_2)^2]^2 - [2 p^0 (k_1^4 + k_2^4)]^2 + M^4\nn
&& - 2 i [(p^0)^2 - (k_1^4 + k_2^4)^2 - (\bm{p} - \bm{k}_1 - \bm{k}_2)^2] [2 p^0 (k_1^4 + k_2^4)]\nonumber
\ee
If we keep tracking with the term carrying imaginary label $i$, we found that such term is always an odd function of $(k_1^4 + k_2^4)$ (we consider $\epsilon = 0$). It is quite straightforward since $i$ always appears associated with $(k_1^4 + k_2^4)$. When we integrate $k_1^4$ and $k_2^4$ from $-\infty$ to $\infty$, we get a zero result. Therefore, the imaginary part of $\mathcal{M}_{11}$ is zero.
Similarly, for $\mathcal{M}_{12}$ and $\mathcal{M}_{21}$ the imaginary part of the integrands are odd functions of $k_1^0$ and $k_2^0$, giving zero as a final result. Let us take $\mathcal{M}_{12}$ as an example. The $i$ terms come from the third and the fourth denominators:
\be
&&
\frac{1}{[(\bar{k}_{2;4}^0)^2 - \bm{k}_2^2 - m^2 +i\epsilon]\{[(\bar{k}_{2;4}^0)^2 - \bm{k}_2^2]^2 +M^4\}}\nonumber \\
&& \qquad\qquad\qquad
= \frac{1}{(p^0 - i k_1^4 - \sqrt{\bm{k}_1^2 + m^2 -i \epsilon})^2 - \bm{k}_2^2 - m^2 +i\epsilon}\nn
&&\qquad\qquad\qquad\quad\times\frac{1}{[(p^0 - i k_1^4 - \sqrt{\bm{k}_1^2 + m^2 -i \epsilon})^2 - \bm{k}_2^2]^2 +M^4}\nonumber
\ee
The key idea to take from this formula is that, since $i$ factors always appear with $k_1^4$, the imaginary part of the integrand is an odd function of $k_1^4$. Therefore, the integral in $d k_1^4$ vanishes and the imaginary part of $\mathcal{M}_{12}$ is zero. Similarly, since the imaginary part of the integrand of $\mathcal{M}_{21}$ is an odd function of $k_2^4$, the whole imaginary part of $\mathcal{M}_{21}$ is zero. 
Finally, the imaginary part of the whole amplitude comes only from $\mathcal{M}_{22}$, and the full two-loop amplitude (\ref{two-loops}) satisfies the following Cutkosky rule (see Appendix \ref{M22C}):
\be
{\rm Disc} \,\mathcal{M} &=& 2i\, {\rm Im} \,\mathcal{M} \nonumber \\
&=& -\frac{\lambda^2}{2}\frac{M^{12}}{(m^4+M^4)^3} \int \frac{id^4 k_1}{(2\pi)^4} \int \frac{id^4 k_2}{(2\pi)^4}(-2\pi i)^3 \delta(k_1^2-m^2)\delta(k_2^2-m^2)\nn
&&\qquad\qquad\times\delta[(p-k_1-k_2)^2-m^2]\, . 
\label{Final2L}
\ee

\subsection{General Cutkosky rules}

Having checked the Cutkosky rules for several amplitudes, we can now state the general result that we will later use to prove unitarity for the scalar field theory with cubic interactions $\phi^3$. The generalization to $\phi^n$ is straightforward. 

The final statement for the Cutkosky rules in an higher-derivative theory with normal real particles and complex pairs is
%
%
%
\be \label{unitarity m 1}
&& \hspace{-1cm} 
\lim_{\epsilon\rightarrow 0} \left[ \mathcal{M}(E_h,\epsilon)-\mathcal{M}(E_h,\epsilon)^* \right] \equiv {\rm Disc} \,\mathcal{M}(E_h) \, ,
 \nonumber \\ 
 && 
 \hspace{-1cm} 
 {\rm Disc} \,\mathcal{M}(E_h)
= - \frac{ \lambda^V  }{S_{\#} } \sum \int_{\Omega_1} \ldots
\int_{\Omega_L}\, \prod_{i=1}^L \frac{i
	\, d^4 k_i}{(2\pi)^4} \prod_{k=1}^N (-2\pi i)\,\delta(Q_k^2-m^2) \sigma(Q^0_k)  \times \nonumber \\
	&& \qquad\qquad\times 
\prod_{j=1}^{I-N} \frac{i M^4 }{\left( Q_j^2 - m^2 + i \epsilon \right) \left[ Q_j^4 + M^4  \right]} \,  
B(k_i,p_h) \, ,
\label{CutkoskyF}
\ee
where the $Q_k$ are the momenta corresponding to internal lines ($I$ is the number of internal lines). Each term in the sum in (\ref{unitarity m 1}) corresponds to a cut diagram in which $N$ propagators of real particles (never complex particles) are on-shell while $I-N$ are not on-shell. $E_h=p_h^0$ are the external energies not involved in the cut. Moreover, for each term, the $i$-th integration region $\Omega_i$ can be $\mathbb{R}^4$ or $\mathcal{I} \times \mathbb{R}^3$, depending on whether the corresponding momenta $k_i$ are contained in the propagator of one of the cut lines or not. Finally, $B(k_i,p_h)$ is a general function of the loop momenta, internal, and external energies. In the $\phi^3$-theory  studied in this paper, the function $B(k_i,p_h)$ is just a constant, but it is non-trivial in the gauge theory (\ref{gauge}) and in the gravitational theory (\ref{gravity}), or even in scalar theories with derivatives in the vertices.

\section{Perturbative unitarity {of the scalar theory}}

The Cutkosky rules are only the first of the two pieces needed to prove unitarity, which requires the S-matrix to satisfy 
\be
S^\dagger S = \mathbbm{1}\,.
\label{SdS}
\ee
In order to properly state the unitarity condition, we need to review a basic result for the scalar field theory. The equations of motion for free particles in the theory (\ref{phin1}) reads
\be
\left[ \left( \frac{\Box}{M^2} \right)^2 + 1 \right] 
\left(\Box +  m^2 \right)\phi = 0 \,,
\ee
whose solutions are plane waves (normal real particles or stable perturbations at classical level),
\be
\hspace{-.6cm}\left(\Box +  m^2 \right)\phi = 0  \quad \Longrightarrow \quad 
\phi (x) = \int \frac{d^3 \bm{p}}{(2 \pi )^3 \sqrt{2 \omega_{ \bm{p} }}} \left(a_{\bm{p}} \, e^{- i p\cdot x} + a^\dagger_{\bm{p}} \, e^{i p\cdot x}  \right), \qquad p^2 = m^2 \, , 
\label{solR}
\ee
and damped or runaway solutions,
\be
\left[ \left( \frac{\Box}{M^2} \right)^2 + 1 \right] \phi_{\rm d, r} = 0 \quad \Longrightarrow \quad 
\phi_{\rm d,r} \sim e^{ i p\cdot x} \, , \quad {\rm for} \quad p^2 = i M^2 \, , 
\ee
which we can discard from the spectrum of the asymptotic states because they cannot be regenerated in the loop amplitudes, as explained in the introduction and according to the Cutkosky rules (\ref{CutkoskyF}). 

{
From the Lagrangian given in (\ref{phin1}), the conjugate momentum to the real field $\phi$ reads
\be
\Pi = \frac{\partial {\cal L}}{\partial \dot \phi} =  \left[ \left( \frac{\Box}{M^2} \right)^2 + 1 \right] \dot{\phi} \, , 
\label{P1}
\ee
which on-shell, making use of the solution (\ref{solR}), turns into 
\be
\Pi =  \frac{m^4  + M^4}{M^4} \, \dot{\phi} \equiv {\rm c} \,  \dot{\phi}  \, , 
\label{P2}
\ee


Imposing the canonical commutation relations
}
\be
[ \phi(\bm{x} ,t) , \, \Pi (\bm{y}, t)] = i \delta^3(\bm{x} - \bm{y}) \, ,
\label{phiPi}
\ee
we find the following commutation relations for $a_{\bm{p}}$ and $a^\dagger_{\bm{p}}$,\footnote{The reader can, for example, check (\ref{phiPi}) assuming the commutation relations (\ref{ACOperators}).}
\be
[ a_{\bm{p}} , \, a^\dagger_{\bm{k}} ] = {\frac{1}{\rm c} } (2 \pi)^3 \delta^3(\bm{p} - \bm{k} ) \, .
\label{ACOperators}
\ee
Therefore, defining the one-particle state of momentum $\bm{p}$ by 
\be
a^\dagger_{\bm{p}} | 0 \rangle = {\frac{1}{\rm c} } \frac{1}{\sqrt{2 \omega_{ \bm{p} }} } | \bm{p} \rangle \, , 
\label{1p}
\ee
and assuming the normalization $\langle 0 | 0 \rangle$ = 1 of the vacuum state, the scalar product in the momentum Hilbert space reads
\be
\langle \bm{p} | \bm{k} \rangle = 2 \sqrt{\omega_{\bf p}}  \sqrt{\omega_{\bf k}} \langle 0 | a_{\bm{p}} a^\dagger_{\bm{k}} | 0 \rangle 
= {{\rm c} }
2 \omega_{ \bm{p} } \, (2 \pi)^3 \delta^3(\bm{p} - \bm{k} ) \, . 
\ee
Finally, the identity operator for one-particle states in momentum space is
\be
 \int \frac{d^3 \bm{p}}{(2 \pi )^3 2 \omega_{ \bm{p} }} | \bm{p} \rangle \langle \bm{p} | \, {\frac{1}{\rm c}= \int \frac{d^3 \bm{p}}{(2 \pi )^3 2 \omega_{ \bm{p} }} | \bm{p} \rangle \langle \bm{p} | \, \frac{M^4}{m^4  + M^4}}= \mathbbm{1} \, .
 \label{completeness}
\ee

We now introduce the definition of amplitude $\mathcal{T}$, namely 
\be 
S = \mathbbm{1}+ i \, \mathcal{T} \, .
\ee
Hence, from the unitarity condition (\ref{SdS}) we get
\be
S^\dagger S = \mathbbm{1} \quad \Longrightarrow \quad \mathcal{T} - \mathcal{T}^\dagger = i \mathcal{T}^\dagger \mathcal{T} \, .
\label{TdT}
\ee
We now take the expectation value of Eq.\ (\ref{TdT}) between an initial and a final state and we plug the completeness relation in between $\mathcal{T}^\dagger$ and $\mathcal{T}$ in (\ref{TdT}), making use of the following short notation for (\ref{completeness}): 
\be
\sum_{\ell = 1}^{n} | \ell \rangle \langle \ell | = \mathbbm{1}\,.
\ee
{Here and in the following, we omit ${\rm c}$ factors and restore them at the end.} We obtain 
\be
&&\langle f | \mathcal{T} | i \rangle  - \langle f | \mathcal{T}^\dagger | i \rangle 
 = i \sum_{\ell = 1}^{n} \langle f | T^\dagger | \ell \rangle \langle \ell \mathcal{T} | i \rangle 
 \qquad \Longrightarrow \nn
&& \mathcal{T}_{f i} - \mathcal{T}^*_{i f}  = i \sum_{\ell = 1}^{n}  \mathcal{T}^*_{\ell f} \mathcal{T}_{\ell i}
 = 
 i \,
  \sum_n \frac{1}{{\rm s}_n}
 \int \prod_{\ell = 1}^n \frac{d^3 \bm{p}_\ell}{(2 \pi)^3 2 E_\ell} \mathcal{T}^*_{\ell f} \mathcal{T}_{\ell i}
  \,,
 \label{TsT}
\ee
where $E_\ell=\om_{\bm{p}_\ell}$. Notice that the sum in (\ref{TsT}) now involves all possible $n$ intermediate states having fixed the initial and final states. 
We also introduced the symmetry factor $1/{\rm s}_n$, which can be reinterpreted as the symmetry factor for identical bosons in the final state. 
Moreover, (\ref{TsT}) can be used as a perturbative relation in the coupling constant and/or the loop expansion in $\hbar$. 

We now introduce the definition of the amplitude $\mathcal{M}$ in terms of $\mathcal{T}$:
\be
\mathcal{T}_{a b} = (2 \pi)^4 \delta^4(p_a - p_b) \mathcal{M}_{ab}.
\ee
Hence, the unitarity condition in four dimensions turns into
\be
&&
(2 \pi)^4 \delta^4(p_f - p_i) \mathcal{M}_{f i} - (2 \pi)^4 \delta^4(p_i - p_f) \mathcal{M}^*_{i f} \nonumber \\
&& \qquad\qquad
= i 
\sum_n \frac{1}{{\rm s}_n}
 \int \left[ \prod_{\ell = 1}^n \frac{d^3 \bm{p}_\ell}{(2 \pi)^3 2 E_\ell} \right] 
 (2 \pi)^4 \, \delta^4 \left( p_i - \sum_{\ell = 1}^n p_\ell \right)\nn
&&\qquad\qquad\quad\times 
(2 \pi)^4 \, \delta^4 \left( p_f - \sum_{\ell = 1}^n p_\ell \right) 
\mathcal{M}^*_{\ell f} \mathcal{M}_{\ell i} \nonumber \\
&& \qquad\qquad
= i 
 \sum \hspace{-0.5cm}\int
  \left[ \prod_{\ell = 1}^n \frac{d^3 \bm{p}_\ell}{(2 \pi)^3 2 E_\ell}  \right] 
 (2 \pi)^4 \delta \left( p^0_i - \sum_{\ell = 1}^n p^0_\ell \right) \delta^3 \! \left( \bm{p}_i - \sum_{\ell = 1}^n \bm{p}_\ell \right)\nn
&&\qquad\qquad\quad\times
(2 \pi)^4 \delta \left( p^0_f - \sum_{\ell = 1}^n p^0_\ell \right) \delta^3 \! \left( \bm{p}_f - \sum_{\ell = 1}^n \bm{p}_\ell \right)
\mathcal{M}^*_{\ell f} \mathcal{M}_{\ell i} , 
\ee
where, for the sake of shortness, we introduced the notation
\be
\sum_n \frac{1}{{\rm s}_n} \int \equiv \sum \hspace{-0.5cm}\int \, .
\ee
Now we integrate in the momentum variable $\bm{p}_n$, which means that we have to replace 
$\bm{p}_n = \bm{p}_i - \sum_{\ell=1}^{n-1} \bm{p}_\ell$ in the Dirac deltas:
\be
&&
(2 \pi)^4 \delta^4(p_f - p_i) \mathcal{M}_{f i} - (2 \pi)^4 \delta^4(p_i - p_f) \mathcal{M}^*_{i f} \nonumber\\
&&\qquad\qquad = i
 \sum \hspace{-0.5cm}\int
 \left[  \prod_{\ell = 1}^{n-1}  \frac{d^3 \bm{p}_\ell}{(2 \pi)^3 2 E_\ell}  
 \right] 
 \int   \frac{d^3 \bm{p}_n}{(2 \pi)^3 2 E_n}\,(2 \pi)^4 \delta \left( p^0_i - \sum_{\ell = 1}^n p^0_\ell \right) 
 \delta^3 \left( \bm{p}_i - \sum_{\ell = 1}^n \bm{p}_\ell \right) \nonumber \\
 &&\qquad\qquad\quad \times (2 \pi)^4 \delta \left( p^0_f - \sum_{\ell = 1}^n p^0_\ell \right) 
 \delta^3 \left( \bm{p}_f - \sum_{\ell = 1}^n \bm{p}_\ell \right) 
\mathcal{M}^*_{\ell f} \mathcal{M}_{\ell i} \nonumber\\
&& \qquad\qquad
= i
 \sum \hspace{-0.5cm}\int 
 \left[ \prod_{\ell = 1}^{n-1}  \frac{d^3 \bm{p}_\ell}{(2 \pi)^3 2 E_\ell}\right]  
  \frac{1}{(2 \pi)^3 2 E_n}  
 (2 \pi)^4 \delta \left( p^0_i - \sum_{\ell = 1}^n p^0_\ell \right) 
 \delta^3 \left( \bm{p}_i -  \bm{p}_f \right)\nonumber\\
&& \qquad\qquad\quad\times
(2 \pi)^4 \delta \left( p^0_f - \sum_{\ell = 1}^n p^0_\ell \right) 
\mathcal{M}^*_{\ell f} \mathcal{M}_{\ell i}.\label{mms}
\ee
Let us now recall the following identity in the space of distributions:
\be
\delta(x-z) \delta (y - z) = \delta (x-y) \delta(x - z) \, .
\label{ddd}
\ee
The proof is simple. Integrating in the $z$ variable the left-hand side of (\ref{ddd}) with the insertion of a fast-decreasing test function $f(z)$, we get
\be
\int d z f(z) \delta(x-z) \delta (y - z) = \delta (x-y) f(y) \, .
\label{dddL}
\ee
On the other hand, integrating in the $z$ variable the right-hand side of (\ref{ddd}) with the insertion of the same test function, we get
\be
\int d z f(z) \delta (x-y) \delta(x - z) = \delta (x-y) f(x) \, .
\label{dddR}
\ee
Therefore, (\ref{dddL})$=$(\ref{dddR}). 

Using (\ref{ddd}) with the identifications
\be
z = \sum_{\ell =1}^n p^0_\ell \, , \qquad x = p_i^0 \, , \qquad y = p^0_f \, , 
\ee
we can rewrite (\ref{mms}) as
\be
&& 
(2 \pi)^4 \delta^4(p_f - p_i) \mathcal{M}_{f i} - (2 \pi)^4 \delta^4(p_i - p_f) \mathcal{M}^*_{i f}
\nonumber\\
&& 
\qquad\qquad= i
 \sum \hspace{-0.5cm}\int
 \left[  \prod_{\ell = 1}^{n-1}  \frac{d^3 \bm{p}_\ell}{(2 \pi)^3 2 E_\ell}   \right] 
  \frac{1}{(2 \pi)^3 2 E_n}  
 (2 \pi)^4 \, \delta \left( p^0_i -  p^0_f \right) \delta^3 \left( \bm{p}_i -  \bm{p}_f \right)\nn
&&\qquad\qquad\quad\times
(2 \pi)^4 \,  \delta \left( p^0_i - \sum_{\ell = 1}^n p^0_\ell \right)
\mathcal{M}^*_{\ell f} \mathcal{M}_{\ell i} \nonumber \\
&&\qquad \qquad
= i
 \sum \hspace{-0.5cm}\int
  \left[ \prod_{\ell = 1}^{n-1}  \frac{d^3 \bm{p}_\ell}{(2 \pi)^3 2 E_\ell}   \right] 
  \frac{1}{(2 \pi)^3 2 E_n}  
 (2 \pi)^4 \, \delta^4 \left( p_i -  p_f \right)\nn
&&\qquad\qquad\quad\times
(2 \pi)^4 \, \delta \left( p^0_i - \sum_{\ell = 1}^n p^0_\ell \right) 
\mathcal{M}^*_{\ell f} \mathcal{M}_{\ell i} 
\nonumber 
\\
&&\qquad\qquad
= i
(2 \pi)^4 \delta^4 \left( p_i -  p_f \right) 
 \sum \hspace{-0.5cm}\int
   \left[ \prod_{\ell = 1}^{n-1}  \frac{d^3 \bm{p}_\ell}{(2 \pi)^3 2 E_\ell}  \right] 
  \frac{1}{(2 \pi)^3 2 E_n} \nn
&&\qquad\qquad\quad\times
(2 \pi)^4 \delta \left( p^0_i - \sum_{\ell = 1}^n p^0_\ell \right) 
\mathcal{M}^*_{\ell f} \mathcal{M}_{\ell i} \, .\label{mms2}
\ee
Factoring out the Dirac delta for energy-momentum conservation on both sides, we finally get
\be
\mathcal{M}_{f i} -  \mathcal{M}^*_{i f} = i
\sum \hspace{-0.5cm}\int
  \left[ \prod_{\ell = 1}^{n-1}  \frac{d^3 \bm{p}_\ell}{(2 \pi)^3 2 E_\ell}  \right] 
  \frac{1}{(2 \pi)^3 2 E_n} 
(2 \pi)^4 \delta \left( p^0_i - \sum_{\ell = 1}^n p^0_\ell \right) 
\mathcal{M}^*_{\ell f} \mathcal{M}_{\ell i} \, .
\label{UnitarityMsM}
\ee

In order to apply the formula (\ref{UnitarityMsM}) to higher-derivative theories with complex ghosts, we must {restore the following factor for each internal line, compatibly with the normalization of the momentum Fock states}:  
\be \nonumber
\frac{M^4}{M^4 + m^4} \, .
\ee
%
Therefore, the unitarity condition becomes 
\be
\mathcal{M}_{f i} -  \mathcal{M}^*_{i f} &=& i
\sum \hspace{-0.5cm}\int
 \left[ \prod_{\ell = 1}^{n-1}  \frac{d^3 \bm{p}_\ell}{(2 \pi)^3 2 E_\ell}\, \frac{M^4}{M^4 + m^4} 
\right] 
  \frac{1}{(2 \pi)^3 2 E_n} \nn
	&&\times \frac{M^4}{M^4 + m^4} \, 
(2 \pi)^4 \delta \left( p^0_i - \sum_{\ell = 1}^n p^0_\ell \right) 
\mathcal{M}^*_{\ell f} \mathcal{M}_{\ell i} \, .
\label{UnitarityMsMG}
\ee

\subsection{Unitarity at one loop} 
Let us now apply (\ref{UnitarityMsMG}) to the one-loop bubble diagram. In this case, we denote the initial and final states by $i=f=1$. Hence, according to (\ref{DiscResR}) the {left-hand side of (\ref{UnitarityMsMG})} reads:
\be
{\rm Disc} \,\mathcal{M}_{11} = 2 i {\rm Im} \,\mathcal{M}_{11} =  i  \frac{\lambda^2}{16\pi}\frac{M^8}{(m^4+M^4)^2}\sqrt{1-\frac{4 m^2}{p_0^2}}\sigma(p_0 - 2m) \, .
\label{DiscResRF}
\ee

Now we work out the {right-hand side of (\ref{UnitarityMsMG})} for $n=2$, corresponding to the two internal lines of the bubble diagram:
\be 
\hspace{-.5cm}i
\frac{1}{\rm s_2} 
\int  \frac{d^3 \bm{p}_1}{(2 \pi)^3 \, 2 E_1}  
\, \frac{M^4}{M^4 + m^4} \, 
  \frac{1}{(2 \pi)^3 \,  2 E_2} \, \frac{M^4}{M^4 + m^4} \, 
(2 \pi)^4 \delta \left( p^0 - E_1 - E_2 \right) 
\mathcal{M}^*_{2 1} \mathcal{M}_{2 1} \, .
\label{UnitarityMsMGB}
\ee
where $p^0 \equiv p^0_i$ is the initial external energy. Moreover, the symmetry factor is $1/{\rm s_2} =1/2$ because we have two identical bosonic particles as a final state of the amplitude $\mathcal{M}_{1 2}$.

In order to complete the proof, we need the scattering amplitude for one particle into two, namely 
\be
S^{(1)}_{12} &=& \mathbbm{1} + i \, \mathcal{T}^{ (1) }_{12} =\mathbbm{1} +  i \frac{(- \lambda)}{3 !}  3 ! (2 \pi)^4 \delta^4 (p_i- p_1 - p_2)\nn
&=& \mathbbm{1} - i \lambda (2 \pi)^4 \delta^4 (p_i - p_1 - p_2) \, , 
\label{S12}
\ee
where $p_i = (p^0 , \, \bm{p} )$. Therefore,
\be
\mathcal{M}_{12} = - \lambda \,, 
\ee
where, again, we omitted the two factors $M^4/(M^4 + m^4)$ coming from the external legs. Such constant cancels the same contribution present on the left hand side of the unitarity condition (\ref{UnitarityMsM}).

In order to simplify the integral in $\bm{p}_1$ in (\ref{UnitarityMsMGB}) we set $\bm{p} = 0$. Hence, the momentum conservation for the process (\ref{S12}) implies $\bm{p} = \bm{p}_1 + \bm{p}_2 = 0$, or $\bm{p}_1 = - \bm{p}_2$, which, together with 
$E_1 = \sqrt{\bm{p}_1^2 + m^2}$ and $E_2 = \sqrt{\bm{p}_2^2 + m^2}$, finally gives  $E_1 = E_2$. 
Now the amplitude (\ref{UnitarityMsMGB}) in spherical coordinates reads
\be
\hspace{-1.2cm}&& i \,
{\frac{1}{2}}
\int  \frac{ d |\bm{p}_1|\,4 \pi |\bm{p}_1|^2 }{(2 \pi)^3 \, 2 E_1}  
\, \frac{M^4}{M^4 + m^4} \, 
  \frac{1}{(2 \pi)^3 \,  2 E_2} \, \frac{M^4}{M^4 + m^4} \, 
(2 \pi)^4 \delta \left( p^0 - E_1 - E_2 \right) 
\mathcal{M}^*_{2 1} \mathcal{M}_{2 1} \nonumber\\
\hspace{-1.2cm}&&\qquad
=  i \,
{\frac{1}{2}}
\left(\frac{M^4}{M^4 + m^4} \right)^2 
\int  \frac{ d |\bm{p}_1|\,4 \pi |\bm{p}_1|^2 }{(2 \pi)^3 \, 2 E_1}  
 \, 
  \frac{1}{(2 \pi)^3 \,  2 E_2} 
   \, 
(2 \pi)^4 \delta \left( p^0 - E_1 - E_2 \right) 
\mathcal{M}^*_{2 1} \mathcal{M}_{2 1} .
\label{UnitarityMsMGC}
\ee
We now make the following change of variables,
\be
E_1^2 = \bm{p}_1^2 + m^2 \, , \quad E_1 \, d E_1 = |\bm{p}_1| \, d |\bm{p}_1| \, , 
\ee
in the integral (\ref{UnitarityMsMGC}), which turns into (notice that $E_2=E_1$) 
\be
\hspace{-1.2cm}&& i 
\,
{\frac{1}{2}}
\left(\frac{M^4}{M^4 + m^4} \right)^2 
 \int  \frac{ d E_1\,4 \pi |\bm{p}_1| E_1}{(2 \pi)^3 \, 2 E_1}  
 \, 
  \frac{1}{(2 \pi)^3 \,  2 E_1}
   \, 
(2 \pi)^4 \delta \left( p^0 - 2 E_1 \right) 
\mathcal{M}^*_{2 1} \mathcal{M}_{2 1} \nonumber\\
\hspace{-1.2cm}&&\qquad
=
i \,
{\frac{1}{2}}
\left(\frac{M^4}{M^4 + m^4} \right)^2 
 \int  \frac{ d E_1\,4 \pi |\bm{p}_1| \cancel{E_1}}{(2 \pi)^3 \, 2 \cancel{E_1}}  
 \, 
  \frac{1}{ \cancel{(2 \pi)^3} \,  2 E_1}
   \, 
(2 \pi)^{\cancel{4}} \delta \left( p^0 - 2 E_1 \right) 
\mathcal{M}^*_{2 1} \mathcal{M}_{2 1} .
\label{UnitarityMsMGD-1}
\ee
Replacing $|\bm{p}_1| = \sqrt{E_1^2 - m^2}$ in (\ref{UnitarityMsMGD-1}), 
\be
&&
i \,
{\frac{1}{2}}
\left(\frac{M^4}{M^4 + m^4} \right)^2 \frac{1}{4 \pi}
 \int    d E_1 
 \, \sqrt{E_1^2 - m^2} \,
  \frac{1}{ E_1}
   \, 
\delta \left( p^0 - 2 E_1  \right) 
\mathcal{M}^*_{2 1} \mathcal{M}_{2 1}
 \nonumber\\
&&\qquad\qquad
=
i \,
{\frac{1}{2}}
\left(\frac{M^4}{M^4 + m^4} \right)^2 \frac{1}{4 \pi}
 \int    d E_1 
 \, \sqrt{E_1^2 - m^2} \,
  \frac{1}{ E_1}
   \, 
\delta \left[ 2 \left( \frac{p^0}{2}  - E_1  \right) \right] 
\mathcal{M}^*_{2 1} \mathcal{M}_{2 1}
\nonumber\\
&&\qquad\qquad
=
i \, 
{\frac{1}{2}}
\left(\frac{M^4}{M^4 + m^4} \right)^2 \frac{1}{4 \pi}
 \int    d E_1 
 \, \sqrt{1 - \frac{m^2}{E_1^2}} \,
   \frac{1}{2} 
   \, 
\delta \left(   \frac{p^0}{2}  - E_1   \right) 
\mathcal{M}^*_{2 1} \mathcal{M}_{2 1}
\nonumber\\
&&\qquad\qquad
=
i \, 
{\frac{1}{2}}
\frac{\lambda^2}{4 \pi}  \frac{1}{2}
\frac{M^8}{(M^4 + m^4)^2} \frac{1}{4 \pi}
  \sqrt{ 1 - \frac{4 m^2}{p_0^2} } 
 \, \sigma(p^0 - 2m)
 \nonumber\\
&&\qquad\qquad
=
i \, 
\frac{\lambda^2}{16 \pi} 
\frac{M^8}{(M^4 + m^4)^2} \frac{1}{4 \pi}
  \sqrt{ 1 - \frac{4 m^2}{p_0^2} } 
 \, \sigma(p^0 - 2m)
  \, ,
\label{UnitarityMsMGD}
\ee
which is identical to the left-hand side of the unitarity condition (\ref{UnitarityMsMG}). That is to say, (\ref{UnitarityMsMGD}) coincides with (\ref{DiscResRF}).

\section{Perturbative unitarity of gauge theories and gravity}\label{section gauge gravity}

The Cutkosky rules (\ref{CutkoskyF}) and the unitarity condition (\ref{UnitarityMsMG}) have been derived for a scalar field theory. However, in order to prove the unitarity of local higher-derivative gauge or gravitational theories, we have to take into account gauge invariance and diffeomorphism invariance, respectively. Therefore, we need to show the cancellation of unphysical cuts, namely, that the unphysical longitudinal and timelike degrees of freedom cannot go on-shell. Actually, such a proof is not so different from the one performed for the two-derivative gauge or gravity case. In the scalar field theory, the tensorial structure is trivial, but in local higher-derivative gauge and gravitational theories some of the components of the propagators seem to violate unitarity at the cuts when the Cutkosky rules are implemented. However, this is not the case because of the presence of the Fadeev--Popov ghosts that also contribute to the cuts and exactly cancel the unphysical degrees of freedom. Hence, when we cut the loop amplitudes only physical states (physical gauge-invariant transverse polarizations) can go on-shell. 

{The difference between a scalar field theory and a gauge or gravitational theory only consists on showing that the unphysical degrees of freedom cancel out with equal contributions coming from the BRST ghosts. For this to happen, it is crucial that the vertices are either local (as is the case of the present paper) or weakly nonlocal \cite{Briscese:2018oyx}. If the latter requirement is not accomplished, then we can have poles also in the vertices and unitarity becomes very hard to prove. Therefore, the following proof does not cover the case of nonlocal theories with other types of operators.}

In order to explicitly prove unitarity, we need to use the Ward identities that have been derived for a general field theory in \cite[Sec.\ 2.7]{Shapirobook} and more recently in \cite{Lavrov:2019nuz}. The cancellation of unphysical cuts relies only on the gauge or diffeomorphism (or their quantum BRST version) invariance of the action, and it is usually stated at the formal level of the path-integral independence of the gauge-fixing term, which, of course, holds here as well.

\subsection{Unitarity in two-derivative gauge theories}\label{unitwoder}

{We first recall the proof of unitarity for local gauge theories at one loop described in \cite{Tom97,Lan16}, which we report here in full since the details will be relevant for its extension to the case of gravity, which we will also discuss. All the diagrams are adapted from \cite{Lan16}. The proof consists in showing that the sum over all the cuts of two-particle intermediate states, i.e., the gauge field states and the two BRST ghosts states, equals the sum over the cuts of physical two-particle intermediate states:}
\be\label{mainlan}
\centering
\hspace{-1.1cm}\parbox{14cm}{\includegraphics[width=0.95\textwidth]{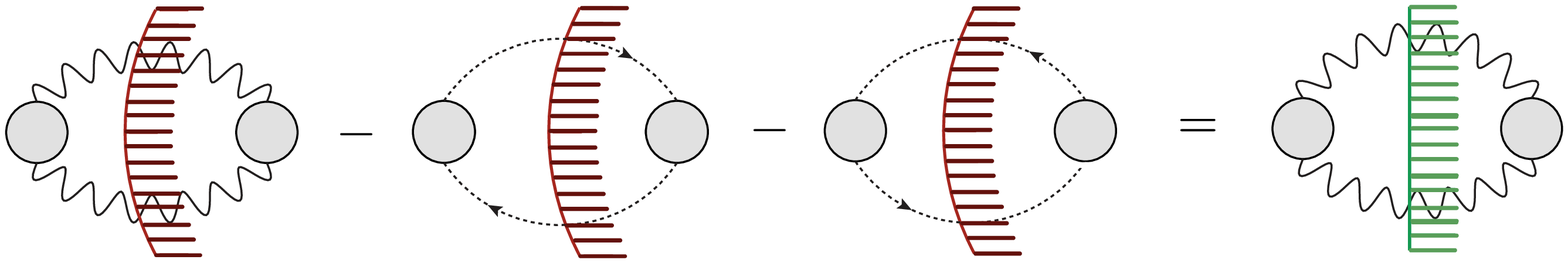}}
\ee
{The comb-like symbols represent the following cut propagators:}
\bea
\parbox{3.5cm}{\includegraphics[height=2.0cm]{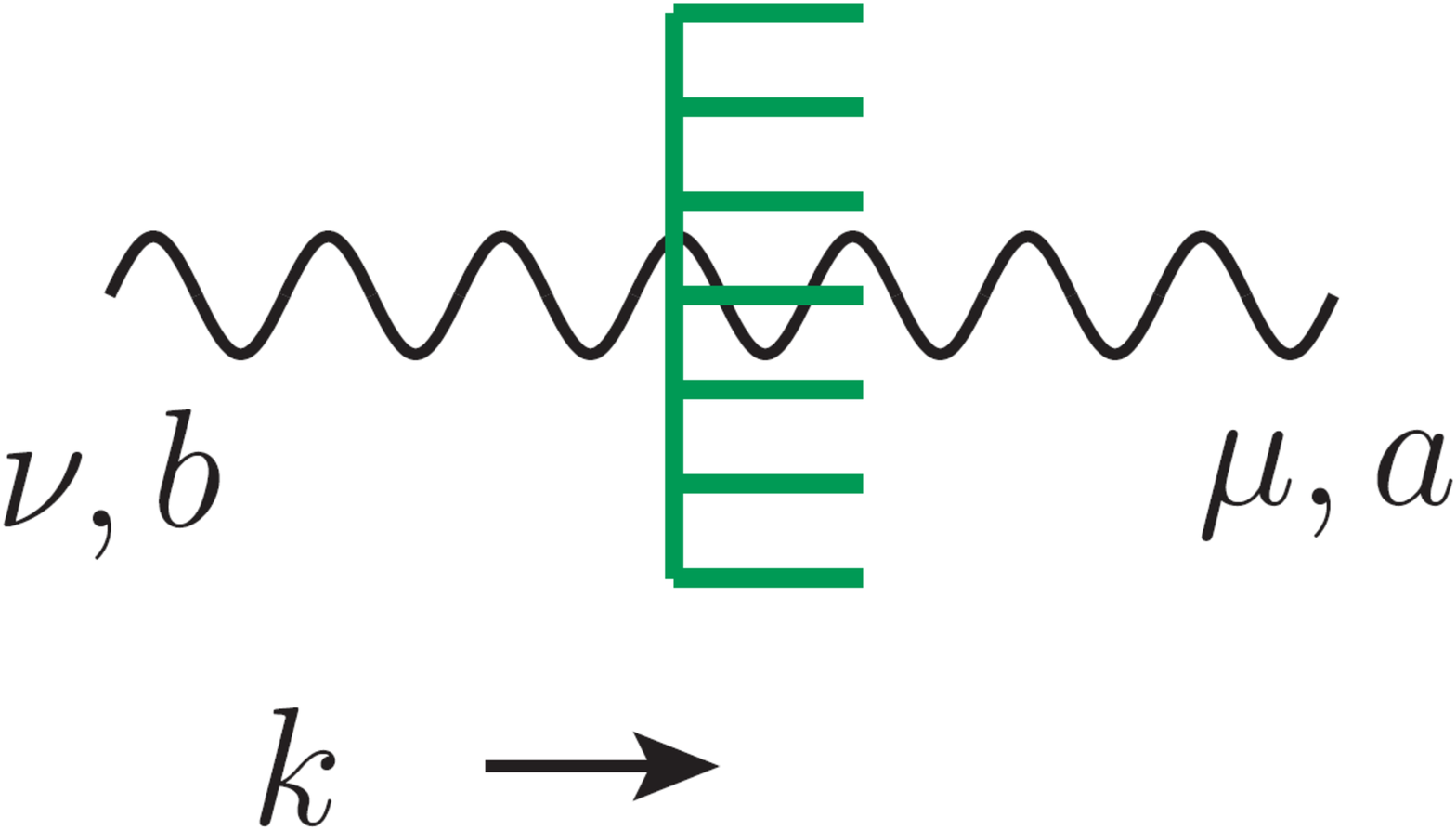}} &=& -\delta_{ab}\left[\eta_{\mu\nu}+\frac{\bar k_\mu k_\nu+k_\mu \bar k_\nu}{2(k\cdot\zeta)^2}\right]D_+(k)\label{propa1}\,,\\
\parbox{3.5cm}{\includegraphics[height=2.0cm]{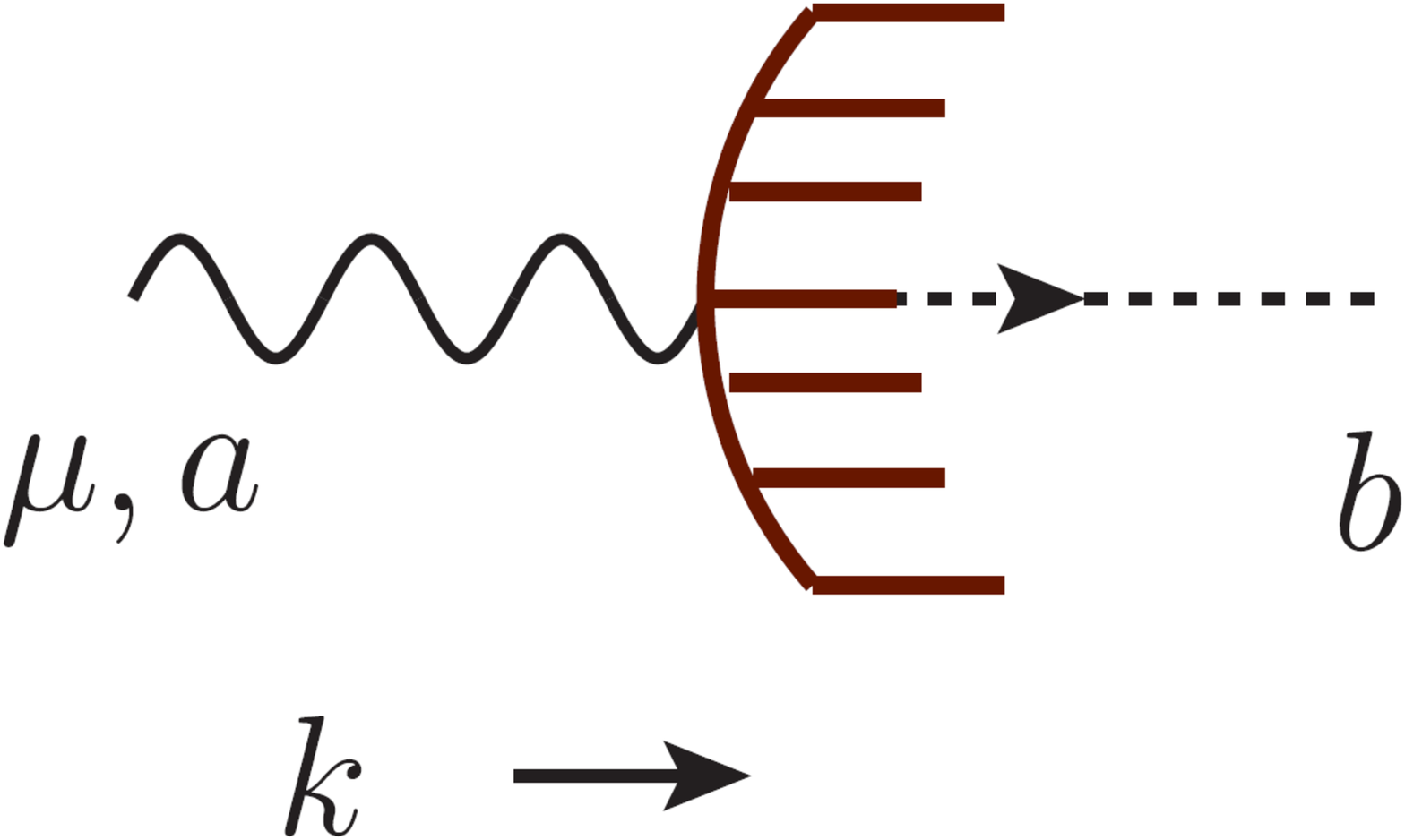}} &=& \delta_{ab}\frac{i\bar k_\mu}{2(k\cdot\zeta)^2}D_+(k)\,, \label{propa2} \\
\parbox{3.5cm}{\includegraphics[height=2.0cm]{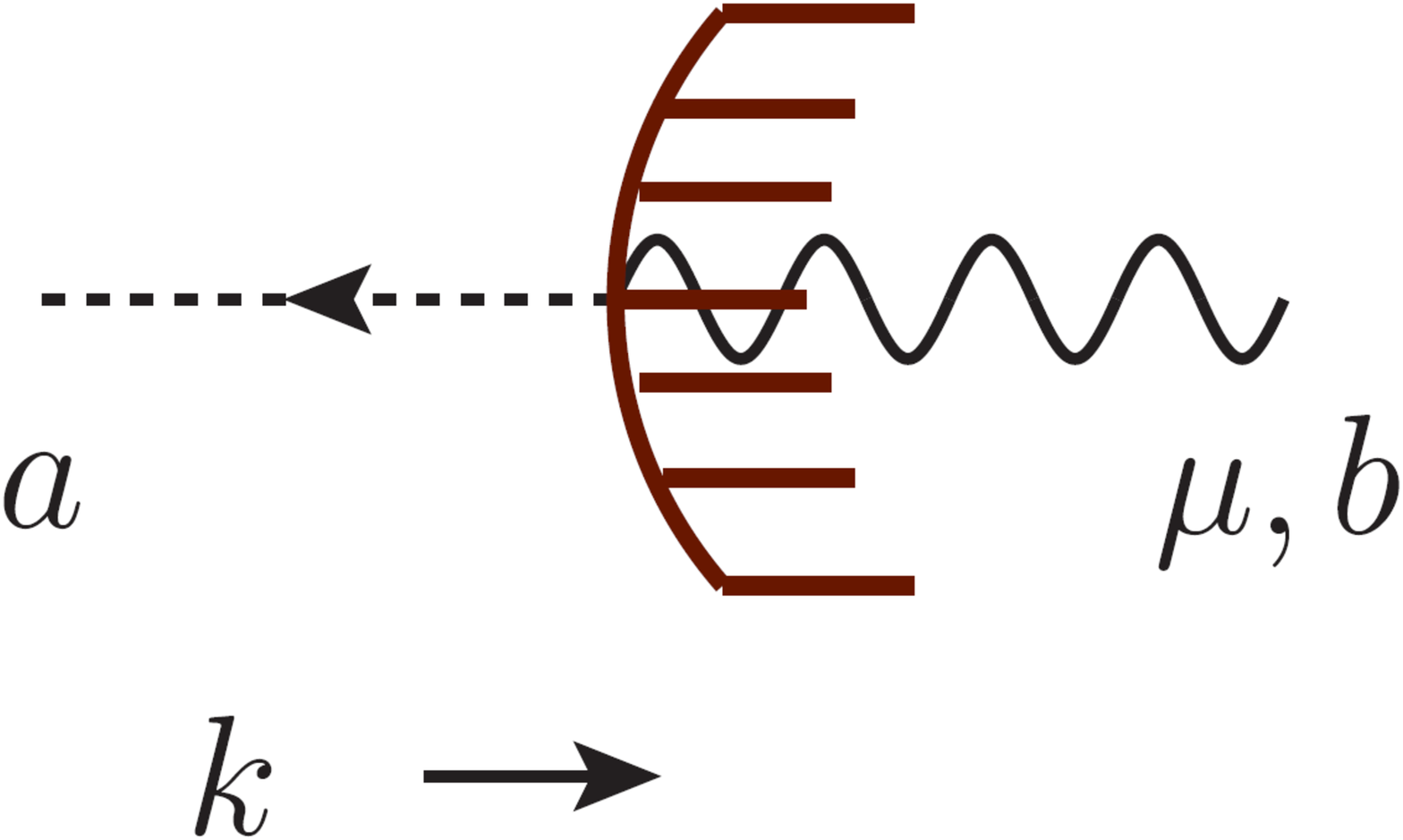}} &=& \delta_{ab}\frac{-i\bar k_\mu}{2(k\cdot\zeta)^2}D_+(k)\,,\label{propa3}
\eea
{where $\bar k_\mu$ and $\zeta_\mu$ are the following vectors:
\be
\bar k_\mu := k_\mu-2(k\cdot\zeta)\zeta_\mu\,,\qquad \zeta_\mu:=(1,0,0,0)_\mu\,,
\ee
and $D_+$ is the cut massless ghost propagator
\be
D_+(k) := 2 \pi\theta(k_0)\delta(k^2)\, .
\label{on-shell}
\ee
Notice that the operator (\ref{propa1}) only propagates physical degrees of freedom.

To reach eq.\ (\ref{mainlan}), as a starting point we need the Ward identities derived for a local or weakly nonlocal general theory in
 \cite{Shapirobook,Tom97}. For the theory (\ref{acym}), such identities in their functional form are
\be
0&=&\int[{\cal D} A] [{\cal D} C] [{\cal D} \bar C]\exp\left[i\int({\cal L}+J\cdot A+\bar\eta C+\eta\bar C)\right]\nonumber\\
&&\times\int\left[J_a^{\mu}D^{ab}_\mu C^b-\frac12 \bar\eta^a f_{abc}C^b C^c+\frac{1}{\xi_{\textsc{ym}}}{\rm P}({\cal D}^2_M)(\partial_\mu A_a^{\mu})\eta^a\right],
\label{WI}
\ee
where $J$, $\eta$ and $\bar\eta$ are external sources for the gauge field and ghosts, $D^{ab}_\mu$ is the gauge-covariant derivative and $f_{abc}$ are the structure constants of the gauge group. Notice that the form of the identities is the same of ordinary two-derivatives theories because it only relies on the BRST invariance of the theory, namely, it is only a consequence of gauge or diffeomorphism invariance of the action. 

From (\ref{WI}), one derives a version of the Ward identities for on-shell gauge vector amplitudes \cite{Tom97}, which we write in the following diagrammatic form:}
\be\label{wardid}
\centering
\parbox{14cm}{\includegraphics[width=0.8\textwidth]{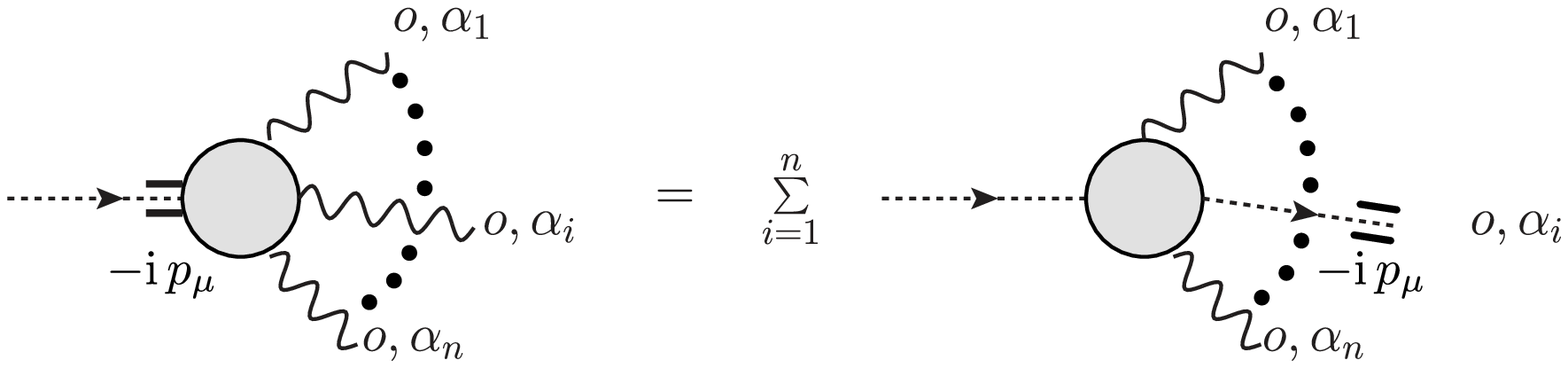}}
\ee
{In this equation, the double short line denotes multiplication by $- i p_\mu$, where $p_\mu$ is the momentum flowing into the vertex the leg is attached to. The label $o$ means that the legs in the diagrams are on-shell, while $\alpha_i$ represents the collection of indices (color, Lorentz, and so on) relative to the $i$-th leg. The wavy and dashed lines denote, respectively, the gluon and the ghost propagator.

Since
\be
-\eta_{\mu\nu}-\frac{\bar k_\mu k_\nu+k_\mu \bar k_\nu}{2(k\cdot\zeta)^2}=-\eta_{\mu\nu}-ik_\nu\frac{-i\bar k_\mu }{2(k\cdot\zeta)^2}-(-ik_\mu) \frac{i\bar k_\nu}{2(k\cdot\zeta)^2},
\ee
we have a relation for the propagators (\ref{propa1})--(\ref{propa3}):}
\be\label{Fig2}
\centering
\parbox{14cm}{\includegraphics[width=0.8\textwidth]{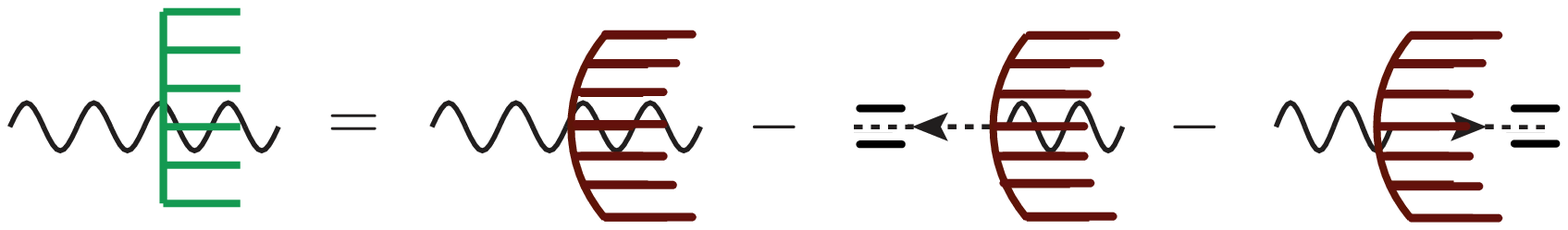}}
\ee
{Applying eq.\ (\ref{Fig2}) to one of the internal lines of a diagram with two intermediate physical particles, we get}
\be\label{Fig4}
\centering
\parbox{14cm}{\includegraphics[width=0.8\textwidth]{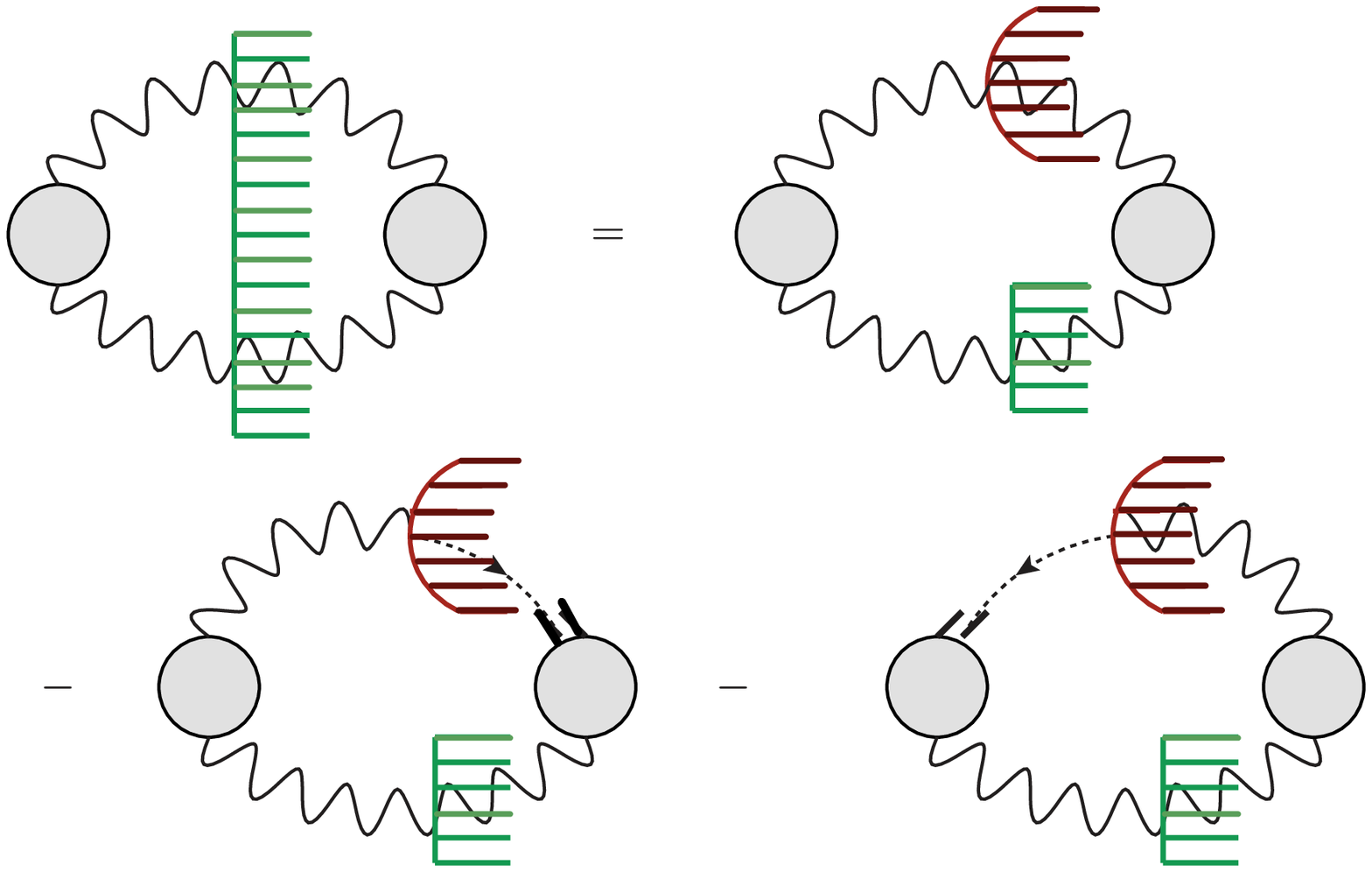}}
\ee
{However, each of the last two diagrams vanish due to the Ward identities (\ref{wardid}). In fact, these identities move the double short line from above to below the bubble, on the gluon line associated with the physical intermediate state. Then, the contraction of the transverse propagator (\ref{propa1}) with $- i k_\mu$ gives zero. Finally, applying equality (\ref{Fig2}) to the first line of (\ref{Fig4}), one obtains}
\be\label{Fig7}
\centering
\parbox{14cm}{\includegraphics[width=0.8\textwidth]{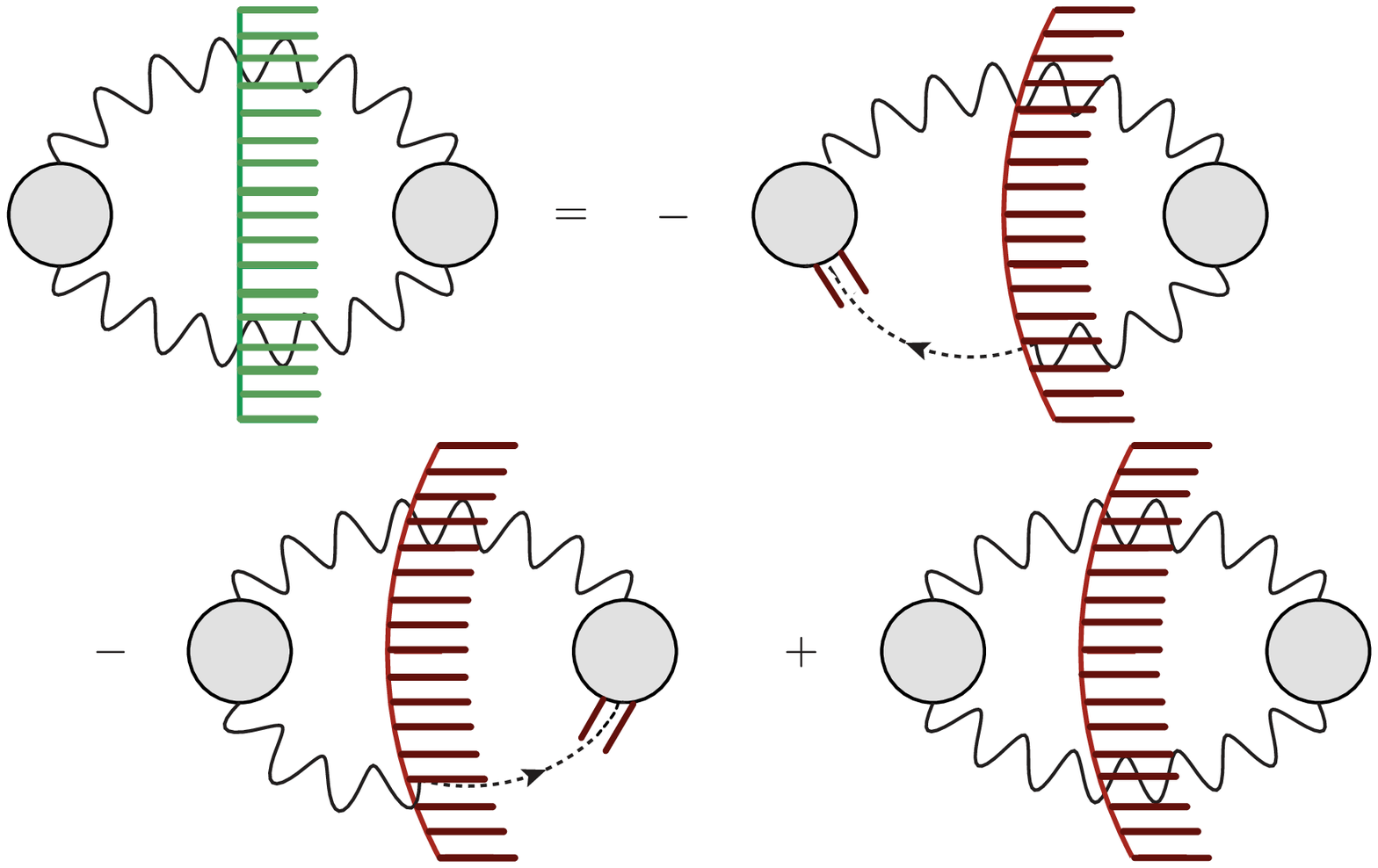}}
\ee
{Iterating the Ward identity (\ref{wardid}) twice, one obtains}
\be\label{Fig3}
\centering
\parbox{14cm}{\includegraphics[width=0.8\textwidth]{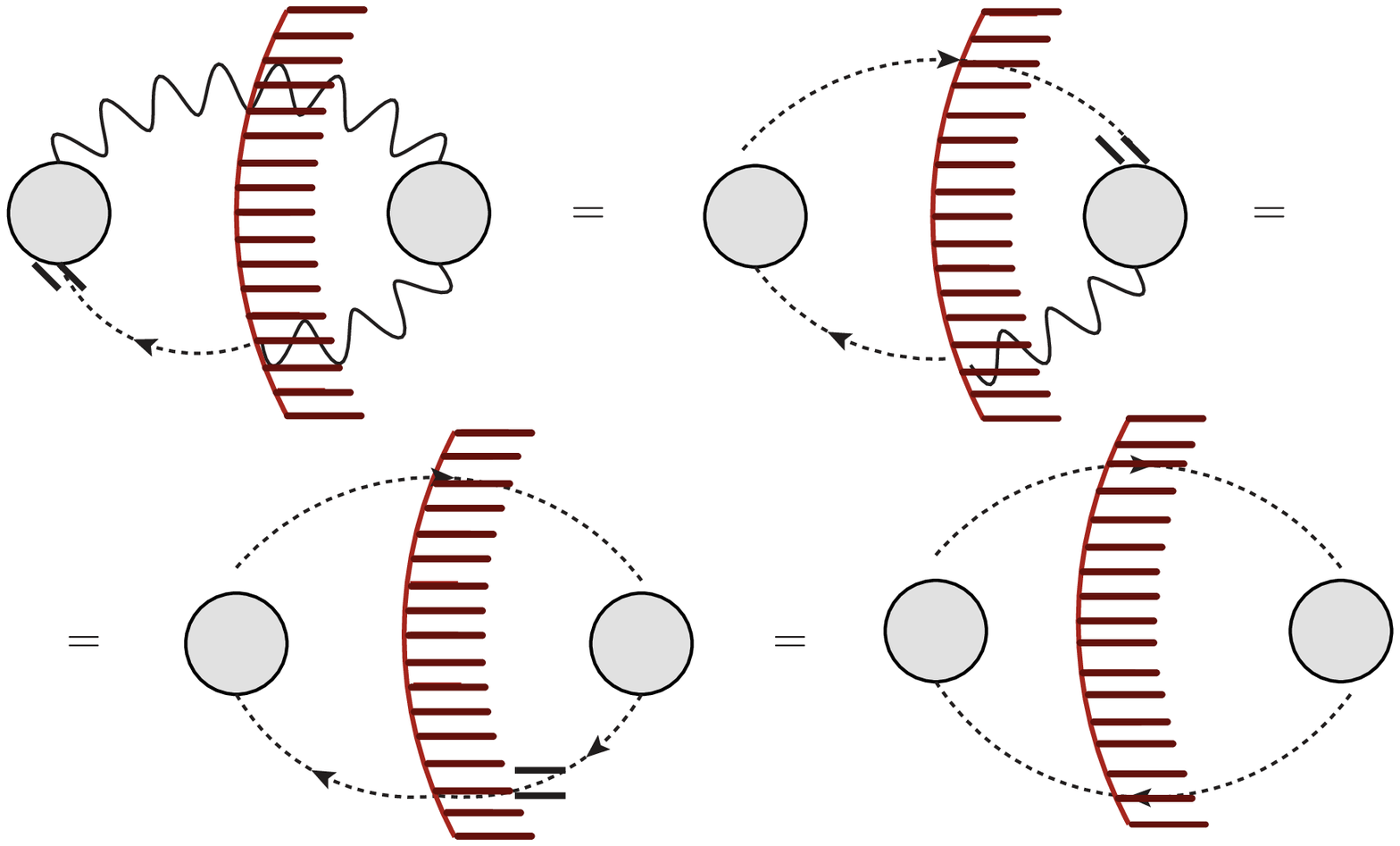}}
\ee
{where the last equality holds because
\be
D_+(k) \frac{i\bar k_\mu}{2(k\cdot\zeta)^2}\,i k^\mu = D_+(k)\,.
\ee
Noting that the first and second diagram on the right-hand side in eq.\ (\ref{Fig7}) coincide with the one in eq.\ (\ref{Fig3}), one gets the main result (\ref{mainlan}):}
\be\label{mainlan2}
\centering
\parbox{14cm}{\includegraphics[width=0.8\textwidth]{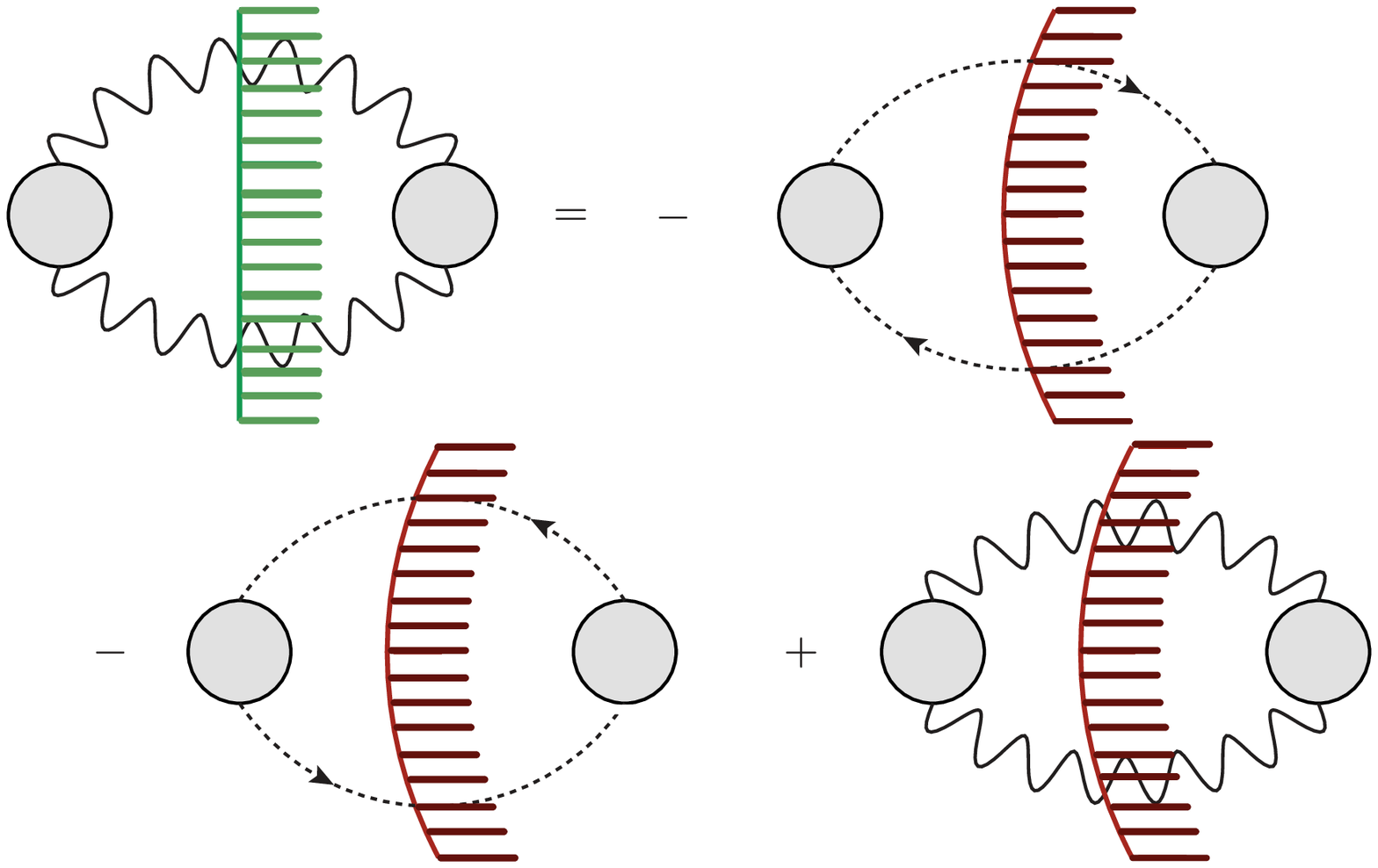}}
\ee

{
\subsection{Unitarity in weakly nonlocal gauge theories}

This conclusion for local gauge theories \cite{Lan16} is straightforwardly extended to weakly nonlocal gauge theories after introducing a nonlocal form factor ${\rm P}({\cal D}^2_M)$ in the denominator of the propagators for both the gauge bosons and the BRST ghosts \cite{Tom97}. Indeed, the form factor ${\rm P}({\cal D}^2_M)$ takes the value $1$ because it is evaluated on-shell according to the Cutkosky rules. Similarly, all the identities derived in the previous subsection are the same because they are obtained starting from the cut diagrams (\ref{propa1})--(\ref{propa3}), which are defined on-shell (see eq.\ (\ref{on-shell})).

\subsection{Unitarity in gauge theories and gravity with complex ghost pairs}

Let us now tackle the case of the theories under consideration in this paper, which are local higher-derivative theories with only complex conjugate ghosts pairs. The reason why we first presented the proof for the two-derivative and the weakly nonlocal cases is because they do not need major readjustments in the presence of complex pairs.

The propagators (\ref{propa1})--(\ref{propa3}) are modified introducing an extra factor $(k^4+M^4)/M^4$ in the denominators, for both gauge fields and BRST ghosts. One can easily see this in the case of BRST ghosts introducing the identity $(\det C)^{1/2}(\det C)^{-1/2}  = 1$ in the path integral and integrating out the $b$ ghost. As a result, the propagators for the gauge boson and the BRST ghosts $\bar C,C$ are the same as above but with an extra overall factor $M^4/(k^4+M^4)$. For gravity we have exactly the same, except that the analysis is slightly more complicated because the graviton is a spin-2 massless particle \cite{vanDam:1970vg}. 

In the higher-derivative case, we only have to take care of the real particles at the cuts, while the complex-conjugate ghosts do not contribute, according to the explicit computations in the first part of this paper. Therefore, in gauge theories, the only modifications to the above proof consists of changing the propagators of the local two-derivative theory to the higher-derivative one, namely, (\ref{NLPym}) and (\ref{GYM}) for the gauge fields and the BRST ghosts, respectively. However, in all the steps followed in section \ref{unitwoder} we are dealing with cut propagators that are not affected by the presence of the polynomial ${\rm P}(k^2)$ because it is replaced by $1$ when evaluated on-shell, i.e., for $k^2=0$ (see, for example, eq.\ (\ref{Pbox})). We can conclude that the above proof of perturbative unitarity applies without modifications to gauge theories with higher derivatives and complex conjugate ghosts.

Once one is convinced that for weakly nonlocal gauge theories and for gauge theories with complex conjugate poles the result is actually the same as for the local theories thanks to the Cutkosky identities, then one can finally assert that also higher-derivative gravity with complex pairs and nonlocal gravity are perturbatively unitary. From a technical point of view, the proof in section \ref{unitwoder} is slightly more involved due to the tensorial structure but, in the end, one gets the same diagrammatic result as in eq.\ (\ref{mainlan}). 

Summarizing, for local higher-derivative gauge or gravitational theories, in the loop amplitudes we have the sum of the contributions coming from gauge bosons or the graviton and the Fadeev--Popov ghosts. However, all the amplitudes are (non)analytically continued in the same way regardless of the particle species and according to the Cutkosky rules derived for the real scalar field up to an overall tensorial structure. Hence, unitarity is guaranteed by the Ward identities that imply the cancellation of the longitudinal degrees of freedom for the gauge bosons or the graviton with equal contributions coming from the BRST ghosts.

For completeness, we should finally mention that we can rewrite the six-derivative purely gravitational theory as the two-derivative Einstein--Hilbert theory coupled to matter fields. Such rewriting mimics the one doable in four-derivative Stelle's theory but more complicated than that because of the two extra derivatives. The outcome is a theory with the usual graviton field and two complex fields, each of them consisting of a massive complex spin-2 field and a complex massive scalar. The latter statement can be checked for the case of our theory implementing the usual expansion given in formula (20) of \cite{HigherDG}.

}

\section{Conclusions}\label{concl}

In this paper, we focused on the unitarity issue in a class of theories with propagators having the usual real poles (normal particles) and also complex conjugate ghostlike poles. At the tree level, the complex conjugate poles do not contribute to the discontinuity of the amplitudes, while in the loop amplitudes we achieved unitarity applying the same technique successfully implemented in nonlocal field theories \cite{Briscese:2021mob}. The latter procedure consists in integrating the internal energies in loop diagrams along special paths (in the complex hyperplane of dimension $L$, where $L$ is the number of loops) resulting from the deformation of the integrals along the imaginary axes when the external energies move from purely imaginary to real values. Making use of the above integration domain in the loop amplitudes, we derived the Cutkosky rules for higher derivative theories with complex conjugate ghosts. 

The key feature of these theories resides in the presence of complex conjugate poles. Indeed, the absence of such compensation would violate unitarity in our (non)analytic deformation. The nonanalyticity shows up when the complex poles overlap at the threshold for the production of the complex ghosts that, however, do not go on-shell because the imaginary part of the amplitudes is zero. The proof given here is very characteristic of complex conjugate poles and it does not work for real ghosts such as the one present in Stelle's theory (\ref{stelle}). 

Unitarity is proved using the Cutkosky rules and removing by hand the complex ghosts from the spectrum of the asymptotic states. 
Although this construction is partially a definition of the theory, its consistency is shown by the Cutkosky rules themselves: Once the ghosts are removed from the asymptotic spectrum, they cannot be generated again in loop amplitudes. 

The results obtained for the scalar field theory can be easily extended to gauge and gravitational theories making use of the Ward identities.

To complement our findings with a physical interpretation and future directions, 
 we compare the behaviour of conjugate ghosts with that of gluons. In quantum Yang--Mills theories and chromodynamics, it is often stated \cite{Sainapha:2019cfn} (and references therein) that the violation of reflection positivity is related to the confinement of gluons in the glueballs or of quarks in hadrons. Such statement is mainly based on the nonperturbative propagator for gluons. It turns out that the classical propagator for a complex pair of ghosts present in the theory studied in this paper is exactly the same of the nonperturbative propagator for the gluons up to the opposite overall sign. Indeed, we can rewrite the two-point function (\ref{ScalarG}) as
\be
G(k) 
= \frac{i M^4}{(k^2-m^2+i\epsilon) \left[ (k^2)^2+M^4 \right] }
=  \frac{i}{(k^2-m^2+i\epsilon) } -  i \frac{k^2}{(k^2)^2 + M^4 - m^2 k^2 } \, .
\label{ScalarGC}
\ee
For $m^2 < M^2$, the second propagator has complex conjugate poles and it exactly looks like the one that emerges in Yang--Mills theory at the nonperturbative level (see \cite{Holdom:2015kbf} and references therein). The only difference is the overall negative sign. The form of the quantum nonperturbative propagator for gluons is usually taken as evidence for the confinement of gluons in glueballs. Therefore, directly exporting such interpretation in higher-derivative theories, one might conjecture that also complex ghosts undergo some sort of confinement into ``\emph{ghostballs}.'' Such a mechanism would exclude the complex pairs from the asymptotic spectrum of free particles as found here and, most importantly, it would not undermine the stability of the vacuum. Clarifying the existence of ghost confinement in this theory of quantum gravity, as it may happen in other nonlocal models \cite{Frasca:2021iip}, will require further study.

\bigskip

\noindent\textbf{Note added.} While our paper was under completion, we became aware of ref.~\cite{Frasca:2022gdz}, where the topic of complex ghosts and their confinement is studied within nonlocal and Lee--Wick theories. These results might help to answer some of the questions posed in our concluding section.


\section*{Acknowledgments}

G.C.\ and L.M.\ are supported by grant PID2020-118159GB-C41 funded by MCIN/AEI/ 10.13039/501100011033. J.L.\ and L.M.\ are supported by the Basic Research Program of the Science, Technology, and Innovation Commission of Shenzhen Municipality (grant no.\ JCYJ20180302174206969).

\appendix 

\section{Triangle and square amplitudes}\label{TriBox}
We complete the unitarity analysis with two other relevant diagrams: the triangle diagram and the square diagram in the $\phi^3$ theory. 

\subsection{Discontinuity of the triangular amplitude}
For the triangular diagram, we assume that the particle with momentum $p_1$ corresponds to the initial state, while the particles with momenta $p_2$ and $p_3$ correspond to the final state. We have three internal propagators carrying momenta $k$, $p_1 - k$, and $p_3 - k$, respectively. Hence, the whole amplitude reads
\be
\mathcal{M}_{\rm triangle}\! &=& \! - \frac{\lambda^3}{2} 
\int_{\mathcal{C} \times \mathbb{R}^3}
\,
\frac{i \, d^4 k}{(2 \pi)^4}
\,
\frac{1}{k^2 - m^2 + i \epsilon}
\,
\frac{1}{k^4 + M^4}
\,
\frac{1}{(p_1 - k)^2 - m^2 + i \epsilon}
\,
\frac{1}{(p_1 - k)^4 + M^4}
\nonumber \\
&& \hspace{.7cm}
\times
\,
\frac{1}{(p_3 - k)^2 - m^2 + i \epsilon}
\,
\frac{1}{(p_3 - k)^4 + M^4} \, .
\label{Triangle}
\ee
The above triangle amplitude (\ref{Triangle}) has eighteen poles in the $k^0$-plane. The poles of all the propagators in (\ref{Triangle}) are located at
\bea
&&{\rm six \,\, poles \,\, of \,\, the \,\, first \,\, propagator} : \quad \left\{ 
\begin{array}{ll} \bar{k}_{1,2}^0 = \pm \sqrt{\bm{k}^2 + m^2 - i \epsilon} \, , \\
k_{1,2}^0 = \sqrt{\bm{k}^2 \pm i M^2} \, , \\
k_{3,4}^0 = - \sqrt{\bm{k}^2 \pm i M^2} \,, 
\end{array}\right.\\ \nonumber\\
&&{\rm six \,\, poles \,\, of \,\, the \,\, second \,\, propagator} : \quad  \left\{ 
\begin{array}{ll} \bar{k}_{3,4}^0 = p_1^0 \pm \sqrt{(\bm{p}_1-\bm{k})^2 + m^2 - i \epsilon} \, , \\
k_{5,6}^0 = p_1^0 + \sqrt{(\bm{p}_1-\bm{k})^2 \pm i M^2} \, ,  \\
k_{7,8}^0 = p_1^0 - \sqrt{(\bm{p}_1-\bm{k})^2 \pm i M^2} \, ,
\end{array}\right.\\ \nonumber\\
&&{\rm six \,\, poles \,\, of \,\, the \,\, third \,\, propagator} : \quad  \left\{ 
\begin{array}{ll} \bar{k}_{5,6}^0 = p_3^0 \pm \sqrt{(\bm{p}_3-\bm{k})^2 + m^2 - i \epsilon} \, ,  \\
k_{9,10}^0 = p_3^0 + \sqrt{(\bm{p}_3-\bm{k})^2 \pm i M^2} \, , \\
k_{11,12}^0 = p_3^0 - \sqrt{(\bm{p}_3-\bm{k})^2 \pm i M^2} \, . 
\end{array}\right. 
\eea

Among these eighteen poles, only six will move across the imaginary axis during the 
deformation of the path. Such poles are $\bar{k}_{4,6}^0$ and $k_{7,8,11,12}^0$. Therefore, the amplitude can be recast into the following form:
\be
\mathcal{M}_{\rm triangle} &=& - \frac{\lambda^3}{2} \Bigg[
\int_{\mathcal{I} \times \mathbb{R}^3} 
\,
\frac{i \, d^4 k}{(2 \pi)^4}
\,
\frac{1}{k^2 - m^2 + i \epsilon}
\,
\frac{1}{k^4 + M^4}
\,
\frac{1}{(p_1 - k)^2 - m^2 + i \epsilon} \nonumber \\
&& \hspace{1.7cm}
\times
\frac{1}{(p_1 - k)^4 + M^4}\,
\frac{1}{(p_3 - k)^2 - m^2 + i \epsilon}
\,
\frac{1}{(p_3 - k)^4 + M^4}\nonumber\\
&&  
+ (2 \pi i) \int_{\mathbb{R}^3}
\,
\frac{i \, d^3 \bm{k}}{(2 \pi)^4}
\,
\frac{\sigma \left[ \Re (\bar{k}_4^0) \right]}{(\bar{k}_4^0)^2 - \bm{k}^2 -m^2 + i \epsilon}
\,
\frac{1}{[(\bar{k}_4^0)^2 - \bm{k}^2]^2 + M^4}
\nonumber \\
&& \hspace{0.5cm}
\times \frac{1}{-2 \sqrt{(\bm{p}_1 - \bm{k})^2 + m^2 - i \epsilon}} 
\,
\frac{1}{m^4 + M^4} 
\,
\frac{1}{(p_3^0 - \bar{k}_4^0)^2 - (\bm{p}_3 - \bm{k})^2 - m^2 + i \epsilon}
\nonumber \\
&& \hspace{0.5cm}
\times 
\frac{1}{\left[ (p_3^0 - \bar{k}_4^0)^2 - (\bm{p}_3 - \bm{k})^2 \right]^2 + M^4}
\nonumber \\
&&   
+ (2 \pi i) \int_{\mathbb{R}^3}
\,
\frac{i \, d^3 \bm{k}}{(2 \pi)^4}
\,
\frac{\sigma \left[ \Re (k_7^0) \right]}{(k_7^0)^2 - \bm{k}^2 -m^2 + i \epsilon}
\,
\frac{1}{[(k_7^0)^2 - \bm{k}^2]^2 + M^4}
\,
\frac{1}{i M^2 - m^2} \nonumber \\
&& \hspace{0.5cm}
\times 
\frac{1}{-4 i M^2 \sqrt{(\bm{p}_1 - \bm{k})^2 + i M^2}}
\,
\frac{1}{(p_3^0 - k_7^0)^2 - (\bm{p}_3 - \bm{k})^2 - m^2 + i \epsilon}
\nonumber \\
&& \hspace{0.5cm}
\times 
\frac{1}{\left[ (p_3^0 - k_7^0)^2 - (\bm{p}_3 - \bm{k})^2 \right]^2 + M^4}
\nonumber \\
&&   
+ (2 \pi i) \int_{\mathbb{R}^3}
\,
\frac{i \, d^3 \bm{k}}{(2 \pi)^4}
\,
\frac{\sigma \left[ \Re (k_8^0) \right]}{(k_8^0)^2 - \bm{k}^2 -m^2 + i \epsilon}
\,
\frac{1}{[(k_8^0)^2 - \bm{k}^2]^2 + M^4}
\,
\frac{1}{- i M^2 - m^2}  \nonumber \\
&& \hspace{0.5cm}
\times \frac{1}{4 i M^2 \sqrt{(\bm{p}_1 - \bm{k})^2 - i M^2}}
\,
\frac{1}{(p_3^0 - k_8^0)^2 - (\bm{p}_3 - \bm{k})^2 - m^2 + i \epsilon}\nonumber \\
&& \hspace{0.5cm}
\times 
\frac{1}{\left[ (p_3^0 - k_8^0)^2 - (\bm{p}_3 - \bm{k})^2 \right]^2 + M^4}
\nonumber \\
&&   
+ (2 \pi i) \int_{\mathbb{R}^3}
\,
\frac{i \, d^3 \bm{k}}{(2 \pi)^4}
\,
\frac{\sigma \left[ \Re (\bar{k}_6^0) \right]}{(\bar{k}_6^0)^2 - \bm{k}^2 -m^2 + i \epsilon}
\,
\frac{1}{[(\bar{k}_6^0)^2 - \bm{k}^2]^2 + M^4}
\nonumber \\
&& \hspace{0.5cm}
\times 
\frac{1}{(p_1^0 - \bar{k}_6^0)^2 - (\bm{p}_1 - \bm{k})^2 - m^2 + i \epsilon}
\,
\frac{1}{\left[ (p_1^0 - \bar{k}_6^0)^2 - (\bm{p}_1 - \bm{k})^2 \right]^2 + M^4}
\nonumber \\
&& \hspace{0.5cm}
\times 
\frac{1}{-2 \sqrt{(\bm{p}_3 - \bm{k})^2 + m^2 - i \epsilon}} 
\,
\frac{1}{m^4 + M^4} 
\nonumber \\
&&   
+ (2 \pi i) \int_{\mathbb{R}^3}
\,
\frac{i \, d^3 \bm{k}}{(2 \pi)^4}
\,
\frac{\sigma \left[ \Re (k_{11}^0) \right]}{(k_{11}^0)^2 - \bm{k}^2 -m^2 + i \epsilon}
\,
\frac{1}{[(k_{11}^0)^2 - \bm{k}^2]^2 + M^4}\nonumber \\
&& \hspace{0.5cm}
\times 
\frac{1}{(p_1^0 - k_{11}^0)^2 - (\bm{p}_1 - \bm{k})^2 - m^2 + i \epsilon}
\,
\frac{1}{\left[ (p_1^0 - k_{11}^0)^2 - (\bm{p}_1 - \bm{k})^2 \right]^2 + M^4}\nonumber \\
&& \hspace{0.5cm}
\times 
\frac{1}{i M^2 - m^2} 
\,
\frac{1}{-4 i M^2 \sqrt{(\bm{p}_3 - \bm{k})^2 + i M^2}}
\nonumber \\
&&   
+ (2 \pi i) \int_{\mathbb{R}^3}
\,
\frac{i \, d^3 \bm{k}}{(2 \pi)^4}
\,
\frac{\sigma \left[ \Re (k_{12}^0) \right]}{(k_{12}^0)^2 - \bm{k}^2 -m^2 + i \epsilon}
\,
\frac{1}{[(k_{12}^0)^2 - \bm{k}^2]^2 + M^4}
\nonumber \\
&& \hspace{0.5cm}
\times 
\frac{1}{(p_1^0 - k_{12}^0)^2 - (\bm{p}_1 - \bm{k})^2 - m^2 + i \epsilon}
\,
\frac{1}{\left[ (p_1^0 - k_{12}^0)^2 - (\bm{p}_1 - \bm{k})^2 \right]^2 + M^4}
\nonumber \\
&& \hspace{0.5cm}
\times 
\frac{1}{-i M^2 - m^2} 
\,
\frac{1}{4 i M^2 \sqrt{(\bm{p}_3 - \bm{k})^2 - i M^2}} \Bigg] \, .
\ee
Notice that the term carrying $k_7^0$ and the term carrying $k_8^0$ are complex conjugate to each other when $\epsilon = 0$. Therefore, since the terms 
\be
&(k_7^0)^2 - \bm{k}^2 -m^2 \, , \quad &(p_3^0 - k_7^0)^2 - (\bm{p}_3 - \bm{k})^2 - m^2 \, , 
\nonumber \\
&(k_8^0)^2 - \bm{k}^2 -m^2 \, ,  \quad &(p_3^0 - k_8^0)^2 - (\bm{p}_3 - \bm{k})^2 - m^2 \, , 
\ee
are never zero for finite values of $M$, we can safely send $\epsilon \rightarrow 0$. Also the term with $k_{11}^0$ and the term with $k_{12}^0$ are complex conjugate. Since the first term of the amplitude $\mathcal{M}_{\rm triangle}$ is real (same arguments of (\ref{Again})), the imaginary part of the amplitude comes only from the second and the fifth term of $\mathcal{M}_{\rm triangle}$. The second term of $\mathcal{M}_{\rm triangle}$ can be expressed as
\be
&&  
(2 \pi i) \int_{\mathbb{R}^3}
\,
\frac{i \, d^3 \bm{k}}{(2 \pi)^4}
\,
\frac{\sigma \left[ \Re (\bar{k}_4^0) \right]}{(\bar{k}_4^0)^2 - \bm{k}^2 -m^2 + i \epsilon}
\,
\frac{1}{((\bar{k}_4^0)^2 - \bm{k}^2)^2 + M^4}
\,
\frac{1}{-2 \sqrt{(\bm{p}_1 - \bm{k})^2 + m^2 - i \epsilon}} \nonumber \\
&& \hspace{1.2cm}
\times
\,
\frac{1}{m^4 + M^4} 
\,
\frac{1}{(p_3^0 - \bar{k}_4^0)^2 - (\bm{p}_3 - \bm{k})^2 - m^2 + i \epsilon}
\,
\frac{1}{\left[ (p_3^0 - \bar{k}_4^0)^2 - (\bm{p}_3 - \bm{k})^2 \right]^2 + M^4}
\nonumber \\
&&
= 
(- 2 \pi i) \int_{\mathbb{R}^4}
\,
\frac{i \, d^4 k}{(2 \pi)^4}
\,
\frac{\sigma (k^0)}{k^2 -m^2 + i \epsilon}
\,
\frac{1}{k^4 + M^4}
\,
\sigma (p_1^0 - k^0)
\,
\delta((p_1 - k)^2 - m^2) \nonumber \\
&&  \hspace{1.8cm}
\times
\,
\frac{1}{m^4 + M^4} 
\,
\frac{1}{(p_3 - k)^2 - m^2 + i \epsilon}
\,
\frac{1}{(p_3 - k)^4 + M^4} \,,
\ee
while the fifth term of $\mathcal{M}_{\rm triangle}$ can be recast into
\be
&& 
(2 \pi i) \int_{\mathbb{R}^3}
\,
\frac{i \, d^3 \bm{k}}{(2 \pi)^4}
\,
\frac{\sigma \left[ \Re (\bar{k}_6^0) \right]}{(\bar{k}_6^0)^2 - \bm{k}^2 -m^2 + i \epsilon}
\,
\frac{1}{((\bar{k}_6^0)^2 - \bm{k}^2)^2 + M^4}\nonumber \\
&& \hspace{1.5cm}
\times
\frac{1}{(p_1^0 - \bar{k}_6^0)^2 - (\bm{p}_1 - \bm{k})^2 - m^2 + i \epsilon} 
\,
\frac{1}{\left[ (p_1^0 - \bar{k}_6^0)^2 - (\bm{p}_1 - \bm{k})^2 \right]^2 + M^4}
\nonumber \\
&& \hspace{1.5cm}
\times
\frac{1}{-2 \sqrt{(\bm{p}_3 - \bm{k})^2 + m^2 - i \epsilon}} 
\,
\frac{1}{m^4 + M^4} 
\nonumber \\
&& = 
(- 2 \pi i) \int_{\mathbb{R}^4}
\,
\frac{i \, d^4 k}{(2 \pi)^4}
\,
\frac{\sigma (k^0)}{k^2 -m^2 + i \epsilon}
\,
\frac{1}{k^4 + M^4}
\,
\frac{1}{(p_1 - k)^2 - m^2 + i \epsilon} \nonumber \\
&& \hspace{1.5cm}
\times
\,
\frac{1}{(p_1 - k)^4 + M^4}
\,
\sigma (p_3^0 - k^0)
\,
\delta((p_3 - k)^2 - m^2)
\,
\frac{1}{m^4 + M^4} \, . 
\ee
Therefore, the imaginary part of the amplitude can be expressed as
\be
\hspace{-.6cm}&&\mathcal{M}_{\rm triangle} - \mathcal{M}_{\rm triangle}^*\nn
\hspace{-.6cm}&&\qquad\qquad=
- \frac{\lambda^3}{2} \Bigg[
(- 2 \pi i)^2 \int_{\mathbb{R}^4}
\,
\frac{i \, d^4 k}{(2 \pi)^4}
\,
\left( \frac{\delta((p_3 - k)^2 - m^2)}{k^2 -m^2 + i \epsilon} + \frac{\delta(k^2 - m^2)}{(p_3 - k)^2 - m^2 + i \epsilon}
\right) \nn
\hspace{-.6cm}&&\qquad\qquad\qquad\times\sigma (k^0)  \,
\,
\sigma (p_1^0 - k^0)
\,
\delta((p_1 - k)^2 - m^2) 
\,
\frac{1}{k^4 + M^4}
\,
\frac{1}{m^4 + M^4} 
\,
\frac{1}{(p_3 - k)^4 + M^4}
\nonumber \\
\hspace{-.6cm}&&\qquad\qquad\quad+ (- 2 \pi i)^2 \int_{\mathbb{R}^4}
\,
\frac{i \, d^4 k}{(2 \pi)^4}
\,
\left( \frac{\delta((p_1 - k)^2 - m^2)}{k^2 -m^2 + i \epsilon} + \frac{\delta(k^2 - m^2)}{(p_1 - k)^2 - m^2 + i \epsilon}
\right) 
\,
\sigma (k^0) 
\nonumber \\
\hspace{-.6cm}&&\qquad\qquad\qquad
\times
\,
\sigma (p_3^0 - k^0)
\,
\delta((p_3 - k)^2 - m^2) 
\,
\frac{1}{k^4 + M^4}
\,
\frac{1}{(p_1 - k)^4 + M^4}
\,
\frac{1}{m^4 + M^4} \Bigg].
\ee
Since the kinematical condition requires that $p_1^2 > p_2^2 + p_3^2$, only the propagators involving the momenta $k$ and $p_1 - k$ can go on-shell together. Hence, finally we get
\be
\hspace{-.5cm}
\mathcal{M}_{\rm triangle} - \mathcal{M}_{\rm triangle}^* & = & - \frac{\lambda^3}{2} 
(- 2 \pi i)^2 \int_{\mathbb{R}^4}
\,
\frac{i \, d^4 k}{(2 \pi)^4}
\,
\frac{\delta(k^2 - m^2)}{(p_3 - k)^2 - m^2 + i \epsilon} 
\,
\delta((p_1 - k)^2 - m^2) \nn 
\hspace{-.5cm}&&\quad\times
\sigma (k^0) 
\,
\sigma (p_1^0 - k^0) 
\,
\frac{1}{k^4 + M^4}
\,
\frac{1}{m^4 + M^4} 
\,
\frac{1}{(p_3 - k)^4 + M^4} \, .
\ee

\subsection{Discontinuity of the square amplitude}
%
%
We assume the initial state to consist of two particles with momenta $p_1$ and $p_2$, while the final state consists of two particles with momenta $p_3$ and $p_4$.
 We have four internal propagators carrying momenta $k$, $p_1 - k$, $p_3 - k$, and $p_1 + p_2 - k$, respectively. Therefore, the whole amplitude reads
\be
\mathcal{M}_{\rm box} &=& - \frac{\lambda^3}{2} 
 \int_{\mathcal{C} \times \mathbb{R}^3}
\,
\frac{i \, d^4 k}{(2 \pi)^4}
\,
\frac{1}{k^2 - m^2 + i \epsilon}
\,
\frac{1}{k^4 + M^4}
\,
\frac{1}{(p_1 - k)^2 - m^2 + i \epsilon}
\,
\frac{1}{(p_1 - k)^4 + M^4}\nonumber \\
&& \qquad
\times
\,
\frac{1}{(p_3 - k)^2 - m^2 + i \epsilon}
\,
\frac{1}{(p_3 - k)^4 + M^4}
\,
\frac{1}{(p_1 + p_2 - k)^2 - m^2 + i \epsilon}
\nonumber \\
&& \qquad
\times
\frac{1}{(p_1 + p_2 - k)^4 + M^4}.
\label{BOX}
\ee
The amplitude (\ref{BOX}) has a total of 24 poles in the $k^0$-plane:
\bea
\hspace{-.5cm}&&{\rm the \,\, six \,\, poles \,\, of \,\, the \,\, first \,\, propagator}: \quad   \left\{ 
\begin{array}{ll} \bar{k}_{1,2}^0 = \pm \sqrt{\bm{k}^2 + m^2 - i \epsilon}\, , \\
k_{1,2}^0 = \sqrt{\bm{k}^2 \pm i M^2} \, , \\
k_{3,4}^0 = - \sqrt{\bm{k}^2 \pm i M^2} \, ,
\end{array}\right. \\
\hspace{-.5cm}&&{\rm six \,\, poles \,\, of \,\, the \,\, second \,\, propagator}: \quad   \left\{ 
\begin{array}{ll} \bar{k}_{3,4}^0 = p_1^0 \pm \sqrt{(\bm{p}_1-\bm{k})^2 + m^2 - i \epsilon} \, , \\
k_{5,6}^0 = p_1^0 + \sqrt{(\bm{p}_1-\bm{k})^2 \pm i M^2} \, , \\
k_{7,8}^0 = p_1^0 - \sqrt{(\bm{p}_1-\bm{k})^2 \pm i M^2} \, , 
\end{array}\right. \\
\hspace{-.5cm}&&{\rm six \,\, poles \,\, of \,\, the \,\, third \,\, propagator} : \quad  \left\{ 
\begin{array}{ll} \bar{k}_{5,6}^0 = p_3^0 \pm \sqrt{(\bm{p}_3-\bm{k})^2 + m^2 - i \epsilon} \, , \\
k_{9,10}^0 = p_3^0 + \sqrt{(\bm{p}_3-\bm{k})^2 \pm i M^2} \, , \\
k_{11,12}^0 = p_3^0 - \sqrt{(\bm{p}_3-\bm{k})^2 \pm i M^2} \, ,
\end{array}\right. \\
\hspace{-.5cm}&&{\rm six \,\, poles \,\, of \,\, the \,\, fourth \,\, propagator}: \quad  \left\{ 
\begin{array}{ll} \bar{k}_{7,8}^0 = p_1^0 + p_2^0 \pm \sqrt{(\bm{p}_1 +\bm{p}_2 - \bm{k})^2 + m^2 - i \epsilon} \, , \\
k_{13,14}^0 = p_1^0 + p_2^0 + \sqrt{(\bm{p}_1 +\bm{p}_2 - \bm{k})^2 \pm i M^2} \, , \\
k_{15,16}^0 = p_1^0 + p_2^0 - \sqrt{(\bm{p}_1 +\bm{p}_2 - \bm{k})^2 \pm i M^2} \, .
\end{array}\right. \nonumber\\
\eea
Out of the 24 poles, eight move across the imaginary axis during the continuation. Such poles are $\bar{k}_{4,6,8}^0$ and $k_{7,8,11,12,15,16}^0$. Hence, the amplitude can be divided into nine terms. Similarly to the triangular amplitude, the terms that come from the same pair are indeed complex conjugate to each other for $\epsilon = 0$, and we can safely send $\epsilon \rightarrow0$ as we stated in the section about the triangle amplitude. Moreover, only the propagators with momenta $k$ and $p_1 + p_2 - k$ can go on-shell together because of the kinematical condition. Therefore, the imaginary part of the amplitude can be recast into
\be
\mathcal{M}_{\rm box} - \mathcal{M}_{\rm box}^* &=& - \frac{\lambda^4}{2} 
(- 2 \pi i)^2 \int_{\mathbb{R}^4}
\,
\frac{i \, d^4 k}{(2 \pi)^4}
\,
\frac{\delta(k^2 - m^2)}{(p_1 - k)^2 - m^2 + i \epsilon} 
\,
\frac{\delta[(p_1 + p_2 - k)^2 - m^2]}{(p_3 - k)^2 - m^2 + i \epsilon} \nonumber \\
&& \qquad
\times \sigma (k^0)
\,
\sigma (p_1^0 + p_2^0 - k^0) 
\,
\frac{1}{k^4 + M^4}
\,
\frac{1}{(p_1 - k)^4 + M^4} 
\nn
&&\qquad
\times
\frac{1}{(p_3 - k)^4 + M^4} 
\,
\frac{1}{m^4 + M^4} \, .
\ee

\section{Explicit derivation of the discontinuity of the amplitude at two-loops} \label{M22C}
In this section, we explicitly compute the discontinuity of the amplitude (\ref{Emme22}).
First of all, we extract the imaginary part of $\mathcal{M}_{22}$. In the limit $\epsilon \rightarrow 0$, only the first propagator
in (\ref{Emme22}) can be singular, namely  
\be\nonumber
\frac{1}{(\bar{k}_{2;3}^0)^2 - \bm{k}_2^2 - m^2 + i \epsilon} \, .
\ee 
Hence, using the Sokhotski--Plemelj formula, we get
\be
\hspace{-1cm}&&\lim_{\epsilon \rightarrow 0} \frac{1}{(\bar{k}_{2;3}^0)^2 - \bm{k}_2^2 - m^2 + i \epsilon}\nn
\hspace{-1cm}&&\qquad\qquad=  \frac{1}{\bar{k}_{2;3}^0 + \sqrt{\bm{k}_2^2 + m^2}} 
\left[{\rm PV} \frac{1}{\bar{k}_{2;3}^0 - \sqrt{\bm{k}_2^2 + m^2}} - i \pi \delta(\bar{k}_{2;3}^0 - \sqrt{\bm{k}_2^2 + m^2}) 
\right] 
\nonumber \\
\hspace{-1cm}&&\qquad\qquad=  \frac{1}{\bar{k}_{2;3}^0 + \sqrt{\bm{k}_2^2 + m^2}} {\rm PV} \frac{1}{\bar{k}_{2;3}^0 - \sqrt{\bm{k}_2^2 + m^2}}  - i \pi \frac{\delta(\bar{k}_{2;3}^0 - \sqrt{\bm{k}_2^2 + m^2})}{2 \sqrt{\bm{k}_2^2 + m^2}} \, .
\ee
Using the above limit in (\ref{Emme22}), the imaginary part of the amplitude reads
\be
&& {\rm Im} \,\mathcal{M}_{22} = 
-\frac{\lambda^2 M^{12}}{2} \int_{\mathbb{R}^3} \frac{id^3 \bm{k}_1}{(2\pi)^4} (2\pi i) \int_{\mathbb{R}^3}  \frac{i d^3 \bm{k}_2}{(2\pi)^4} (2\pi i) 
 (- \pi) \frac{\delta(\bar{k}_{2;3}^0 - \sqrt{\bm{k}_2^2 + m^2})}{2 \sqrt{\bm{k}_2^2 + m^2}}
\nonumber \\
&& 
\hspace{4cm} \times 
 \frac{1}{[(\bar{k}_{2;3}^0)^2-\bm{k}_2^2]^2+M^4}
\,\,
 \frac{1}{-2\sqrt{\bm{k}_1^2 +m^2 -i \epsilon}} \nonumber  \\
&& \hspace{4cm} 
\times 
 \frac{1}{\left[\left(p^0 -\bar{k}_{2;3}^0 -\sqrt{(\bm{k}_1+\bm{k}_2)^2 +m^2 -i\epsilon}\right)^2 - \bm{k}_1^2\right]^2 +M^4} \nonumber \\
&& 
\hspace{4cm} 
\times 
\frac{1}{-2\sqrt{(\bm{k}_1+\bm{k}_2)^2 +m^2 -i\epsilon}}
\,\,
\frac{\sigma[\Re(\bar{k}_{2;3}^0)]}{m^4 +M^4} \, .
\ee
Replacing $\bar{k}_{2;3}^0$ in the Dirac delta distribution above, 
\be
\hspace{-.8cm}{\rm Im} \,\mathcal{M}_{22} &=& 
-\frac{\lambda^2 M^{12}}{2} \int_{\mathbb{R}^3} \frac{i { d^3 \bm{k}_1} }{(2\pi)^4} \int_{\mathbb{R}^3}  \frac{i d^3 \bm{k}_2}{(2\pi)^4} (- 2\pi i)^2 
 (- \pi) \nn
\hspace{-.8cm}&&\quad\times\frac{\delta \!\left[p^0 - \sqrt{\bm{k}_1^2 + m^2} - \sqrt{(\bm{k}_1+\bm{k}_2)^2 + m^2} - \sqrt{\bm{k}_2^2 + m^2} \right]}{2 \sqrt{\bm{k}_2^2 + m^2}} 
 \nn
\hspace{-.8cm}&&\quad\times
\frac{1}{m^4 +M^4}  \,\, 
{\frac{1}{2\sqrt{\bm{k}_1^2 +m^2 -i\epsilon}} }
\,\, \frac{1}{m^4 +M^4} \nn
\hspace{-.8cm}&&\quad\times
\frac{1}{2\sqrt{(\bm{k}_1+\bm{k}_2)^2 +m^2 -i\epsilon}}
\,\,
\frac{\sigma\! \left[p^0 - \sqrt{\bm{k}_1^2 + m^2} - \sqrt{(\bm{k}_1+\bm{k}_2)^2 + m^2} \right]}{m^4 +M^4}. 
\ee
We can now rewrite the integral $d^3\bm{k}_1/(2\sqrt{ \bm{k}_1^2 +m^2 -i\epsilon})$ as $d^4 k_1 \sigma(k_1^0) \delta(k_1^2 - m^2)$, so that the imaginary part becomes
\be
\hspace{-.8cm}{\rm Im} \,\mathcal{M}_{22} &=&
-\frac{\lambda^2 }{2}  \frac{M^{12}}{(m^4 + M^4)^3}  \int_{\mathbb{R}^4} \frac{ i { d^4 k_1} }{(2\pi)^4} \int_{\mathbb{R}^3}  \frac{i d^3 \bm{k}_2}{(2\pi)^4} (- 2\pi i)^2(- \pi) \nn
\hspace{-.8cm}&&\qquad\times
  \frac{\delta\! \left[p^0 - k_1^0 - \sqrt{(\bm{k}_1+\bm{k}_2)^2 + m^2} - \sqrt{\bm{k}_2^2 + m^2} \right]}{2 \sqrt{\bm{k}_2^2 + m^2}}  \,\,
\sigma(k_1^0)\, \delta(k_1^2 - m^2) \nn
\hspace{-.8cm}&&\qquad\times
\frac{1}{2\sqrt{(\bm{k}_1+\bm{k}_2)^2 +m^2 -i\epsilon}}
\,\,
\sigma\! \left[p^0 - k_1^0 - \sqrt{(\bm{k}_1+\bm{k}_2)^2 + m^2} \right]\, .
\ee
Similarly, we rewrite the three-dimensional integral 
\be\nonumber
d^3 \bm{k}_2 \,\, \frac{1}{2\sqrt{ (\bm{k}_1+\bm{k}_2)^2  +m^2 -i \epsilon}} 
\, \sigma \left[ p^0 - k_1^0 - \sqrt{(\bm{k}_1+\bm{k}_2)^2 + m^2} \right]
,
\ee
as the-dimensional integral
\be\nonumber
  d^4 k_2 
\,\, \sigma(p^0 - k_1^0 -k_2^0) 
\,
\delta[(p - k_1 - k_2)^2 - m^2]
 \, \sigma \left[p^0 - k_1^0 - (p^0 - k_1^0 - k_2^0) \right] \, ,
\ee
 and the imaginary part turns into
\be
{\rm Im} \,\mathcal{M}_{22} &=&
-\frac{\lambda^2 }{2}  \frac{M^{12}}{(m^4 + M^4)^3}  \int_{\mathbb{R}^4} \frac{id^4 k_1}{(2\pi)^4} \int_{\mathbb{R}^4}  \frac{i d^4 k_2}{(2\pi)^4} (- 2\pi i)^2 
(- \pi)\nn
&&\qquad\times \frac{\delta \left[p^0 - k_1^0 -  (p^0 - k_1^0 - k_2^0) - \sqrt{\bm{k}_2^2 + m^2} \right]}{2 \sqrt{\bm{k}_2^2 + m^2}}
\,\,
\sigma(k_1^0) 
\,\, 
\delta(k_1^2 - m^2)\nn
&&\qquad\times\,
\sigma(p^0 - k_1^0 -k_2^0) \,\,\delta[(p - k_1 - k_2)^2 - m^2]
\,\,
\sigma\!\left[p^0 - k_1^0 - (p^0 - k_1^0 - k_2^0) \right]
\nonumber  \\
&=&
-\frac{\lambda^2 }{2}  \frac{M^{12}}{(m^4 + M^4)^3}  \int_{\mathbb{R}^4} \frac{id^4 k_1}{(2\pi)^4} \int_{\mathbb{R}^4}  \frac{i d^4 k_2}{(2\pi)^4} (- 2\pi i)^2 
{(- \pi) \delta(k_2^2 - m^2) }
\nonumber  \\
&& \qquad\times \,\sigma(k_1^0) 
\,\, 
\delta(k_1^2 - m^2)
\,\,
\sigma(p^0 - k_1^0 -k_2^0) 
\,\, 
\delta[(p - k_1 - k_2)^2 - m^2]
\,\,
\sigma(k_2^0) \, . 
\ee
Finally, we have
\be
{\rm Disc} \,\mathcal{M} &=& 2i\, {\rm Im} \,\mathcal{M} 
= 2i\, {\rm Im} \,\mathcal{M}_{22} \nonumber \\
&=& -\frac{\lambda^2 }{2}  \frac{M^{12}}{(m^4 + M^4)^3}  \int_{\mathbb{R}^4} \frac{id^4 k_1}{(2\pi)^4} \int_{\mathbb{R}^4}  \frac{i d^4 k_2}{(2\pi)^4} (- 2\pi i)^3 \,\,
\sigma(k_1^0) 
\,\, 
\delta(k_1^2 - m^2)
 \nonumber  \\
&& \qquad
\times 
\,
\sigma (k_2^0)
\,\, 
\delta(k_2^2 - m^2)
\,\,
\sigma(p^0 - k_1^0 -k_2^0) 
\,\, 
\delta[(p - k_1 - k_2)^2 - m^2]\,  . 
\ee



\begin{thebibliography}{99}

%
%
%
%
%
%
%
%
%

%
%
%
%
%
%





\bibitem{Stelle} \au{K.S}{Stelle}, \tia{Renormalization of higher-derivative quantum gravity} \doin{10.1103/PhysRevD.16.953}{Phys.\ Rev.}{D}{16}{953}{1977}.







\bibitem{frts82}
\au{E.S}{Fradkin} and \au{A.A}{Tseytlin}, \tia{Renormalizable asymptotically free quantum theory of gravity} \doin{10.1016/0550-3213(82)90444-8}{Nucl.\ Phys.}{B}{201}{469}{1982}

\bibitem{avbar86} \au{I.G}{Avramidi} and \au{A.O}{Barvinsky}, \tia{Asymptotic freedom in higher-derivative quantum gravity} \doin{10.1016/0370-2693(85)90248-5}{Phys.\ Lett.}{B}{159}{269}{1985}.

\bibitem{Avr86} \au{I.G}{Avramidi}, \tia{Asymptotic behavior of the quantum theory of gravity with higher derivatives} Sov.\ J.\ Nucl.\ Phys.\ {\bf 44} (1986) 255.



%
%
%
%
%
%
%






\bibitem{Shapirobook} \au{I.L}{Buchbinder}, \au{S.D}{Odintsov} and \au{I.L}{Shapiro}, \book{Effective Action in Quantum Gravity}{CRC Press}{Bristol}{UK}{1992}.



 



%
%
%
%
%
%
%
%
%
%
%
%


  







\bibitem{CLOP}
\au{R.E}{Cutkosky}, \au{P.V}{Landshoff}, \au{D.I}{Olive} and \au{J.C}{Polkinghorne}, \tia{A non-analytic $S$-matrix} \doin{10.1016/0550-3213(69)90169-2}{Nucl.\ Phys.}{B}{12}{281}{1969}.

\bibitem{LW1} 
\au{T.D}{Lee} and \au{G.C}{Wick}, \tia{Negative metric and the unitarity of the $S$-matrix} \doin{10.1016/0550-3213(69)90098-4}{Nucl.\ Phys.}{B}{9}{209}{1969}.

\bibitem{LW2} 
\au{T.D}{Lee} and \au{G.C}{Wick}, \tia{Finite theory of quantum electrodynamics} \doin{10.1103/PhysRevD.2.1033}{Phys.\ Rev.}{D}{2}{1033}{1970}.



\bibitem{AnselmiPiva1}
\au{D}{Anselmi} and \au{M}{Piva}, \tia{A new formulation of Lee--Wick quantum field theory} \doij{10.1007/JHEP06(2017)066}{JHEP}{1706}{066}{2017} [\arX{1703.04584}].

\bibitem{AnselmiPiva2}
\au{D}{Anselmi} and \au{M}{Piva}, \tia{Perturbative unitarity of Lee--Wick quantum field theory} \doin{10.1103/PhysRevD.96.045009}{Phys.\ Rev.}{D}{96}{045009}{2017} [\arX{1703.05563}].

\bibitem{AnselmiPiva3}
\au{D}{Anselmi}, \tia{Fakeons and Lee--Wick models} \doij{10.1007/JHEP02(2018)141}{JHEP}{1802}{141}{2018} [\arX{1801.00915}].



\bibitem{ClassicalPrescription1} \au{D}{Anselmi} \tia{Fakeons, microcausality and the classical limit of quantum gravity} \doinn{10.1088/1361-6382/ab04c8}{Class.\ Quantum Gravity}{36}{065010}{2019} [\arX{1809.05037}].

\bibitem{ClassicalPrescription2} \au{D}{Anselmi} and \au{A}{Marino},
\tia{Fakeons and microcausality: light cones, gravitational waves and the Hubble constant} \doinn{10.1088/1361-6382/ab78d2}{Class.\ Quantum Grav.}{37}{095003}{2020} [\arX{1909.12873}].





\bibitem{LM-Sh} \au{L}{Modesto} and \au{I.L}{Shapiro}, \tia{Super-renormalizable quantum gravity with complex ghosts} \doin{10.1016/j.physletb.2016.02.021}{Phys.\ Lett.}{B}{755}{279}{2016} [\arX{1512.07600}].

\bibitem{ModestoLW} \au{L}{Modesto}, \tia{Super-renormalizable or finite Lee--Wick quantum gravity} \doin{10.1016/j.nuclphysb.2016.06.004}{Nucl.\ Phys.}{B}{909}{584}{2016} [\arX{1602.02421}].

\bibitem{shapiro3}
\au{M}{Asorey}, \au{J.L}{L\'opez} and \au{I.L}{Shapiro}, \tia{Some remarks on high derivative quantum gravity} \doin{10.1142/S0217751X97002991}{Int.\ J.\ Mod.\ Phys.}{A}{12}{5711}{1997} [\oarX{hep-th/9610006}].

\bibitem{ConCos1} \au{L}{Modesto} and \au{G}{Calcagni}, \tia{Early universe in quantum gravity} \arX{2206.06384}.
\bibitem{ConCos2} \au{G}{Calcagni} and \au{L}{Modesto}, \tia{Testing quantum gravity with primordial gravitational waves} \arX{2206.07066}.



















\bibitem{Briscese:2018oyx} 
\au{F}{Briscese} and \au{L}{Modesto}, \tia{Cutkosky rules and perturbative unitarity in Euclidean nonlocal quantum field theories} \doin{10.1103/PhysRevD.99.104043}{Phys.\ Rev.}{D}{99}{104043}{2019} [\arX{1803.08827}].

\bibitem{Briscese:2021mob}
\au{F}{Briscese} and \au{L}{Modesto}, \tia{Non-unitarity of Minkowskian non-local quantum field theories} \doin{10.1140/epjc/s10052-021-09525-7}{Eur.\ Phys.\ J.}{C}{81}{730}{2021} [\arX{2103.00353}].









\bibitem{Krasnikov}
\au{N.V}{Krasnikov}, \tia{Nonlocal gauge theories} \doinn{10.1007/BF01017588}{Theor.\ Math.\ Phys.}{73}{1184}{1987} [\ndoinn{http://www.mathnet.ru/php/archive.phtml?wshow=paper&jrnid=tmf&paperid=5624&option_lang=eng}{Teor.\ Mat.\ Fiz.}{73}{235}{1987}].

\bibitem{kuzmin}
\au{Yu.V}{Kuz'min}, \tia{The convergent nonlocal gravitation} Sov.\ J.\ Nucl.\ Phys.\ {\bf 50} (1989) 1011 [Yad.\ Fiz.\ {\bf 50} (1989) 1630].



\bibitem{modesto}
\au{L}{Modesto}, \tia{Super-renormalizable quantum gravity} \doin{10.1103/PhysRevD.86.044005}{Phys.\ Rev.}{D}{86}{044005}{2012} [\arX{1107.2403}].
  
  
\bibitem{review} 
\au{L}{Modesto} and \au{L}{Rachwa\l}, \tia{Nonlocal quantum gravity: a review} \doin{10.1142/S0218271817300208}{Int.\ J.\ Mod.\ Phys.}{D}{26}{1730020}{2017}.


   \bibitem{modestoLeslaw}
\au{L}{Modesto} and \au{L}{Rachwa\l}, \tia{Super-renormalizable and finite gravitational theories} \doin{10.1016/j.nuclphysb.2014.10.015}{Nucl.\ Phys.}{B}{889}{228}{2014} [\arX{1407.8036}].






\bibitem{Calcagni:2014vxa}
\au{G}{Calcagni} and \au{L}{Modesto}, \tia{Nonlocal quantum gravity and M-theory} \doin{10.1103/PhysRevD.91.124059}{Phys.\ Rev.}{D}{91}{124059}{2015} [\arX{1404.2137}].






\bibitem{Universally} 
\au{L}{Modesto} and \au{L}{Rachwa\l}, \tia{Universally finite gravitational and gauge theories} \doin{10.1016/j.nuclphysb.2015.09.006}{Nucl.\ Phys.}{B}{900}{147}{2015} [\arX{1503.00261}].


\bibitem{FiniteGaugeTheory} 
\au{L}{Modesto}, \au{M}{Piva} and \au{L}{Rachwa\l}, \tia{Finite quantum gauge theories} \doin{10.1103/PhysRevD.94.025021}{Phys.\ Rev.}{D}{94}{025021}{2016} [\arX{1506.06227}].






\bibitem{HigherDG}
\au{A}{Accioly}, \au{A}{Azeredo} and \au{H}{Mukai}, \tia{Propagator, tree-level unitarity and effective nonrelativistic potential for higher-derivative gravity theories in $D$ dimensions} \doinn{10.1063/1.1415743}{J.\ Math.\ Phys.\ (N.Y.)}{43}{473}{2002}.






\bibitem{Rachwal:2021bgb}
\au{L}{Rachwa\l}, \au{L}{Modesto}, \au{A}{Pinzul} and \au{I.L}{Shapiro}, \tia{Renormalization group in six-derivative quantum gravity} \doin{10.1103/PhysRevD.104.085018}{Phys.\ Rev.}{D}{104}{085018}{2021} [\arX{2104.13980}].


%


%
%

%
%







\bibitem{Anselmi:2022lbw}
\au{D}{Anselmi}, \tia{Purely virtual particles in quantum gravity, inflationary cosmology and collider physics} \doinn{10.3390/sym14030521}{Symmetry}{14}{521}{2022} [\arX{2203.02516}].


\bibitem{Anselmi:2021hab}
\au{D}{Anselmi}, \tia{Diagrammar of physical and fake particles and spectral optical theorem} \doij{10.1007/JHEP11(2021)030}{JHEP}{2111}{030}{2021} [\arX{2109.06889}].



\bibitem{Lavrov:2019nuz}
\au{P.M}{Lavrov} and \au{I.L}{Shapiro}, \tia{Gauge invariant renormalizability of quantum gravity} \doin{10.1103/PhysRevD.100.026018}{Phys.\ Rev.}{D}{100}{026018}{2019} [\arX{1902.04687}].


\bibitem{Tom97} \au{E.T}{Tomboulis}, \tia{Super-renormalizable gauge and gravitational theories} \oarX{hep-th/9702146}.

\bibitem{Lan16} \au{S}{Lanza}, \href{https://etd.adm.unipi.it/t/etd-06202016-152710}{\cob\emph{Renormalizability and Finiteness of Nonlocal Quantum Gravity}}, M.Sc.\ thesis (U.\ Pisa, Italy, 2016).


\bibitem{vanDam:1970vg} \au{H}{van Dam} and \au{M.J.G}{Veltman}, \tia{Massive and massless Yang--Mills and gravitational fields} \doin{10.1016/0550-3213(70)90416-5}{Nucl.\ Phys.}{B}{22}{397}{1970}.

\bibitem{Sainapha:2019cfn} \au{T}{Sainapha}, \tia{Gribov ambiguity} \arX{1910.11659}.

\bibitem{Holdom:2015kbf} \au{B}{Holdom} and \au{J}{Ren}, \tia{QCD analogy for quantum gravity} \doin{10.1103/PhysRevD.93.124030}{Phys.\ Rev.}{D}{93}{124030}{2016} [\arX{1512.05305}].


\bibitem{Frasca:2021iip} \au{M}{Frasca}, \au{A}{Ghoshal} and \au{N}{Okada}, \tia{Confinement and renormalization group equations in string-inspired nonlocal gauge theories} \doin{10.1103/PhysRevD.104.096010}{Phys.\ Rev.}{D}{104}{096010}{2021} [\arX{2106.07629}].
\bibitem{Frasca:2022gdz} \au{M}{Frasca}, \au{A}{Ghoshal} and \au{A.S}{Koshelev}, \tia{Confining the complex ghosts out} \arX{2207.06394}.
\end{thebibliography}
\end{document}